\documentclass[12pt,a4paper]{article}    
\usepackage{jheppub}
\makeatletter
\usepackage[svgnames,table]{xcolor}
\usepackage{booktabs,caption}
\usepackage{tabulary}
\newcolumntype{K}[1]{>{\centering\arraybackslash}p{#1}}
\usepackage{fontenc}
\usepackage[T1]{fontenc} 
\usepackage{float}
\usepackage{units}
\usepackage{amsmath}
\usepackage{amssymb}
\usepackage{graphicx}
\usepackage{color}
\usepackage{esint}
\usepackage{youngtab}
\newcommand{\pd}[2]{\frac{\partial #1}{\partial #2}}

\usepackage{wrapfig}
\usepackage{slashed}
\usepackage{epsfig}
\usepackage{tensor}

\def\be{\begin{eqnarray}}
\def\ee{\end{eqnarray}}
\newcommand{\nn}{\nonumber}

\def\Dslash{\,\,{\raise.15ex\hbox{/}\mkern-12mu D}}
\def\Dbarslash{\,\,{\raise.15ex\hbox{/}\mkern-12mu {\bar D}}}
\def\delslash{\,\,{\raise.15ex\hbox{/}\mkern-9mu \partial}}
\def\delbarslash{\,\,{\raise.15ex\hbox{/}\mkern-9mu {\bar\partial}}}
\def\pslash{\,\,{\raise.15ex\hbox{/}\mkern-9mu p}}
\def\calDslash{\,\,{\raise.15ex\hbox{/}\mkern-12mu {\cal D}}}

\newcommand{\D}{{\partial}}

\newcommand{\ket}{\rangle}
\newcommand{\Tr}{{\rm Tr}}

\def\ket#1{|{#1}\rangle}

\def\lae{\mathrel{\mathop{\smash{\lower .5 ex \hbox{$\stackrel<\sim$}}}}}
\def\lae{\mathrel{\mathop{\smash{\lower .5 ex \hbox{$\stackrel>\sim$}}}}}

\usepackage[useregional]{datetime2}

\usepackage{float}
\usepackage{tikz}
\usepackage{pgfplots}
\pgfplotsset{compat=1.16}
\usetikzlibrary{calc,arrows,cd}
\usetikzlibrary{shapes,snakes}
\usepackage{makecell}
\pgfdeclarelayer{nodelayer}
\pgfdeclarelayer{edgelayer}
\pgfsetlayers{edgelayer,nodelayer,main}
\tikzstyle{ghost}=[fill={rgb,255: red,140; green,76; blue,150}, draw=black, shape=circle,scale=0.6]
\tikzstyle{real_ghost}=[fill=none, draw=none, shape=circle]
\tikzstyle{none}=[fill=none, draw=none, shape=circle]
\tikzstyle{Red_Circle}=[fill=none, draw=red, shape=circle]

\tikzstyle{BlackLine}=[-, draw=black, fill=white]
\tikzstyle{Arrow}=[<-, thick]
\tikzstyle{thin_black_line}=[-, fill=blue]
\tikzstyle{thin_black_line_red}=[-, fill=red]
\tikzstyle{thin_black_line_semi_purple}=[-, fill={rgb,255: red,1; green,1; blue,255}]
\tikzstyle{thin_black_line_purple}=[-, fill={rgb,255: red,200; green,1; blue,255}]
\tikzstyle{thin_black_line_turquoise}=[-, fill={rgb,255: red,1; green,200; blue,200}]
\tikzstyle{thin_black_line_gray}=[-, fill=gray]
\tikzstyle{thin_black_line_null}=[-, fill=orange]
\tikzstyle{Double_Arrow}=[<->]
\tikzstyle{Dashed_arrow}=[->, dashed, draw=red]
\tikzstyle{Dashed_arrow_gray}=[->, dashed, draw=gray]
\tikzstyle{BlackLine_dash}=[-, draw=red, fill=white, dashed]
\tikzstyle{Arrow_order}=[<-, draw=red]

\title{\bf 
Nonrelativistic Approximations of Closed Bosonic String Theory

}

\author[]{Jelle Hartong,}
\author[]{Emil Have}

\affiliation[]{School of Mathematics and Maxwell Institute for Mathematical Sciences,\\
 University of Edinburgh, Peter Guthrie Tait Road, Edinburgh EH9 3FD, UK}

\emailAdd{j.hartong@ed.ac.uk}
\emailAdd{emil.have@ed.ac.uk}

\abstract{We further develop the string $1/c^2$ expansion of closed bosonic string theory, where $c$ is the speed of light. The expansion will be performed up to and including the next-to-next-to-leading order (NNLO). We show that the next-to-leading order (NLO) theory is equal to the Gomis--Ooguri string, generalised to a curved target space, provided the target space geometry admits a certain class of co-dimension-2 foliations. We compute the energy of the string up to NNLO for a flat target space with a circle that must be wound by the string, and we show that it agrees with the $1/c^2$ expansion of the relativistic energy. We also compute the algebra of Noether charges for a flat target space and show that this matches order-by-order with an appropriate expansion of the Poincar\'e algebra, which at NLO gives the string Bargmann algebra. Finally, we expand the phase space action, which allows us to perform the Dirac procedure and pass to the quantum theory. It turns out that the Poisson brackets change at each order, and we show that the normal ordering constant of the relativistic theory, which does not depend on $c$, can be reproduced by the NLO and NNLO theories. 

}

\usepackage{todonotes}
\usepackage{mathrsfs}
\usepackage{amscd}

\usepackage{amsthm}

\theoremstyle{remark}

\usepackage{tcolorbox}
\usepackage{cancel}
\usepackage{xcolor}

\usepackage{tikz}

\DeclareFontFamily{U}{skulls}{}
\DeclareFontShape{U}{skulls}{m}{n}{ <-> skull }{}

\renewcommand{\i}{\text{i}}
\begin{document}
\pagestyle{plain} \setcounter{page}{1}
\newcounter{bean}
\baselineskip16pt \setcounter{section}{0}
\maketitle
\flushbottom
\section{Introduction}

Following the pioneering work of \cite{Gomis:2000bd,Danielsson:2000gi}, nonrelativistic (NR) string theory has developed into an active field of research. There is a growing class of string theories whose worldsheet or target space geometry is non-Lorentzian. Such settings are relevant for non-Lorentzian versions of holographic dualities, to better understand certain limits of relativistic string/M-theory, and to study non-Lorentzian theories of quantum gravity. The nonrelativistic string studied here belongs to this larger class of string theories. In this paper, following the foundations laid out in \cite{Hartong:2021ekg}, we further develop the $1/c$ expansion of closed relativistic bosonic strings.

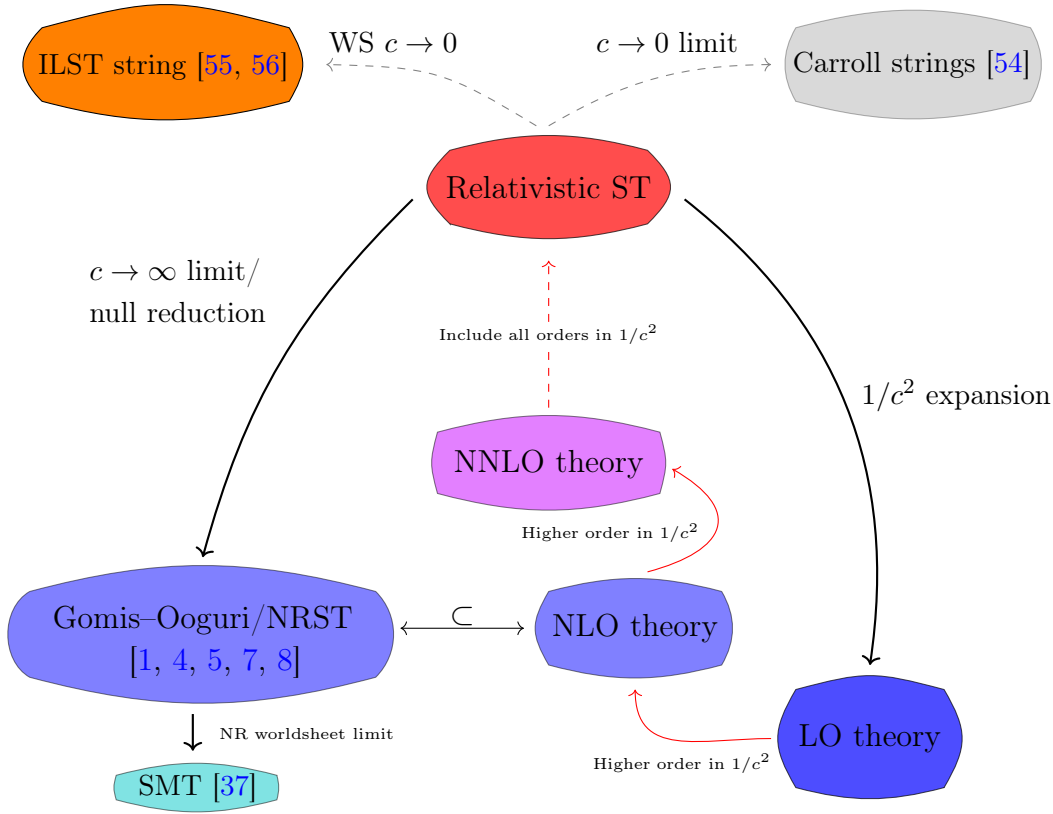
\begin{figure}
    \centering
    \begin{tikzpicture}[scale=0.65]
 	\begin{pgfonlayer}{nodelayer}
 		\node [style=none] (4) at (-2, 6) {};
 		\node [style=none] (5) at (2, 6) {};
 		\node [style=none] (6) at (-2, 4.5) {};
 		\node [style=none] (7) at (2, 4.5) {};
 		\node [style={real_ghost}] (8) at (0, 5.25) {Relativistic ST};
 		\node [style=none] (9) at (2.75, 5) {};
 		\node [style=none] (10) at (6.5, -4.5) {};
 		\node [style=none] (11) at (5, -5) {};
 		\node [style=none] (12) at (8, -5) {};
 		\node [style=none] (13) at (5, -7) {};
 		\node [style=none] (14) at (8, -7) {};
 		\node [style={real_ghost}] (15) at (6.5, -6) {LO theory};
 		\node [style={real_ghost}] (16) at (8.25, 1) {\small{$1/c^2$ expansion}};
 		\node [style=none] (17) at (-2.75, 5) {};
  		\node [style=none] (18) at (-7, -2.3) {};
 		\node [style=none] (19) at (4.5, -6) {};
 		\node [style=none] (20) at (1.75, -5) {};
 		\node [style=none] (21) at (2, -2.6) {};
 		\node [style=none] (23) at (-3.5, -3) {};
 		\node [style=none] (24) at (-10.5, -4.75) {};
 		\node [style=none] (25) at (-3.5, -4.75) {};
 		\node [style=none] (26) at (-10.5, -3) {};
 		\node [style={real_ghost}] (27) at (-7, -4) {\makecell[l]{Gomis--Ooguri/NRST\\\hspace{1.0cm}\cite{Gomis:2000bd,Danielsson:2000mu,Harmark:2017rpg,Harmark:2018cdl,Bergshoeff:2018yvt}}};
 		\node [style=none] (28) at (0, -3) {};
 		\node [style=none] (29) at (3.5, -3) {};
 		\node [style=none] (30) at (0, -4.5) {};
 		\node [style=none] (31) at (3.5, -4.5) {};
 		\node [style={real_ghost}] (32) at (1.75, -3.75) {NLO theory};
 		\node [style=none] (33) at (-3, -3.75) {};
 		\node [style=none] (34) at (-0.5, -3.75) {};
 		\node [style=none] (35) at (-1.75, -3.5) {$\subset$};
 		\node [style=none] (36) at (-2.25, 0.25) {};
 		\node [style=none] (37) at (2.25, 0.25) {};
 		\node [style=none] (38) at (-2.25, -1) {};
 		\node [style=none] (39) at (2.25, -1) {};
 		\node [style={real_ghost}] (40) at (0, -0.4) {NNLO theory};
 		\node [style=none] (41) at (2.5, -0.5) {};
 		\node [style=none] (42) at (0, 3.75) {};
 		\node [style=none] (43) at (0, 2.5) {};
 		\node [style={real_ghost}] (44) at (-7.5, 3.1) {\makecell[l]{\small{$c\rightarrow \infty$ limit/}\\\small{null reduction}}};
 		\node [style={real_ghost}] (45) at (2.7, -6.5) {\tiny{Higher order in $1/c^2$}};
 		\node [style={real_ghost}] (46) at (0, 2.25) {\tiny{Include all orders in $1/c^2$}};
 		\node [style={real_ghost}] (47) at (1.25, -1.8) {\tiny{Higher order in $1/c^2$}};
 		\node [style=none] (48) at (0, 0.75) {};
 		\node [style=none] (49) at (0, 2) {};
 		\node [style=none] (50) at (-7.2, -5.5) {};
 		\node [style=none] (51) at (-7.2, -6.25) {};
 		\node [style=none] (52) at (-8.75, -6.75) {};
 		\node [style=none] (53) at (-5.5, -6.75) {};
 		\node [style=none] (54) at (-8.75, -7.25) {};
 		\node [style=none] (55) at (-5.5, -7.25) {};
 		\node [style=none] (56) at (-7.1, -7) {\small SMT~\cite{Harmark:2020vll}};
 		\node [style=none] (57) at (0, 6.5) {};
 		\node [style=none] (58) at (4.5, 7.75) {};
 		\node [style=none] (59) at (2.4, 8.2) {\small{$c\rightarrow 0$ limit}};
 		\node [style=none] (60) at (5, 8.5) {};
 		\node [style=none] (61) at (9.75, 8.5) {};
 		\node [style=none] (62) at (5, 7) {};
 		\node [style=none] (63) at (9.75, 7) {};
 		\node [style=none] (64) at (7.35, 7.75) {\small Carroll strings~\cite{Cardona:2016ytk} };
 		\node [style=none] (65) at (-4.9, -5.9) {\tiny{NR worldsheet limit}};
 		\node [style=none] (66) at (-0.25, 6.5) {};
 		\node [style=none] (67) at (-4.5, 7.75) {};
 		\node [style=none] (68) at (-10.25, 8.5) {};
 		\node [style=none] (69) at (-5.25, 8.5) {};
 		\node [style=none] (70) at (-5.25, 7) {};
 		\node [style=none] (71) at (-10.25, 7) {};
 		\node [style=none] (72) at (-7.8, 7.7) {\small ILST string~\cite{Isberg:1993av,Bagchi:2015nca} };
 		\node [style=none] (73) at (-3.15, 8.2) {\small{WS $c\rightarrow 0$}};
 		\end{pgfonlayer}
 	\begin{pgfonlayer}{edgelayer}
		\draw [style={thin_black_line_red}, opacity = 0.7] (6.center)
 			 to [bend left=45, looseness=1.50] 
 			 (4.center)
 			 to [in=165, out=15] 
 			 (5.center)
 			 to [bend left=45, looseness=1.50] 
 			 (7.center)
 			 to [bend left=15]
 			 cycle;
 		\draw [style=Arrow, bend right] (10.center) to (9.center);
 		\draw [style={thin_black_line}, opacity = 0.7] (14.center)
 			 to [bend right, looseness=1.25] (12.center)
 			 to [bend right=15, looseness=1.25] (11.center)
 			 to [bend right, looseness=1.25] (13.center)
 			 to [bend right=15] cycle;
 		\draw [style=Arrow, bend left=15] (18.center) to (17.center);
 		\draw [style={thin_black_line_semi_purple}, opacity = 0.5] (24.center)
 			 to [bend right=15] (25.center)
 			 to [bend right=45] (23.center)
 			 to [bend right=15] (26.center)
 			 to [bend right=60] cycle;
 		\draw [style={thin_black_line_semi_purple}, opacity = 0.5] (29.center)
 			 to [bend right=15] (28.center)
 			 to [bend right, looseness=1.25] (30.center)
 			 to [bend right=15] (31.center)
 			 to [bend right] cycle;
 		\draw [style=Arrow_order, in=-180, out=-90, looseness=1.25] (20.center) to (19.center);
 		\draw [style={Double_Arrow}] (33.center) to (34.center);
 		\draw [style={thin_black_line_purple}, opacity = 0.5] (39.center)
 			 to [bend left=15] (38.center)
 			 to [bend left=15, looseness=1.25] (36.center)
 			 to [bend left=15] (37.center)
 			 to [bend left=15, looseness=1.25] cycle;
 		\draw [style=Arrow_order, in=15, out=-30, looseness=2.00] (41.center) to (21.center);
 		\draw [style={Dashed_arrow}] (43.center) to (42.center);
 		\draw [style=BlackLine_dash] (48.center) to (49.center);
 		\draw [style=Arrow] (51.center) to (50.center);
 		\draw [style={thin_black_line_turquoise}, opacity = 0.5] (53.center)
 			 to [bend right=15] (52.center)
 			 to [bend right=15, looseness=1.25] (54.center)
 			 to [bend right=15] (55.center)
 			 to [bend right=15] cycle;
 		\draw [style={Dashed_arrow_gray}, bend left=15] (57.center) to (58.center);
\draw [style={thin_black_line_gray}, opacity=0.3] (62.center)
 			 to [bend right=15] (63.center)
 			 to [bend right] (61.center)
 			 to [bend right=15] (60.center)
 			 to [bend right] cycle;
 		\draw [style={Dashed_arrow_gray}, bend right=15, looseness=1.25] (66.center) to (67.center);
 		\draw [style={thin_black_line_null}] (71.center)
 			 to [bend right=15] (70.center)
 			 to [bend right, looseness=1.25] (69.center)
 			 to [bend right=15, looseness=1.25] (68.center)
 			 to [bend right=45, looseness=1.25] cycle;
 	\end{pgfonlayer}
 \end{tikzpicture}
     \caption{Relations between relativistic string theory and various non-Lorentzian avatars. Starting from relativistic string theory, taking a string $c\rightarrow \infty$ limit or performing a null reduction leads to the Gomis--Ooguri string \cite{Gomis:2000bd,Danielsson:2000gi,Danielsson:2000mu}, which is contained in the NLO theory we get from a string $1/c^2$ expansion~\cite{Harmark:2017rpg,Harmark:2018cdl,Bergshoeff:2018yvt}. Taking a worldsheet NR limit of the Gomis--Ooguri string leads to spin matrix theory (SMT) strings~\cite{Harmark:2017rpg,Harmark:2018cdl,Harmark:2019upf,Harmark:2020vll} (see also~\cite{Harmark:2014mpa}). If we instead take a Carrollian worldsheet limit of relativistic string theory where the worldsheet speed of light goes to zero, we get the ILST (or tensionless) string~\cite{Isberg:1993av,Bagchi:2015nca}. This is classically equivalent to the ambitwistor string~\cite{Mason:2013sva,Casali:2016atr}. We also note that one may instead take an ultralocal limit in target space, where the speed of light goes to zero. This was considered in a phase space formulation in~\cite{Cardona:2016ytk}. }
     \label{fig:Diagram-of-strings}
 \end{figure}

The generalisation of \cite{Gomis:2000bd,Danielsson:2000gi,Danielsson:2000mu} to curved target space geometries was considered in~\cite{Harmark:2017rpg,Kluson:2018egd,Bergshoeff:2018yvt,Harmark:2018cdl,Gallegos:2019icg,Harmark:2019upf,Bidussi:2021ujm}, and these geometries were further studied in~\cite{Bergshoeff:2018vfn,Bergshoeff:2021bmc,Yan:2021lbe,Bergshoeff:2022fzb}, in part based on earlier work in \cite{Andringa:2012uz}. The beta functions were studied in~\cite{Gallegos:2019icg,Gomis:2019zyu,Bergshoeff:2019pij,Yan:2019xsf}, while a Hamiltonian perspective was taken in~\cite{Kluson:2018egd,Kluson:2018grx,Kluson:2018vfd,Kluson:2019qgj,Kluson:2019xuo,Kluson:2019ajy,Kluson:2021qqv}. Open NR strings and DBI-like actions for NR D$p$-branes were described in~\cite{Gomis:2004pw,Brugues:2004an,Gomis:2005bj,Brugues:2006yd,Gomis:2020fui,Gomis:2020izd,Kluson:2019avy,Roychowdhury:2019qmp,Kluson:2020aoq,Ebert:2021mfu}, and the connection to limits of the AdS/CFT correspondence was investigated in~\cite{Harmark:2017rpg,Harmark:2018cdl,Harmark:2019upf,Gomis:2005pg,Harmark:2020vll,Fontanella:2021hcb,Fontanella:2021btt,Roychowdhury:2021wte,Fontanella:2022fjd,Fontanella:2022pbm}. Supersymmetric NR strings were considered in~\cite{Gomis:2005pg,Kim:2007pc,Blair:2019qwi,Bergshoeff:2022pzk}, and the relation between double field theory and NR strings was established in~\cite{Ko:2015rha,Berman:2019izh,Cho:2019ofr,Blair:2020ops,Park:2020ixf,Gallegos:2020egk,Blair:2020gng}. For a recent review of NR strings, see~\cite{Oling:2022fft}. To put the framework we develop into the context of these (and other) already established theories, we have included Figure~\ref{fig:Diagram-of-strings}.

The idea of perturbatively expanding in an appropriately small parameter is a central tenet of physics: in perturbative QFT, the expansion parameter is some small coupling constant, while post-Newtonian expansions assume both weak fields and small velocities and thus can be considered expansions in both $G$ and $1/c$ (at each order in $1/c$ the $G$ expansion is finite). Further afield, the gradient expansion of hydrodynamics corresponds to an expansion in small values of the wave number, or, in other words, a long wavelength expansion. When we assume analyticity in the dimensionful parameter $1/c$, as we do here, there must exist some reference velocity $v$ so that ultimately we are expanding in the dimensionless parameter $v/c$. While the $1/c$ expansion can be done off shell and in full generality the interpretation of $v$ is context dependent and requires going on shell. 

\begin{figure}[H]
\vspace{-2.5cm}
\centering
\begin{tikzpicture}[scale=0.4]
	\begin{pgfonlayer}{nodelayer}
		\node [style=ghost] (0) at (0, 0) {};
		\node [style=ghost] (1) at (15, 0) {};
		\node [style=ghost] (2) at (15, 15) {};
		\node [style=ghost] (3) at (0, 15) {};
		\node [style=ghost] (4) at (9, 18) {};
		\node [style=ghost] (5) at (24, 18) {};
		\node [style=ghost] (6) at (24, 3) {};
		\node [style={real_ghost}] (7) at (-1, 16.7) {NR Quantum Gravity};
		\node [style={real_ghost}] (9) at (-1.6, -1) {Quantum Mechanics};
		\node [style={real_ghost}] (10) at (17.6, -1) {Classical Mechanics};
		\node [style={real_ghost}] (11) at (28.4, 2.5) {Special Relativity};
		\node [style={real_ghost}] (12) at (20.5, 13.75) {Nonrelativistic gravity};
		\node [style={real_ghost}] (13) at (28.5, 17) {General Relativity};
		\node [style={real_ghost}] (14) at (9, 18.9) {Quantum Gravity};
		\node [style=ghost] (15) at (9, 3) {};
		\node [style={real_ghost}] (16) at (6.9, 3.7) {QFT};
		\node [style={real_ghost}] (17) at (11.75, -0.75) {};
		\node [style={real_ghost}] (18) at (4.25, -0.75) {};
		\node [style={real_ghost}] (19) at (8, -1.75) {$\hbar$};
		\node [style={real_ghost}] (20) at (-0.75, 12) {};
		\node [style={real_ghost}] (21) at (-0.75, 4.25) {};
		\node [style={real_ghost}] (22) at (-1.5, 8.25) {$G$};
		\node [style={real_ghost}] (23) at (18, 0.25) {};
		\node [style={real_ghost}] (24) at (23, 2) {};
		\node [style={real_ghost}] (25) at (21.25, 0.5) {$1/c$};
		\node [style={Red_Circle}] (26) at (17.25, 15.75) {};
		\node [style={real_ghost}] (27) at (19, 19) {};
		\node [style={real_ghost}] (28) at (17.25, 16) {};
		\node [style={real_ghost}] (29) at (22.75, 19.7) {NR gravity as developed in \cite{VandenBleeken:2017rij,Hansen:2018ofj,VandenBleeken:2019gqa,Hansen:2019vqf,Hansen:2020pqs,Ergen:2020yop}};
	\end{pgfonlayer}
	\begin{pgfonlayer}{edgelayer}
		\draw [style=BlackLine] (0) to (1);
		\draw [style=BlackLine] (2) to (1);
		\draw [style=BlackLine] (3) to (0);
		\draw [style=BlackLine, in=180, out=0, looseness=1.25] (3) to (2);
		\draw [style=BlackLine] (3) to (4);
		\draw [style=BlackLine] (5) to (2);
		\draw [style=BlackLine] (4) to (5);
		\draw [style=BlackLine] (1) to (6);
		\draw [style=BlackLine] (5) to (6);
		\draw [style=BlackLine] (0) to (15);
		\draw [style=BlackLine] (15) to (6);
		\draw [style=BlackLine] (4) to (15);
		\draw [style=Arrow] (18) to (17);
		\draw [style=Arrow] (20) to (21);
		\draw [style=Arrow] (24) to (23);
		\draw [style=Arrow, in=-165, out=105] (28) to (27);
	\end{pgfonlayer}
\end{tikzpicture}
\vspace{-2cm}
\caption{Bronstein's cube of physical theories. A perturbative expansion around a vertex takes us along the edges of the cube, which we illustrate for the $1/c$ expansion of gravity developed in~\cite{VandenBleeken:2017rij,Hansen:2018ofj,VandenBleeken:2019gqa,Hansen:2019vqf,Hansen:2020pqs,Ergen:2020yop}. Nonrelativistic gravity contains Newton--Cartan gravity and thus Newtonian gravity as a special case, but is more general than that.}
\label{fig:Bronstein-cube}
\end{figure}
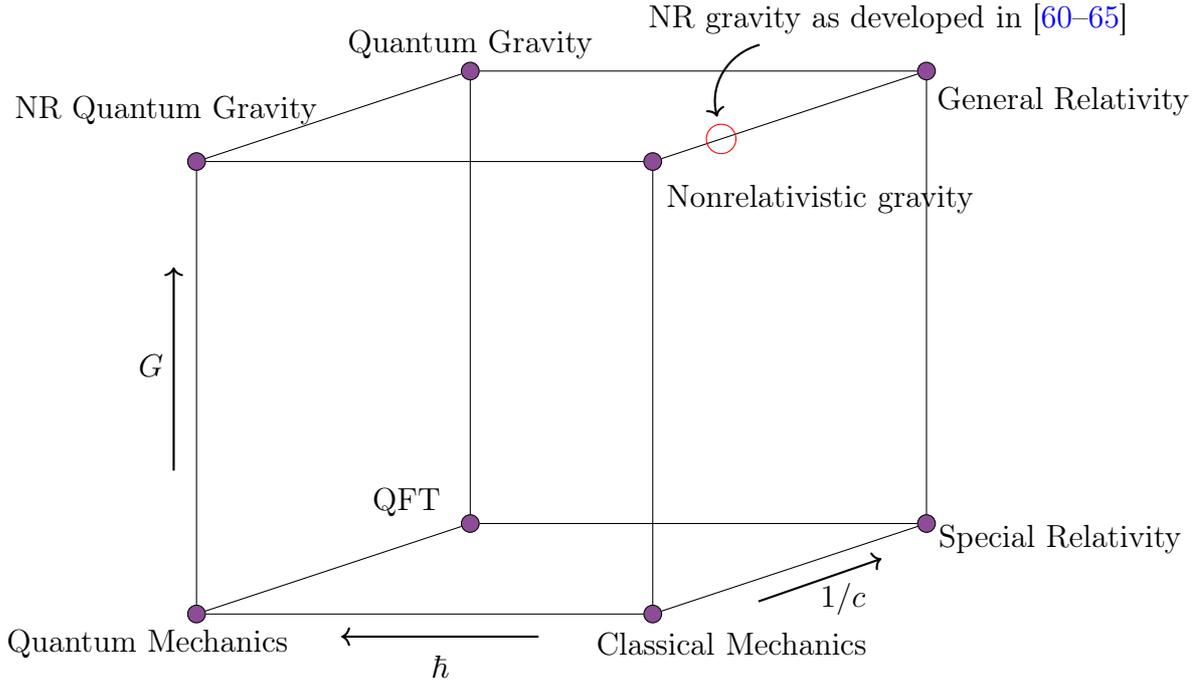

One advantage of performing a $1/c$ expansion, as opposed to taking the $c\to\infty$ limit, is that we can, in principle, go to any order we like.
Furthermore, the $c\to\infty$ limit often requires fine-tuning certain fields to ensure that the limit of the Lagrangian does not blow up, while the $1/c$ expansion does not require any such fine-tuning. 
Consider the Bronstein cube in Figure~\ref{fig:Bronstein-cube}. The $1/c$ expansion allows us to probe the edges of the cube (along the $1/c$ axis). This is exemplified below for the case of gravity.

In this work, whenever we perform a $1/c$ expansion we will always restrict to even powers only, so all expansions are $1/c^2$ expansions. We will next discuss two different kinds of $1/c^2$ expansions: a particle and a string one. The string $1/c^2$ expansion singles out both the time direction and a distinguished space direction, which form the so-called longitudinal directions; the remaining directions are called transverse. This stands in contrast to the particle $1/c^2$ expansion of \cite{VandenBleeken:2017rij,Hansen:2018ofj,VandenBleeken:2019gqa,Hansen:2019vqf,Hansen:2020pqs,Ergen:2020yop}, where only time is singled out. As was shown in these works, the particle $1/c^2$ expansion of the metric up to $\mathcal{O}(c^{-2})$ leads to type II torsional Newton--Cartan geometry, and reduces to ordinary (type I) Newton--Cartan geometry when the torsion is zero. As we will show, the string $1/c^2$ expansion of a Lorentzian geometry with metric $G_{MN}$ gives rise to what we call type II string Newton--Cartan (SNC) geometry. The expansion of the metric $G_{MN}$ up to order $c^{-2}$ is\footnote{Type II SNC involves an additional field that appears in the expansion of the transverse part of the metric.}
\begin{equation}
\label{eq:metric-exp-intro}
    G_{MN} = c^2(-\tau_M{^0}\tau_N{^0} + \tau_M{^1}\tau_N{^1}) + H_{MN} + \mathcal{O}(c^{-2})\,,
\end{equation}
and this reduces to (ordinary or type I) SNC geometry if the strong foliation constraint
\begin{equation}
\label{eq:Foliation-intro}  
    d\tau^A = \varepsilon^A{_B}\omega \wedge \tau^B
\end{equation}
is satisfied for the $1$-forms $\tau^A$ that appear at order $c^2$ in the expansion of the metric~\eqref{eq:metric-exp-intro}. This is in complete analogy with the particle case. In this expression, $\omega$ is a $1$-form determined by this equation, and $A,B=0,1$ are longitudinal tangent space indices. The reason we add the qualifier ``strong'' to the condition~\eqref{eq:Foliation-intro} is that the $1/c^2$ expansion of string theory (via the worldsheet beta functions for a target space with only a metric) comes with the general foliation constraint
\begin{equation}
\label{eq:frobenius-intro}
    d\tau^A = \alpha^A{_B}\wedge\tau^B\,,
\end{equation}
for arbitrary $1$-forms $\alpha^A{_B}$. This condition~\eqref{eq:frobenius-intro} is nothing but the Frobenius integrability condition, which guarantees that the 1-forms $\tau^A$ define a co-dimension-$2$ foliation. Ignoring the dilaton and the Kalb--Ramond field, the beta functions are $R_{MN} = 0$ to leading order in $\alpha'$, and the $1/c^2$ expansion of this equation leads to~\eqref{eq:frobenius-intro} at leading order. Choosing $\alpha^A{_B}$ to be proportional to $\varepsilon^A{_B}$ reduces~\eqref{eq:frobenius-intro} to the strong foliation constraint~\eqref{eq:Foliation-intro}.

At this point, one might wonder why we must use the string $1/c^2$ expansion rather than the particle $1/c^2$ expansion. Had we performed a particle $1/c^2$ expansion, the resulting leading order string theory would be the Galilean string \cite{Batlle:2016iel}. These strings suffer from the problem that they do not admit oscillations, which follows from the fact that the particle $1/c^2$ expansion places the kinetic and potential terms at different orders in $1/c^2$. This precludes the exchange of energy between the potential and kinetic terms and thus does not lead to string oscillations, and as such these ``string theories'' describe rigid extended objects with only center-of-mass motion. 

It was already observed in \cite{Gomis:2000bd,Danielsson:2000gi} that the longitudinal spatial direction that we single out in the string $1/c^2$ expansion must be compact and wound by the string for the resulting string theories to have a non-trivial spectrum. One way to understand this is to observe that the compact direction provides an additional length scale, say $R$, which due to the presence of the intrinsic string length scale $\ell_{\text{s}}$ is required to form a dimensionless parameter in terms of which we can perform the $1/c^2$ expansion. This is because, at the end of the day, the $1/c^2$ expansion must be an expansion in terms of a dimensionless parameter formed by quantities already present in the theory (e.g., $c$, $\hbar$ etc.~as well as other characteristic quantities) and which is small when $c$ is large. For string theory, the parameter is $\ell_{\text{s}}/R$, which, as we will show, corresponds to an expansion in $1/c^2$ in the sense that the centre of mass velocity in the compact direction is much smaller than the speed of light. Since $\ell_{\text{s}}/R$ is small when $R$ is large, this also has the interpretation as an expansion around a decompactification limit.

We show that the leading order string theory only probes the longitudinal target space directions. The equation of motion for the embedding scalars vanish identically upon using the Frobenius constraint~\eqref{eq:frobenius-intro} if we assume that the 1-forms $\alpha^A{}_B$ are traceless.

At the next order in $1/c^2$ -- the next-to-leading order (NLO) -- the theory becomes much richer. If the target space is such that the otherwise arbitrary $1$-forms in~\eqref{eq:frobenius-intro} are traceless, $\alpha^A{_A} = 0$, the Lagrangian of the theory reduces to that of the SNC string~\cite{Harmark:2017rpg,Bergshoeff:2018yvt,Harmark:2019upf}, but with a slightly more general target space geometry as we will see in section \ref{sec:Matching-with-SNC-string}. This is still more general than the strong foliation constraint~\eqref{eq:Foliation-intro}, where $\alpha^A{_B} \sim \varepsilon^A{_B}$.\footnote{Recently, other studies have also proposed relaxing the strong foliation constraint motivated by the quantum theory~\cite{Yan:2021lbe,Bergshoeff:2021bmc}.} When the target space is flat, the NLO theory is the Gomis--Ooguri string~\cite{Gomis:2000bd,Danielsson:2000gi}. An important extra ingredient to add is the Kalb--Ramond field $B_{MN}$, which is included in the theory via a Wess--Zumino (WZ) term. By tuning the Kalb--Ramond field to a critical value, we can at least classically remove the need to impose a foliation constraint on $\tau^A$.

The string theory at next-to-next-to-leading order (NNLO) is more complicated due to the proliferation of terms that the $1/c^2$ expansion gives rise to. We show that the gauge fixed NNLO theory on flat space has a spectrum that corresponds to the expansion of the spectrum of the relativistic string. 

The string $1/c^2$ expansion in particular involves expanding the embedding scalars $X^M = x^M + c^{-2}y^M + \cdots$. The subleading components of these make sure that the string theory at any given order ``remembers'' the string theories at all previous orders. For example, the dynamics of the $x^M$ fields in the LO theory are imposed in the NLO theory by $y^M$.

In parallel to the string $1/c^2$ expansion of the geometry, the underlying symmetry algebra, the Poincar\'e algebra, also gets expanded~\cite{Harmark:2019upf} (see~\cite{Hansen:2020pqs} for the particle equivalent). In the string theory, this shows for example at the level of the Noether charges of the string theory when the target space is flat, which have Poisson brackets that match order-by-order with the string $1/c^2$ expansion of the Poincar\'e algebra. 

A natural next step to consider is the quantisation of the $1/c^2$ expanded string theories. It will turn out that the Poisson brackets change at each order. We work out the phase space versions of the $1/c^2$ expanded string theories and perform the Dirac procedure to find the gauge-fixed Poisson brackets (for a gauge choice that is similar but not identical to light cone gauge), which allow us to pass to the quantum theory. As a consistency check we show that the normal ordering constant, which does not depend on $c$, is reproduced both at NLO and NNLO in the expansion.\\

The paper is organised as follows. In Section~\ref{sec:expansion-setup}, we describe the string $1/c^2$ expansion that lies at the heart of the expansion of string theory that this paper investigates, and we  discuss the interpretation of the string $1/c^2$ expansion as an expansion around a decompactification limit. We expand the spectrum of a closed relativistic bosonic string on a background with a compact circle that is wound by the string. This is followed by a discussion of the string $1/c^2$ expansion of Lorentzian geometry in Section~\ref{sec:string-1/c^2-exp}. Then, in Section \ref{sec:string-expansions}, we expand both the Nambu--Goto and Polyakov actions up to NNLO. Furthermore, we employ the string $1/c^2$ expansion of gravity and demonstrate that the LO part of Einstein's equations imposes a two-dimensional foliation structure in the sense of Frobenius on the longitudinal target space. We also discuss the LO equation of motion for the embedding fields and show that when $\alpha^A{_B}$ in~\eqref{eq:frobenius-intro} is traceless, this equation of motion is automatically satisfied. In Section \ref{sec:Matching-with-SNC-string}, we explicitly demonstrate the equivalence between the Gomis--Ooguri string generalised to a curved background, and the NLO theory when the background satisfies~\eqref{eq:frobenius-intro} with $\alpha^A{_A} = 0$. Following this, we discuss the role of the WZ term in Section~\ref{Sec:WZ} and show that we can cancel the LO theory (and thereby remove the need to consider various foliation constraints) by fine-tuning the Kalb--Ramond field. We also discuss St\"uckelberg symmetries between the subleading longitudinal geometric fields and the Kalb--Ramond field. We then go on to consider the spectrum on flat space in Section~\ref{sec:spectrum}, which involves fixing the residual gauge redundancies. At NLO, this reproduces the spectrum of the Gomis--Ooguri string, while the spectrum of the NNLO theory matches the result obtained by $1/c^2$ expanding the relativistic spectrum in Section~\ref{Sec:Decompact}. In Section~\ref{sec:target-space-symmetries}, we consider the target space symmetries of the $1/c^2$ expanded string theories and show that the symmetry algebra corresponds to the string $1/c^2$ expansion of the Poincar\'e algebra. In Section~\ref{sec:phase-space}, we develop the phase space formulation of the LO, NLO and NNLO string theories. Concretely, this is achieved by $1/c^2$ expanding the relativistic phase space action, and we go through the Dirac procedure and find the Dirac brackets at each order in Section~\ref{sec:dirac-brackets}. We then quantise the theories in Section~\ref{sec:quantisation} by deriving the commutators and writing down the normal ordering constant. Finally, we conclude with a discussion in Section \ref{sec:discussion}. In addition to the main text, we have included three appendices: in Appendix \ref{sec:gauge-structure}, we consider in detail the gauge structure of type II SNC geometry, while in Appendix \ref{app:energymomentumexp} we show how to expand quantities such as momentum and energy that involves computing derivatives with respect to fields that are themselves expanded. Finally, in appendix \ref{sec:noether-details} we discuss properties of the expanded Poincar\'e algebra.

\section{The string $1/c^2$ expansion}
\label{sec:expansion-setup}
In this section, we develop the string $1/c^2$ expansion that forms the basis of the remainder of this paper. First introduced in~\cite{Hartong:2021ekg}, we extend the formulation to NNLO which gives rise to the stringy counterpart of what was called type II TNC geometry in~\cite{Hansen:2018ofj,Hansen:2020pqs}. In Section~\ref{sec:string-1/c^2-exp}, we dub this construction type II SNC geometry.

\subsection{Nonrelativistic expansion of the string spectrum}\label{Sec:Decompact}
In order to be able to define a nonrelativistic sector of string theory, which has an intrinsic length scale given by the string length $\ell_{\text{s}} \sim \sqrt{\hbar/(cT)}$, where $T$ is the tension of the string, we need an additional length scale to form a dimensionless expansion parameter. We achieve this by considering flat target space with a compact direction, i.e. $\mathbb{R}^{1,24}\times S^1_R$ where the radius of the circle is $R$. We introduce coordinates $x^M = \{t,x^i,v \}$ on this space, where $\{t,x^i\}$ for $i = 1,\dots,24$ are coordinates on $\mathbb{R}^{1,24}$ and $v$ is the coordinate on $S^1_R$ which we take to have dimensions of time (although $v$ remains a spatial direction). This means that $v$ is periodically identified according to
\begin{equation}
    v \sim v + 2\pi R_{\text{eff}}\,, 
\end{equation}
where we defined the effective radius with dimensions of time as
\begin{equation}
\label{eq:Reff}
    R_{\text{eff}} = R/c\,,
\end{equation}
which we assume is independent of $c$. The directions $t$ and $v$, which both have dimensions of time, will be referred to as the longitudinal directions, while the directions $x^i$, which have dimensions of length, will be called transverse.

The line element on $\mathbb{R}^{1,24}$ in the coordinates $\{t,x^i,v \}$ is given by
\begin{equation}
  ds^2 =  \eta_{MN}dx^M dx^N = c^2\left(-dt^2 + dv^2 \right) + dx^idx^i\,,
\end{equation}
where the components of the Minkowski metric explicitly are given by $\eta_{MN} = c^2\left(-\delta^t_M\delta^t_N + \delta^v_M\delta^v_N \right) + \delta_M^i\delta^i_N$.

The nonrelativistic expansion, as we will see, corresponds to an expansion where $R \gg \ell_{\text{s}}$, which, as discussed in~\cite{Hartong:2021ekg}, thus admits an interpretation as a decompactification limit. Closed relativistic bosonic strings on $\mathbb{R}^{1,24}\times S^1_R$ with metric $\eta_{MN}$ are described by the Polyakov Lagrangian
\be 
\label{eq:Polyakov-Lagrangian}
L_{\text{P}} = -\frac{cT}{2}\oint d \sigma^1 \sqrt{-\gamma}\gamma^{\alpha\beta} \D_\alpha X^M \D_\beta X^N \eta_{MN}\,,
\ee 
where the string embedding fields $X^M(\sigma^0,\sigma^1) = \{X^t(\sigma^0,\sigma^1),X^v(\sigma^0,\sigma^1),X^i(\sigma^0,\sigma^1)\}$ split into longitudinal components $\{X^t,X^v\}$ corresponding to the embedding fields in the time direction and in the compact direction $v$, while the $X^i$ for $i=1,\dots,24$ are the transverse embedding fields. Like $t$ and $v$, both $X^t$ and $X^v$ too have dimensions of time. The transverse embedding fields have dimensions of length, while the worldsheet coordinates $\sigma^\alpha =(\sigma^0,\sigma^1)$ are dimensionless. We take the worldsheet metric $\gamma_{\alpha\beta}$ to be dimensionless, and $\gamma = \det(\gamma_{\alpha\beta})$ is its  determinant. Finally, $T$ is the string tension with dimensions of mass/length. The combination $\D_\alpha X^M \D_\beta X^N \eta_{MN}$ has dimensions of length squared.

The invariant mass squared of a closed bosonic string in such a target space is given by
\be 
\label{eq:inv-mass-squared}
M^2 = \frac{\hbar^2 n^2}{c^2R^2} + \frac{w^2R^2}{{\alpha'}^2} + \frac{2}{\alpha'c}(N + \tilde{N} - 2\hbar)\,,
\ee 
where $\alpha'=\frac{1}{2\pi T}$,  and $w$ and $n$ are, respectively, the winding number and momentum mode in the compact direction. The winding number $w$ counts the number of time the closed string winds around the circle $S^1_R$, while the momentum mode $n$ comes from the quantised centre of mass momentum of the string in the $v$-direction. The number operators $N$ and $\tilde N$, which satisfy $N - \tilde N = \hbar nw$, have dimensions of energy $\times$ time. The relativistic dispersion relation that relates the invariant mass squared to the Noether charges corresponding to energy $E$ and spatial momentum $p_i$ is
\be 
E^2 = M^2c^4 + p^2c^2\,,\label{eq:rel-dispersion}
\ee 
where $p^2 = \vec p \cdot \vec p$ is the norm squared of the spatial momentum. In addition to~\eqref{eq:Reff}, we define the (by assumption) $c$-independent combination
\be 
\alpha'_{\text{eff}} = \frac{\alpha'}{c}\,,
\ee 
which we may equivalently express in terms of the effective tension $T_{\text{eff}} = cT$, which is related to $\alpha'_{\text{eff}}$ as
\be 
\alpha'_{\text{eff}} = \frac{1}{2\pi T_{\text{eff}}}\,.
\ee 
The effective string tension $T_{\text{eff}}$ has dimensions of mass/time, while $\alpha'_{\text{eff}}$ has dimensions of time/mass. In terms of the quantities introduced above, we may write down the following dimensionless parameter
\be 
\label{eq:epsilon-parameter}
\epsilon = \frac{\alpha'\hbar}{cR^2} = \frac{\alpha'_{\text{eff}}\hbar }{c^2R^2_{\text{eff}}}\,,
\ee 
in terms of which the energy as defined in~\eqref{eq:rel-dispersion} can be written as
\begin{equation}
\label{eq:energy-pre-expansion}
    E = \frac{c^2wR_{\text{eff}}}{\alpha'_{\text{eff}}} \sqrt{1 +  \frac{2\epsilon}{ w^2 }\left(\frac{N+\tilde{N}}{\hbar} - 2\right) + \frac{{\alpha'_{\text{eff}}}\epsilon}{\hbar w^2} p^2 + \frac{\epsilon^2 n^2}{w^2}}\,.
\end{equation}
The expansion in $c^{-2}$ is the same as the expansion in $\epsilon$. If we define
\begin{equation}
\label{eq:energy-expansion}
E = c^2E_{\text{LO}} + E_{\text{NLO}} + c^{-2}E_{\text{NNLO}} + \mathcal{O}(c^{-4})\,,
\end{equation}
then we find
\begin{subequations}
\begin{eqnarray}
E_{\text{LO}} & = & \frac{w R_{\text{eff}}}{\alpha'_{\text{eff}}}\,, \label{eq:LO-energy-sec-2} \\
E_{\text{NLO}} & = & \frac{1}{w R_{\text{eff}}}\big(N_{(0)} + \tilde N_{(0)} - 2\hbar\big) + \frac{\alpha'_{\text{eff}}}{2w R_{\text{eff}}}(p_{(0)})^2\,,\label{eq:NLO-energy-sec-2}\\
E_{\text{NNLO}} & = & \frac{1}{w R_{\text{eff}}}\big(N_{(2)} + \tilde N_{(2)} +\alpha'_{\text{eff}}p_{(0)i}p_{(2)i}\big) + \frac{\alpha'_{\text{eff}}\hbar^2 n^2}{2wR^3_{\text{eff}}}\nn\\
&& - \frac{\alpha'_{\text{eff}}}{2w^3R^3_{\text{eff}}}\Big(N_{(0)} + \tilde N_{(0)} - 2\hbar+ \frac{\alpha'_{\text{eff}}}{2}(p_{(0)})^2\Big)^2\,,\label{eq:NNLO-energy-sec-2}
\end{eqnarray}
\end{subequations}
where we expanded
\be 
\begin{aligned}
N &= N_{(0)} + c^{-2}N_{(2)} +\mathcal{O}(c^{-4})\,,\qquad \tilde N = \tilde N_{(0)} +c^{-2}\tilde N_{(2)} + \mathcal{O}(c^{-4})\,,\\
p_i &= p_{(0)i} + c^{-2}p_{(2)i}+ \mathcal{O}(c^{-4})\,.
\end{aligned}
\ee 
Note that we did not expand the momentum mode $n$ and the winding number $w$ since they are integer-valued, although we could have done so abstractly in which case they would also lead to subleading contributions in the same way as $N,\tilde N$, and $p$ above. As we pointed out in~\cite{Hartong:2021ekg}, the nonrelativistic limit that we are considering corresponds to an expansion in the dimensionless quantity $\epsilon$, which we see can equivalently be thought of as a $1/c^2$ expansion or a $1/R_{\text{eff}}^2$ expansion; that is to say, an expansion around a decompactification limit.

The expansion in $\epsilon\ll 1$ can be viewed as $cR_{\text{eff}}\gg \sqrt{\alpha'_{\text{eff}}\hbar}$ which means that the radius of the circle is much larger than the string length. Alternatively we can view it as saying that $\frac{\alpha'_{\text{eff}}\hbar}{c R^2_{\text{eff}}}\ll c$ which means that the velocity of the centre of mass momentum mode $p_{25}$ along $X^{25}=cX^v$ is much smaller than the speed of light.

An important ingredient in string theory is the Kalb--Ramond 2-form field $B_{MN}$, which together with the metric and the dilaton forms the universal massless sector of closed string theory. The coupling between the string embedding fields and the Kalb--Ramond field is described by the Wess--Zumino Lagrangian
\be 
L_{\text{WZ}}=-\frac{cT}{2}\oint d \sigma^1\varepsilon^{\alpha\beta} \partial_\alpha X^M \partial_\beta X^N B_{MN}(X)\,.
\ee 
As we show in Section \ref{Sec:WZ}, if we add a constant Kalb--Ramond $B$-field with legs only in the timelike and compact directions of the form $B_{MN} = 2c^2B\delta_{[M}^{t}\delta_{N]}^{v}$, the energy, which is defined by $E=-\oint  d \sigma^1\frac{\partial\mathcal{L}}{\partial\partial_0 X^t}$, takes the same form except that the LO energy is now
\begin{equation}
    E_{\text{LO}}= \frac{w R_{\text{eff}} }{\alpha'_{\text{eff}}}\left( 1 - B \right)\,,\label{eq:LO-modification-due-to-WZ}
\end{equation}
where we point out that such a Kalb--Ramond 2-form with constant components along $t$ and $v$ is not globally pure gauge because $v$ is periodic.

By tuning the $B$-field, we may for example remove the leading order term entirely, while choosing $B = 1/2$ leads to the spectrum of the Gomis--Ooguri string \cite{Gomis:2000bd} when truncating at $\mathcal{O}(c^0)$.

\subsection{Longitudinal T-duality and the $1/c^2$ expansion}
The spectrum of the relativistic closed string~\eqref{eq:inv-mass-squared} is invariant under T-duality in the $v$-direction, which amounts to the exchanges
\begin{equation}
\label{eq:longitudinal-T-duuality}
    R \leftrightarrow \frac{\hbar \alpha'}{cR}=:\tilde{R}\qquad\text{and}\qquad w\leftrightarrow n\,.
\end{equation}
Adopting the terminology of~\cite{Bergshoeff:2018yvt}, this is a longitudinal spatial T-duality. To explore the role of T-duality in the context of the string $1/c^2$ expansion, it is useful to recast the spectrum~\eqref{eq:inv-mass-squared} in the form
\begin{equation}
    M^2 = \frac{n^2\tilde R^2}{\alpha'^2} + \frac{w^2R^2}{\alpha'^2} + \frac{2}{\alpha'c}( N + \tilde N - 2\hbar )\,,
\end{equation}
which is manifestly invariant under~\eqref{eq:longitudinal-T-duuality}. This suggests, in addition to the expansion set up in~\eqref{eq:energy-pre-expansion} in terms of the dimensionless expansion parameter $\epsilon$ defined in~\eqref{eq:epsilon-parameter}, another dual expansion in terms of the dual dimensionless expansion parameter
\begin{equation}
    \tilde\epsilon = \frac{\alpha'\hbar}{c\tilde R^2}\,.
\end{equation}
The parameter $\tilde\epsilon$ is the T-dual of~\eqref{eq:epsilon-parameter}; i.e., longitudinal T-duality~\eqref{eq:longitudinal-T-duuality} sends $\epsilon\leftrightarrow \tilde{\epsilon}$. The starting point for the expansion in terms of $\tilde\epsilon$ is thus
\begin{equation}
    E = \frac{n\hbar}{R_{\text{eff}}}\sqrt{1 +  \frac{2\tilde\epsilon}{n^2}\left( \frac{N + \tilde N}{\hbar} - 2\right) + \frac{\alpha'_{\text{eff}} \tilde{\epsilon}  }{\hbar n^2}p^2 + \frac{w^2\tilde\epsilon^2}{n^2} }\,,
\end{equation}
which has an expansion around $\tilde{\epsilon} = 0$ of the form 
\begin{equation}
\label{eq:T-dual-expansion}
E = c^2\tilde E_{\text{LO}} + \tilde E_{\text{NLO}} + c^{-2}\tilde E_{\text{NNLO}} + \mathcal{O}(c^{-4})\,,
\end{equation}
where now
\begin{subequations}
\begin{eqnarray}
\tilde E_{\text{LO}} & = & \frac{n \tilde R_{\text{eff}}}{\alpha'_{\text{eff}}}\,, \label{eq:LO-energy-sec-3} \\
\tilde E_{\text{NLO}} & = & \frac{1}{n \tilde R_{\text{eff}}}\big(N_{(0)} + \tilde N_{(0)} - 2\hbar\big) + \frac{\alpha'_{\text{eff}}}{2n \tilde R_{\text{eff}}}(p_{(0)})^2\,,\label{eq:NLO-energy-sec-3}\\
\tilde E_{\text{NNLO}} & = & \frac{1}{n \tilde R_{\text{eff}}}\big(N^{(2)} + \tilde N^{(2)} +\alpha'_{\text{eff}}p_{(0)i}p_{(2)i}\big) + \frac{\alpha'_{\text{eff}}\hbar^2 w^2}{2n\tilde R^3_{\text{eff}}}\nn\\
&& - \frac{\alpha'_{\text{eff}}}{2n^3\tilde R^3_{\text{eff}}}\Big(N_{(0)} + \tilde N_{(0)} - 2\hbar+ \frac{\alpha'_{\text{eff}}}{2}(p_{(0)})^2\Big)^2\,,\label{eq:NNLO-energy-sec-3}
\end{eqnarray}
\end{subequations}
where we defined the $c$-independent combination $\tilde R_{\text{eff}} = \tilde R/c$ in the T-dual picture. Thus, longitudinal T-duality switches between the two expansions~\eqref{eq:energy-expansion} and~\eqref{eq:T-dual-expansion}, and although the relativistic energy~\eqref{eq:energy-pre-expansion} remains T-duality invariant, the $1/c^2$ expansion is no longer longitudinal T-duality invariant order by order. Instead, $\tilde E_{\text{N$^k$LO}}$ transforms into $E_{\text{N$^k$LO}}$ under the replacements~\eqref{eq:longitudinal-T-duuality}. We stress that both expansions correspond to decompactification limits in terms of the radii $R$ and $\tilde R$.

\subsection{SNC geometry from the $1/c^2$ expansion}
\label{sec:string-1/c^2-exp}
Using the methods developed in~\cite{Hansen:2018ofj,Hansen:2019vqf,Hansen:2020pqs}, we show how the string $1/c^2$ expansion of a $(d+2)$-dimensional Lorentzian geometry on a manifold $M$ leads to various notions of SNC geometry on $M$. First, we write the Lorentzian metric and its inverse as
\begin{equation}
\label{eq:metricdecomp}
    \begin{split}
        G_{MN} &= c^2\left(-T_M{^0}T_N{^0} + T_M{^1}T_N{^1} \right) + \Pi_{MN}^\perp = c^2\eta_{AB}T_M{^A}T_N{^B} +\Pi_{MN}^\perp\,,\\
        G^{MN} &= c^{-2}\left(-T^M{_0}T^N{_0} + T^M{_1}T^N{_1}\right) + \Pi^\perp{^{MN}} =c^{-2}\eta^{AB}T^M{_A}T^N{_B} + \Pi^\perp{^{MN}}\,,
    \end{split}
\end{equation} 
where $M,N=0,1,\dots,d+1$ are spacetime indices and $A,B=0,1$ are longitudinal two-dimensional tangent space indices. Here, $\eta_{AB}$ denotes the two-dimensional longitudinal Minkowski metric, $\eta_{AB} = \text{diag}(-1,1)$. The fields in the decompositions in~\eqref{eq:metricdecomp} satisfy the relations
\begin{align}
\label{eq:conditions}
\begin{aligned}
    T_A{^M}\Pi_{MN}^\perp = 0\,,\qquad T_A{^M}T_M{^B} = \delta^B_A\,,\qquad T_M{^A}\Pi^\perp{^{MN}} = 0\,,
\end{aligned}
\end{align}
and the completeness relation 
\begin{equation}
    \delta_M^N = \Pi^\perp_{ML}\Pi^{\perp}{^{LN}} + T_M{^A}T^N{_A}\,.
\end{equation}
The fields in the decompositions in~\eqref{eq:metricdecomp} still depend on $c$, and we assume that they admit a Taylor expansion in $1/c^2$ of the form
\be 
\label{eq:type-II-SNC-expansions}
\begin{aligned}
T_M{^A} &= \tau_M{^A}+c^{-2}m_M{^A}+c^{-4}B_M{^A}+\mathcal{O}(c^{-6})\,,\\
\Pi_{MN}^\perp &= H_{MN}^\perp + c^{-2}\phi^\perp_{MN}+\mathcal{O}(c^{-4})\,.
\end{aligned}
\ee 
Note that the sub-subleading field $B_M{^A}$ that appears in the expansion of the longitudinal vielbein $T_M{^A}$ is unrelated to the Kalb--Ramond $B$-field. Plugging these into the expression for $G_{MN}$ above, we get
\be \label{eq:metricexp}
G_{MN}=c^2\tau_{MN} + H_{MN} + c^{-2}\phi_{MN}+\mathcal{O}(c^{-4})\,,
\ee 
where
\be 
\begin{aligned}
\tau_{MN}&=\eta_{AB}\tau_M{^A}\tau_N{^B}\,,\qquad H_{MN} = H^\perp_{MN} + 2\eta_{AB} \tau_{(M}{^A}m_{N)}{^B}\,,\\
\phi_{MN} &= \phi^\perp_{MN} + \eta_{AB}m_M{^A}m_N{^B}+2\eta_{AB} \tau_{(M}{^A}B_{N)}{^B}\,.
\end{aligned}
\ee 
We refer to Appendix~\ref{sec:gauge-structure} for additional details about the string $1/c^2$ expansion of Lorentzian geometry, including the gauge transformations of the fields. The set of fields $\{ \tau_M{^A}, m_M{^A},H^\perp_{MN},\phi^\perp_{MN} \}$ that arises by truncating the expansions in~\eqref{eq:type-II-SNC-expansions} at order $c^{-2}$ gives rise to a geometry that forms the direct generalisation of ``type II Newton--Cartan geometry'' in~\cite{Hansen:2018ofj,Hansen:2019vqf,Hansen:2020pqs}, where a ``particle'' $1/c^2$ expansion is performed which only singles out the time direction. For this reason, we dub the geometry defined by the set of fields $\{ \tau_M{^A}, m_M{^A},H^\perp_{MN},\phi^\perp_{MN} \}$ ``type II string Newton--Cartan geometry'', or type II SNC geometry for short. As we will see, the string theory that emerges at NLO couples to type II SNC geometry, but it does not couple to $\phi^\perp_{MN}$. The string theory at NNLO couples to the geometry defined by truncating the expansion~\eqref{eq:type-II-SNC-expansions} at order $c^{-4}$, but again it does not couple to the field that arises at this order in the expansion of $\Pi^\perp_{MN}$, and so we have refrained from writing it in~\eqref{eq:type-II-SNC-expansions}. The NNLO string does, however, couple to $B_M{^A}$. 

A natural condition to impose on the geometries that arise from the $1/c^2$ expansion is to demand that the LO longitudinal $1$-forms $\tau_M{^A}$ give rise to a co-dimension-$2$ foliation, which by Frobenius' theorem means that they satisfy
\begin{equation}
\label{eq:Frobenius-thm}
    d\tau^A = \alpha^A{_B}\wedge \tau^B\,,
\end{equation}
where the $\alpha^A{_B}$ are arbitrary $1$-forms. As we will see in Section~\ref{sec:codim-2-foliations}, this condition arises from the string $1/c^2$ expansion of Einstein's equations.

If we remove the field $\phi^\perp_{MN}$ from the description, type II SNC geometry as constructed above reduces to (type I) SNC geometry if we impose the strong foliation constraint of \cite{Bergshoeff:2018vfn,Bergshoeff:2019ctr}
\begin{equation}
\label{eq:Foliation1}  
    d\tau^A = \varepsilon^A{_B} \omega \wedge \tau^B\,,
\end{equation} 
corresponding to the special case $\alpha^A{_B} = \omega \varepsilon^A{_B}$ for some $1$-form $\omega$. When \eqref{eq:Foliation1} holds, the transformation properties of the field $m_M{^A}$ reduce to those of type I SNC geometry (see Appendix \ref{sec:gauge-structure} for details). In other words, when the strong foliation constraint is satisfied, the data $(\tau_M{^A},H^\perp_{MN},m_M{^A})$ describes SNC geometry. This is entirely analogous to the situation in torsional Newton--Cartan geometry \cite{Hansen:2018ofj,Hansen:2019vqf,Hansen:2020pqs}, where type II TNC reduces to Newton--Cartan geometry when the clock form is exact and the field $\Phi_{\mu\nu}$ (the particle $1/c$ expansion analogue of $\phi_{MN}^\perp$) is removed from the description. We have collected the various notions of SNC geometry in Table~\ref{table:SNC-geometries}.

\begin{table}
	\centering
				\begin{tabular}{|K{5cm}|K{9cm}|}\toprule
				 Geometry & Field content \\
				\midrule
				String $1/c^2$ expansion of Lorentzian geometry & LO: $\tau_M{^A}$, $H^\perp_{MN}$ \hspace{6cm} up to NLO: $\tau_M{^A}$, $H^\perp_{MN}$, $m_M{^A}$, $\phi^\perp_{MN}$\hspace{3cm}
				up to NNLO: $\tau_M{^A}$, $H^\perp_{MN}$, $m_M{^A},~\phi^\perp_{MN},~B_M{^A}$ \\
				&etc.\\
				\midrule
				Type II SNC & $\tau_M{^A}$, $H^\perp_{MN}$, $m_M{^A}$, $\phi^\perp_{MN}$ (LO and NLO fields from above) \\
				\midrule
				Type I SNC  & $\tau_M{^A}$, $H^\perp_{MN}$, $m_M{^A}$\\
				\bottomrule
			\end{tabular}
	\caption{Overview of geometries of SNC type and how they arise from the string $1/c^2$ expansion of Lorentzian geometry. All of these may be subjected to foliation constraints: the most general such constraint is that $\tau^A$ defines a co-dimensions-$2$ foliation, i.e., $d\tau^A = \alpha^A{_B}\wedge \tau^B$ for arbitrary $1$-forms $\alpha^A{_B}$. Important special cases include $\alpha^A{_A} =0$ and $\alpha^A{_B} = \varepsilon^A{_B} \omega$ for some $1$-form $\omega$. The latter is known as the strong foliation constraint. When the strong foliation constraint holds the gauge transformations of $\tau_M{^A}$, $H^\perp_{MN}$, $m_M{^A}$ for type II SNC agree with those of type I SNC.}\label{table:SNC-geometries}
\end{table}

\section{Expanding the Nambu--Goto and Polyakov actions}\label{sec:string-expansions}

\subsection{General properties of nonrelativistic string expansions}
Let $\mathcal{L}[X;c]$ be the string Lagrangian---either Nambu--Goto or Polyakov. As indicated, this Lagrangian depends on the embedding scalars $X^M$, which are expanded as in \eqref{eq:EmbeddingExp} below, as well as explicitly on $c$. Based on various assumptions made about how fields and constants depend on $c$ it will turn out that the string Lagrangian starts at $\mathcal{O}(c^2)$, so we can expand it as
\be 
\mathcal{L}[X;c] &=& c^2 \overset{(-2)}{\mathcal{L}}(X) + \overset{(0)}{\mathcal{L}}(X) + c^{-2}\overset{(2)}{\mathcal{L}}(X) + \mathcal{O}(c^{-4})\,.
\ee 
We now (functionally) Taylor expand these Lagrangians to get 
\be 
\mathcal{L}[X;c] &=& c^2 \overset{(-2)}{\mathcal{L}}(x) + \left[\overset{(0)}{\mathcal{L}}(x) + y^M\frac{\delta \overset{(-2)}{\mathcal{L}}(x)}{\delta x^M} \right]\nn\\
&&+c^{-2}\left[ \overset{(2)}{\mathcal{L}}(x) + z^M\frac{\delta \overset{(-2)}{\mathcal{L}}(x)}{\delta z^M} + \frac{1}{2}y^My^N\frac{\delta^2 \overset{(-2)}{\mathcal{L}}(x)}{\delta x^M\delta x^N} +  y^M\frac{\delta \overset{(0)}{\mathcal{L}}(x)}{\delta x^M} \right] + \mathcal{O}(c^{-4})\nn\\
&=& c^2\mathcal{L}_{\text{LO}} + \mathcal{L}_{\text{NLO}} + c^{-2}\mathcal{L}_{\text{NNLO}} + \mathcal{O}(c^{-4})\,,
\ee 
where, e.g., $\frac{\delta \overset{(-2)}{\mathcal{L}}(x)}{\delta x^M}$ is the first variation of $\overset{(-2)}{\mathcal{L}}(x)$ with respect to $x^M$. We define
\begin{subequations}
\be 
\mathcal{L}_{\text{LO}} &=& \overset{(-2)}{\mathcal{L}}(x)\,,\\
\mathcal{L}_{\text{NLO}} &=& \overset{(0)}{\mathcal{L}}(x) + y^M\frac{\delta \mathcal{L}_{\text{LO}}}{\delta x^M}\,,\label{eq:NLO-lag}\\
\mathcal{L}_{\text{NNLO}} &=& \overset{(2)}{\mathcal{L}}(x) + z^M\frac{\delta \mathcal{L}_{\text{LO}}}{\delta x^M} + y^M\frac{\delta \mathcal{L}_{\text{NLO}}}{\delta x^M} - \frac{1}{2}y^M y^N\frac{\delta^2 \mathcal{L}_{\text{LO}}}{\delta x^M\delta x^N}\,.\label{eq:NNLO-lag}
\ee 
\end{subequations}
Note that the minus sign in the last term in~\eqref{eq:NNLO-lag} comes from the second term in~\eqref{eq:NLO-lag} via the term $y^M\frac{\delta \mathcal{L}_{\text{NLO}}}{\delta x^M}$. In this way, the role of the subleading embedding fields is to impose the equations of motion of the Lagrangians that appear at previous orders. Explicitly, the first and second variational derivatives that appear above are given by
\begin{subequations}
\be 
y^M\frac{\delta {\mathcal{L}}_{\text{NLO}} }{\delta x^M} &=& y^M\frac{\D \mathcal{L}_{\text{LO}}}{\D x^M} +   \frac{\D \mathcal{L}_{\text{LO}}}{\D(\D_\alpha x^M)} \D_\alpha y^M \,,\\
y^M y^N\frac{\delta^2 {\mathcal{L}}_{\text{LO}}}{\delta x^M\delta x^N} &=& y^My^N\frac{\D^2 \mathcal{L}_{\text{LO}}}{\D x^M \D x^N} + 2y^M \D_\alpha y^N  \frac{\D^2 \mathcal{L}_{\text{LO}} }{\D x^M \D(\D_\alpha x^N)} \nn\\
&&+ \D_\alpha y^M\D_\beta y^N \frac{\D^2 \mathcal{L}_{\text{LO}}}{\D(\D_\alpha x^M) \D(\D_\beta x^N)} \,.
\ee
\end{subequations}
These expressions are only defined up to total derivatives and will play an important role in our considerations below.

\subsection{Nambu--Goto action}
We now apply the framework developed above to the Nambu--Goto (NG) action. This was also considered in~\cite{Hartong:2021ekg} (see also~\cite{Harmark:2019upf}), but only up to NLO. The relativistic NG action is
\be 
S_{\text{NG}}[X;c] = \int_\Sigma  d ^2 \sigma\,\mathcal{L}_{\text{NG}}[X;c] = -cT\int_\Sigma  d^2 \sigma \sqrt{-\det G_{\alpha\beta}(X)}\,,
\ee 
where $G_{\alpha\beta}(X)$ is the pullback of the relativistic target space metric in terms of the relativistic embedding fields as defined in \eqref{eq:X-pullback-of-rel-TS-metric}, i.e.,
\be 
G_{\alpha\beta}(X) = \D_\alpha X^M\D_\beta X^N G_{MN}(X)\,,\label{eq:X-pullback-of-rel-TS-metric}
\ee 
where the argument ``$X$'' on the left-hand side indicates that the pullback is with respect to the embedding field $X^M$. Expanding the embedding fields according to
\be 
X^M = x^M + c^{-2}y^M + c^{-4}z^M + \mathcal{O}(c^{-6})\,,\label{eq:EmbeddingExp}
\ee 
the $1/c^2$ expansion of $G_{\alpha\beta}(X)$ takes the form
\be 
\label{eq:exp-of-pulled-back-rel-metric}
G_{\alpha\beta}(X) = c^2\tau_{\alpha\beta}(x) + H_{\alpha\beta}(x,y) + c^{-2}\phi_{\alpha\beta}(x,y,z) + \mathcal{O}(c^{-4})\,,
\ee 
where we need to take into account both terms coming from the $1/c^2$ expansion of the pullback maps $\D_\alpha X$ and those that arise from a Taylor expansion of $G_{MM}(X)$, so that we find
\begin{subequations}
\be 
\tau_{\alpha\beta}(x) &=& \D_\alpha x^M\D_\beta x^N\tau_{MN}\,,\\
H_{\alpha\beta}(x,y) &=& H_{\alpha\beta}(x) + 2\tau_{MN}(x)\D_{(\alpha} x^M \D_{\beta)}y^N + \D_\alpha x^M\D_\beta x^N y^L \D_L \tau_{MN}(x)\,,\label{eq:HSNC}\\
\phi_{\alpha\beta}(x,y,z) &=& \Phi_{\alpha\beta}(x) + \D_\alpha y^M\D_\beta y^N \tau_{MN}(x) + 2\D_{(\alpha} x^M\D_{\beta)}z^N \tau_{MN}(x)\nn\\
&& + 2\D_{(\alpha} x^M \D_{\beta)}y^N H_{MN}(x)+2\D_{(\alpha} x^M\D_{\beta)} y^N y^L \D_L \tau_{MN}(x)\label{eq:PHISNC}\\
&&+ \D_\alpha x^M\D_\beta x^N\left( z^L\D_L\tau_{MN}(x) + \frac{1}{2}y^L y^K\D_L\D_K\tau_{MN}(x) + y^L\D_L H_{MN}(x) \right)\,.\nn
\ee 
\end{subequations}
Here $H_{\alpha\beta}(x)$ and $\phi_{\alpha\beta}(x)$ are the pullbacks of $H_{MN}$ and $\phi_{MN}$ which appear in the expansion of the target space metric \eqref{eq:metricexp}.
These expressions are unwieldy, but as we will show, the subleading embedding fields encode information about dynamics at previous orders, so their appearance is entirely dictated by this data, which makes them easy to handle.

By assumption, the pull-back $\tau_{\alpha\beta}(x) =\D_\alpha x^M\D_\beta x^N\tau_{MN}(x)$ is a two-dimensional Lorentzian metric, and so admits an inverse that we denote $\tau^{\alpha\beta}(x)$---this is a condition on the embedding of the worldsheet in target space. This inverse satisfies 
\be 
\tau_{\alpha\delta}(x)\tau^{\delta\beta}(x) = \delta_\alpha^\beta\,,
\ee 
and we can explicitly write it as
\be 
\tau^{\alpha\beta}(x) = \frac{\varepsilon^{\alpha\alpha'}\varepsilon^{\beta\beta'}\tau_{\alpha'\beta'}(x)}{\det(\tau_{\gamma\delta}(x))}\,.
\ee 
This implies that we can write
\be 
 G_{\alpha\beta}(X) &=&c^2 \tau_{\alpha\gamma}(x)\left(\delta^\gamma_\beta + \frac{1}{c^2}\tau^{\gamma\delta}(x)H_{\delta\beta}(x,y) + \frac{1}{c^4}\tau^{\gamma\delta}(x)\phi_{\delta\beta}(x,y,z) \right) +\mathcal{O}(c^{-4})\nn\\
&=:& c^2 \tau_{\alpha\gamma}(x)\left(\delta^\gamma_\beta + \frac{1}{c^2}M^\gamma{_\beta} \right)\,,\nn
\ee 
where the last equality defines the $c$-dependent matrix $M$. The determinant of the term in parenthesis expands as follows
\be
\det(1 + c^{-2}M ) = 1 + c^{-2} \Tr[M] + \frac{c^{-4}}{2}\left(\left(\Tr[M]\right)^2 - \Tr[M^2] \right)\,,
\ee 
where $\Tr[M] = M^\alpha{_\alpha}$. This means that 
\be 
\sqrt{-\det G_{\alpha\beta}(X)} 
&=& c^2\sqrt{-\tau}\left(1 + \frac{1}{2 c^2}\Tr[M] + \frac{1}{8c^4}\left(\left(\Tr[M]\right)^2 - 2\Tr[M^2] \right) + \mathcal{O}(c^{-6})\right)\,,\nn\\
\ee 
where we defined $\tau=\det \tau_{\alpha\beta}$, and where all higher-order terms involve powers of traces of $M$ or traces of powers of $M$. Writing out $M$ explicitly, we get
\be 
\mathcal{L}_{\text{NG}}[X;c] &=& -c^2 T_{\text{eff}}\sqrt{-\tau(x)} - \frac{T_{\text{eff}}}{2}\sqrt{-\tau(x)}\tau^{\alpha\beta}(x)H_{\alpha\beta}(x,y)\nn\\
&& - c^{-2}\frac{T_{\text{eff}}}{8}\sqrt{-\tau(x)}\Big[4\tau^{\alpha\beta}(x)\phi_{\alpha\beta}(x,y,z) + (\tau^{\alpha\beta}(x)H_{\alpha\beta}(x,y))^2 \nonumber\\
&&-2\tau^{\alpha\beta}(x)\tau^{\gamma\delta}(x)H_{\alpha\gamma}(x,y)H_{\delta\beta}(x,y) \Big] + \mathcal{O}(c^{-4})\,.
\ee 
Based on our previous considerations, we find that
\begin{subequations}
\be 
\mathcal{L}_{\text{NG-LO}} &=& -T_{\text{eff}} \sqrt{-\tau(x)}\,,\label{eq:LONGLagrangian}\\
\mathcal{L}_{\text{NG-NLO}} &=& -\frac{T_{\text{eff}}}{2} \sqrt{-\tau(x)}\tau^{\alpha\beta}(x)H_{\alpha\beta}(x) + y^M\frac{\delta \mathcal{L}_{\text{NG-LO}}}{\delta x^M} \,,\label{eq:NLONGLagrangian}\\
\mathcal{L}_{\text{NG-NNLO}} &=& -\frac{T_{\text{eff}}}{8} \sqrt{-\tau (x)}\Big[4\tau^{\alpha\beta}(x)\phi_{\alpha\beta}(x) + (\tau^{\alpha\beta}(x)H_{\alpha\beta}(x))^2 \nn\\
&&- 2\tau^{\alpha\beta}(x)\tau^{\gamma\delta}(x)H_{\alpha\gamma}(x)H_{\delta\beta}(x) \Big]\nn\\
&&+z^M\frac{\delta \mathcal{L}_{\text{NG-LO}}}{\delta x^M} + y^M\frac{\delta \mathcal{L}_{\text{NG-NLO}}}{\delta x^M} - \frac{1}{2}y^M y^N\frac{\delta^2 \mathcal{L}_{\text{NG-LO}}}{\delta x^M \delta x^N}\,.\label{eq:NNLONGLagrangian}
\ee 
\end{subequations}

\subsection{The leading order equation of motion}
We now turn our attention to the equation of motion of the NG-LO Lagrangian~\eqref{eq:LONGLagrangian}, which is imposed by $y^M$ in the NG-NLO Lagrangian~\eqref{eq:NLONGLagrangian}. Working up to total derivatives, we find that 
\be 
y^M\frac{\delta \mathcal{L}_{\text{NG-LO}}}{\delta x^M}
 &=& -T_{\text{eff}}\D_\alpha y^M \sqrt{-\tau}\tau^{\alpha\beta}\D_\beta x^N\tau_{MN} - \frac{T_{\text{eff}}}{2}\sqrt{-\tau}\tau^{\alpha\beta}\D_\alpha x^N\D_\beta x^L y^M \D_M\tau_{NL}\nn\\
&=& -T_{\text{eff}}\varepsilon^{\alpha\beta}\varepsilon_{AB}y^M\D_\alpha x^K\D_\beta x^L\left[\tau_M{^A}\partial_K\tau_L{^B}-\tau_L{^B}\left(\partial_{M}\tau_{K}{^A}-\partial_{K}\tau_{M}{^A}\right)\right]\nn\\
&=& -\frac{1}{2}T_{\text{eff}}\varepsilon^{\alpha\beta}\varepsilon_{AB} y^M \D_\alpha x^K \D_\beta x^L\nn\\
&&\times\left[ \tau_M{^A}(\D_K\tau_L{^B} - \D_L\tau_K{^B}) + \tau_L{^B}( \D_K\tau_M{^A} - \D_M\tau_K{^A} ) \right.\nn\\
&&\left.- \tau_K{^B}( \D_L\tau_M{^A} - \D_M \tau_L{^A} ) \right]\,,\nn
\ee
where we used the relation for the inverse longitudinal vielbeine
\be
\tau^\alpha{_A} &=& \frac{1}{\sqrt{-\tau}}\varepsilon^{\alpha\beta} \tau_\beta{^B}\varepsilon_{BA}\,.\label{eq:invtau}
\ee 
Defining the quantity 
\begin{equation}
\label{eq:def-of-F}
    F_{MN} := \varepsilon_{AB}\tau_M{^A}\tau_N{^B}\,,
\end{equation}
we can write the LO equation of motion as
\begin{equation}
    y^M\frac{\delta \mathcal{L}_{\text{NG-LO}}}{\delta x^M} = -\frac{1}{2}T_{\text{eff}}\varepsilon^{\alpha\beta} y^M \D_\alpha x^K \D_\beta x^L\left[ \D_K F_{ML} + \D_L F_{KM} + \D_M F_{LK} \right]\,.
\end{equation}
The equation of motion for $x$ obtained from the NG-LO Lagrangian can thus be written as
\begin{equation}
\label{eq:finalLOxEOM}
   \varepsilon^{\alpha\beta}\D_\alpha x^K\D_\beta x^L (dF)_{MKL} =0\,,
\end{equation}
where $F$ is the 2-form
\begin{equation}
    F=\frac{1}{2}\varepsilon_{AB}\tau^A\wedge \tau^B\,.
\end{equation}
If we contract this with  $\partial_\gamma x^M$, the fact that the worldsheet indices $\alpha,\beta,\gamma$ only take two values implies that the pullback vanishes identically. The fact that the pullback of the LO equation of motion is identically satisfied is a consequence of two-dimensional reparameterisation invariance of the Lagrangian.

\subsection{Co-dimension-$2$ foliations from the string beta function}
\label{sec:codim-2-foliations}
Conformal invariance of the relativistic quantum string theory requires that the $\beta$-functions vanish. In particular, the $\beta$-function for $G_{MN}$ vanishes, which to leading order in $\alpha'$ is equivalent to the vacuum Einstein equations
\begin{equation}
    0=\beta_{MN}(G) = \alpha' R_{MN} + \mathcal{O}(\alpha'^2)\,,
\end{equation}
where $R_{MN}$ is the Ricci tensor of $G_{MN}$, and where we ignored the Kalb--Ramond $B$-field and the dilaton. In this section, we demonstrate that the LO part of the vacuum Einstein equation implies that the $1/c^2$ expanded geometry admits a co-dimension-$2$ foliation defined by $\tau^A$. To see this, we use equations \eqref{eq:metricdecomp}, and study the behaviour of the Christoffel symbols and the Ricci tensor components to leading order in $1/c^2$ using that the fields on the RHS of \eqref{eq:metricdecomp} admit Taylor expansions. This leads to  
\begin{subequations}
\begin{eqnarray}
G_{MN} & = & c^2 T_{MN}+\Pi^\perp_{MN}\,,\\
G^{MN} & = & c^{-2} T^{MN}+\Pi^\perp{}^{MN}\,,\\
\Gamma^P_{MN} & = & \frac{c^2}{2}\Pi_\perp^{PQ}\left(\partial_M T_{NQ}+\partial_N T_{MQ}-\partial_Q T_{MN}\right)+\mathcal{O}(1)\,,\\
R_{MN} & = & -\Gamma^Q_{MR}\Gamma^R_{QP}+\mathcal{O}(c^2)\,,
\end{eqnarray}
\end{subequations}
where we defined $T_{MN}=\eta_{AB}\tau_M{}^A\tau_N{}^B$ and where $\Gamma^P_{MN}$ are the Christoffel symbols of the Levi-Civita connection of $G_{MN}$. The LO vacuum Einstein equations $R_{MN}=0$ give
\begin{equation}
H^\perp{}^{QS}H^\perp{}^{RT}\eta_{AB}\eta_{CD}\tau_M{^A}\tau_P{^C}\left(d\tau^B\right)_{RS}\left(d\tau^D\right)_{QT}=0\,,
\end{equation}
where $H^\perp{}^{MN}$ is the leading order term in the expansion of $\Pi^{\perp MN}$ which obeys $H^\perp{}^{MN}\tau_M{}^A=0$ since $\Pi^{\perp MN}T_M{}^A=0$.

Contracting with $\tau^M{}_E\tau^P{}_F$ and dropping the invertible $\eta_{AB}$ metrics we obtain
\begin{equation}
H^\perp{}^{QS}H^\perp{}^{RT}\left(d\tau^B\right)_{RS}\left(d\tau^D\right)_{QT}=0\,. 
\end{equation}
This is a sum of squares for $B=D=0,1$ and so it is equivalent to 
\begin{equation}
H^\perp{}^{QS}H^\perp{}^{RT}\left(d\tau^A\right)_{RS}=0\,. 
\end{equation}
This in turn is equivalent to
\begin{equation}\label{eq:Frobenius}
    d\tau^A=\alpha^A{}_B\wedge \tau^B\,,
\end{equation}
for arbitrary $1$-forms $\alpha^A{}_B$. We recognise this as the Frobenius integrability condition~\eqref{eq:Frobenius-thm} for a co-dimension-$2$ foliation of $d$-dimensional Riemannian leaves with normal $1$-forms $\tau_M{}^A$.{}\footnote{We thank Jos\'e Figueroa-O'Farrill for useful discussions on this point.} 

If we assume that~\eqref{eq:Frobenius} holds, $dF$ in the LO equation of motion for $x^M$, equation \eqref{eq:finalLOxEOM}, becomes
\begin{equation}\label{eq:dF}
    dF=-\tau^0\wedge\tau^1\wedge \alpha^A{}_A\,.
\end{equation}
We are interested in finding a condition on the target space geometry (that is independent of the embedding maps) such that \eqref{eq:finalLOxEOM} vanishes identically for all embedding maps $x^M$. This condition is $dF=0$ (see also \cite{Harmark:2019upf}) and by equation \eqref{eq:dF} this will be the case if and only if 
\begin{eqnarray}
\alpha^A{}_A=\tau^A X_A\,,
\end{eqnarray}
for some 0-form $X_A$. A simple sufficient conditions is to take $\alpha^A{}_B$ to be traceless.

When we add a $B$-field the $\beta$-functions of the relativistic string sigma model change and then the results depend on how we expand the $B$-field (see Section~\ref{Sec:WZ} for more details).

\subsection{Polyakov action}
The relativistic Polyakov action is
\be 
\label{eq:rel-polyakov}
S_{\text{P}}[X;c] = \int_\Sigma d ^2\sigma\mathcal{L}_{\text{P}}[X;c] =- \frac{cT}{2}\int_\Sigma d ^2\sigma \sqrt{-\gamma}\gamma^{\alpha\beta}\D_\alpha X^M \D_\beta X^N G_{MN}(X)\,.
\ee 
In accordance with previous results, this leads to the expansion
\be 
\mathcal{L}_{\text{P}} &=& -\frac{T_{\text{eff}}}{2}\sqrt{-\gamma}\left[c^2\gamma^{\alpha\beta}\tau_{\alpha\beta}(x) + \gamma^{\alpha\beta}H_{\alpha\beta}(x,y)+c^{-2}\gamma^{\alpha\beta} \phi_{\alpha\beta}(x,y,z)+\mathcal{O}(c^{-4})\right]\,.\nn\\
&&\label{eq:rel-Pol-Lagrangian}
\ee 
In the relativistic string theory, the fiducial worldsheet metric $\gamma_{\alpha\beta}$ is on-shell equivalent (up to a local rescaling) to the pullback of the target space metric $G_{\alpha\beta}$, which expands according to \eqref{eq:exp-of-pulled-back-rel-metric}. Therefore, we assume a similar expansion for $\gamma_{\alpha\beta}$, namely
\be 
\gamma_{\alpha\beta} =  \gamma_{(0)\alpha\beta} +c^{-2} \gamma_{(2)\alpha\beta} + c^{-4}\gamma_{(4)\alpha\beta} + \cdots\,,
\ee 
where the LO component $\gamma_{(0)\alpha\beta}$ is a Lorentzian metric,
while the subleading components $\gamma_{(2)\alpha\beta}$ and $\gamma_{(4)\alpha\beta}$ are symmetric tensors. The inverse worldsheet metric expands as
\begin{equation}\label{eq:ExpInvWSmetric}
\gamma^{\alpha\beta} =
\gamma_{(0)}^{\alpha\beta} - c^{-2}\gamma_{(2)}^{\alpha\beta} + c^{-4}\left[ \gamma_{(2)}^{\alpha\gamma}\gamma_{(2)\gamma}{^\beta}  - \gamma_{(4)}^{\alpha\beta} \right] + \mathcal{O}(c^{-6})\,,
\end{equation}
where we raised indices on $\gamma_{(2)}$ and $\gamma_{(4)}$ using $\gamma_{(0)}^{\alpha\beta}$. This means that the expansion of the worldsheet metric determinant becomes
\begin{equation}\label{eq:ExpOfMetricDet}
\sqrt{-\gamma} = \sqrt{-\gamma_{(0)}}\left[1 + c^{-2}\frac{1}{2}\gamma_{(2)\alpha}^\alpha\nn+ c^{-4}\frac{1}{8}\left[4 \gamma_{(4)\alpha}^\alpha + (\gamma_{(2)\alpha}^\alpha)^2 - 2\gamma_{(2)}^{\alpha\beta}\gamma_{(2)\alpha\beta} \right] \right]+ \mathcal{O}(c^{-6})\,.
\end{equation}
Expanding all field quantities appropriately in $1/c^2$, the Polyakov Lagrangian acquires the following expansion
\be 
\mathcal{L}_{\text{P}} = c^2 \mathcal{L}_{\text{P-LO}} + \mathcal{L}_{\text{P-NLO}} + c^{-2}\mathcal{L}_{\text{P-NNLO}} + \mathcal{O}(c^{-4})\,,
\ee 
where the leading order Polyakov Lagrangian density is given by
\be 
\mathcal{L}_{\text{P-LO}} = -\frac{T_{\text{eff}}}{2}\sqrt{-\gamma_{(0)}} \,\gamma_{(0)}^{\alpha\beta} \tau_{\alpha\beta}(x)\,.\label{eq:LOPLagrangian}
\ee 
The equation of motion for $x^M$ of $\mathcal{L}_{\text{P-LO}}$ is
\be 
0 &=& \frac{1}{\sqrt{-\gamma_{(0)}}}\D_\alpha\left(\sqrt{-\gamma_{(0)}}\gamma^{\alpha\beta}_{(0)}\D_\beta x^N\tau_{MN}(x) \right) - \frac{1}{2}\gamma^{\alpha\beta}_{(0)}\D_\alpha x^L\D_\beta x^N \D_M \tau_{LN}(x)\,,\label{eq:LOPEOM}
\ee 
while the (Virasoro) constraint from integrating out $\gamma_{(0)\alpha\beta}$ is
\be 
T^{(0)}_{\alpha\beta} = -\frac{2}{T_{\text{eff}}\sqrt{-\gamma_{(0)}}}\frac{\delta \mathcal{L}_{\text{P-LO}} }{\delta \gamma_{(0)}^{\alpha\beta}} = \tau_{\alpha\beta} - \frac{1}{2}\gamma_{(0)}^{\gamma\delta}\tau_{\gamma\delta}\gamma_{(0)\alpha\beta} = 0\,.\label{eq:LOVirasoro}
\ee 
Substituting this into the P-LO Lagrangian density \eqref{eq:LOPLagrangian} gives the NG-LO Lagrangian density \eqref{eq:LONGLagrangian}.

At NLO, the Polyakov Lagrangian becomes
\be
\mathcal{L}_{\text{P-NLO}} &=& -\frac{T_{\text{eff}}}{2}\sqrt{-\gamma_{(0)}} \,\gamma_{(0)}^{\alpha\beta} H_{\alpha\beta}(x,y) + \frac{T_{\text{eff}}}{2}\sqrt{-\gamma_{(0)}} \, \tau_{\alpha\beta}\gamma_{(2)\gamma\delta}\left[\gamma_{(0)}^{\alpha\gamma}\gamma_{(0)}^{\delta\beta} - \frac{1}{2}\gamma_{(0)}^{\alpha\beta}\gamma_{(0)}^{\gamma\delta} \right]\nn\\
&=&-\frac{T_{\text{eff}}}{2}\sqrt{-\gamma_{(0)}} \left(\gamma_{(0)}^{\alpha\beta} H_{\alpha\beta}(x) - \frac{1}{2}G_{(0)}^{\alpha\beta\gamma\delta} \tau_{\alpha\beta}\gamma_{(2)\gamma\delta}\right) + y^M\frac{\delta\mathcal{L}_{\text{P-LO}}}{\delta x^M}\,,\label{eq:NLOPLagrangian}
\ee 
where the second equality is up to total derivatives and where we introduced the Wheeler--DeWitt (WDW) metric
\be 
G_{(0)}^{\alpha\beta\gamma\delta} = \gamma_{(0)}^{\alpha\gamma}\gamma_{(0)}^{\delta\beta} + \gamma_{(0)}^{\alpha\delta}\gamma_{(0)}^{\gamma\beta}- \gamma_{(0)}^{\alpha\beta}\gamma_{(0)}^{\gamma\delta}\,,\label{eq:WdWdef}
\ee 
which has the following symmetries
\be 
G_{(0)}^{\alpha\beta\gamma\delta} =G_{(0)}^{\beta\alpha\gamma\delta}\,,\qquad G_{(0)}^{\alpha\beta\gamma\delta} = G_{(0)}^{\alpha\beta\delta\gamma}\,,\qquad G_{(0)}^{\alpha\beta\gamma\delta} = G_{(0)}^{\gamma\delta\alpha\beta}\,.
\ee 
Varying $\gamma_{(2)\alpha\beta}$ in the NLO Lagrangian produces the LO Virasoro constraint \eqref{eq:LOVirasoro}, while varying $\gamma_{(0)\alpha\beta}$ leads to the NLO Virasoro constraint,
\be 
\label{eq:NLOVirasoro}
\begin{aligned}
T^{(2)}_{\alpha\beta} =& -\frac{2}{T_{\text{eff}}\sqrt{-\gamma_{(0)}}}\frac{\delta \mathcal{L}_{\text{P-NLO}} }{\delta \gamma_{(0)}^{\alpha\beta}}\nn\\
=& H_{\alpha\beta}(x,y) - \frac{1}{2}\gamma_{(0)}^{\gamma\delta} H_{\gamma\delta}(x,y)\gamma_{(0)\alpha\beta} + \frac{1}{4}\tau_{\alpha'\beta'}\gamma_{(2)\gamma\delta}G_{(0)}^{\alpha'\beta'\gamma\delta}\gamma_{(0)\alpha\beta}\nn\\
&-2\tau_{\gamma(\alpha}\gamma_{(2)\beta)}{^\gamma} + \frac{1}{2}\tau_{\alpha\beta}\gamma^\gamma_{(2)\gamma} + \frac{1}{2}\gamma_{(2)\alpha\beta}\gamma_{(0)}^{\gamma\delta}\tau_{\gamma\delta}
= 0\,.
\end{aligned}
\ee 
If we contract this with $\gamma_{(0)}^{\alpha\beta}$, we get
\be 
\gamma^{\alpha\beta}_{(0)}T^{(2)}_{\alpha\beta} &=& 
-\gamma_{(2)}^{\alpha\beta} T_{\alpha\beta}^{(0)}\,.
\ee 
As we will see later in \eqref{eq:NLO-Weyl-WI}, this is the Ward identity corresponding to Weyl symmetry of the P-NLO action.

Repeating this exercise at NNLO leads to
\be \label{eq:LagPNNLO}
\mathcal{L}_{\text{P-NNLO}} &=& 
-\frac{T_{\text{eff}}}{2}\sqrt{-\gamma_{(0)}} \bigg(\gamma_{(0)}^{\alpha\beta} \phi_{\alpha\beta}(x,y,z) - \frac{1}{2}G_{(0)}^{\alpha\beta\gamma\delta}\left[\tau_{\alpha\beta}(x)\gamma_{(4)\gamma\delta} +H_{\alpha\beta}(x,y)\gamma_{(2)\gamma\delta} \right]\nn\\
&&+\frac{1}{2}\tau_{\alpha\beta}(x)\gamma_{(2)\gamma\delta}\gamma_{(2)\epsilon\eta}\left[\gamma_{(0)}^{\alpha\epsilon}G_{(0)}^{\eta\beta\gamma\delta} - \frac{1}{4}\gamma_{(0)}^{\alpha\beta}G_{(0)}^{\gamma\delta\epsilon\eta} \right]\bigg)\nn\\
&=&-\frac{T_{\text{eff}}}{2}\sqrt{-\gamma_{(0)}} \bigg(\gamma_{(0)}^{\alpha\beta} \phi_{\alpha\beta}(x) - \frac{1}{2}G_{(0)}^{\alpha\beta\gamma\delta}\left[\tau_{\alpha\beta}(x)\gamma_{(4)\gamma\delta} +H_{\alpha\beta}(x)\gamma_{(2)\gamma\delta} \right]\nn\\
&&+\frac{1}{2}\tau_{\alpha\beta}(x)\gamma_{(2)\gamma\delta}\gamma_{(2)\epsilon\eta}\left[\gamma_{(0)}^{\alpha\epsilon}G_{(0)}^{\eta\beta\gamma\delta} - \frac{1}{4}\gamma_{(0)}^{\alpha\beta}G_{(0)}^{\gamma\delta\epsilon\eta} \right]\bigg)\nn\\
&&+z^M\frac{\delta \mathcal{L}_{\text{P-LO}}}{\delta x^M} + y^M\frac{\delta \mathcal{L}_{\text{P-NLO}}}{\delta x^M} - \frac{1}{2}y^M y^N\frac{\delta^2 \mathcal{L}_{\text{P-LO}}}{\delta x^M \delta x^N}\,.
\ee 
Integrating out $\gamma_{(4)}$ now gives the LO Virasoro constraint \eqref{eq:LOVirasoro}, while integrating out $\gamma_{(2)}$ gives the NLO Virasoro constraint \eqref{eq:NLOVirasoro}, and a lengthy calculation shows that using these returns the NG-NNLO Lagrangian \eqref{eq:NNLONGLagrangian}. Integrating out $\gamma_{(0)}$ in the P-NNLO Lagrangian gives the NNLO Virasoro constraint
\be \label{eq:NNLOVirasoro}
\hspace{-0.5cm}T^{(4)}_{\alpha\beta} &=& \phi_{\alpha\beta}(x,y,z) - \frac{1}{2}\gamma_{(0)\alpha\beta}\gamma_{(0)}^{\gamma\delta} \phi_{\gamma\delta}(x,y,z) + \text{terms involving $\gamma_{(2)}$ and $\gamma_{(4)}$}\,,
\ee 
where the terms involving $\gamma_{(2)}$ and $\gamma_{(4)}$ will not be required and we refrain from writing them.

Some general comments about the structure of the constraints are in order. First of all, we note that we could equivalently have expanded the relativistic Virasoro constraints 
\begin{equation}
    G_{\alpha\beta}-\frac{1}{2}\gamma_{\alpha\beta}\gamma^{\alpha'\beta'}G_{\alpha'\beta'}=0\,,
\end{equation}
to obtain the LO, NLO, and NNLO constraints above. Secondly, as we saw, the most subleading Virasoro constraint, i.e., the one obtained by integrating out $\gamma_{(0)}$ in the P-NLO and P-NNLO actions, is not required to obtain the corresponding NG actions. This is because imposing all previous constraints (e.g., the LO constraint at NLO) reduces the most subleading constraint, i.e., the one obtained by integrating out $\gamma_{(0)}$, to an equation for the most subleading $\gamma_{(n)}$, which is a Lagrange multiplier for the LO Virasoro constraint which in turn disappears from the action when imposing all previous constraints.

The worldsheet gauge symmetries of the relativistic Polyakov action are diffeomorphisms generated by $\xi^\alpha$ and Weyl transformations with local parameter $\omega$, which act infinitesimally on the worldsheet metric $\gamma_{\alpha\beta}$ as
\be 
\delta \gamma_{\alpha\beta} = \pounds_\xi \gamma_{\alpha\beta} + 2\omega \gamma_{\alpha\beta}\,.\label{eq:generalWStrafo}
\ee 
We expand these gauge parameters according to
\be 
\begin{aligned}
\xi^\alpha &= \xi_{(0)}^\alpha + c^{-2}\xi_{(2)}^\alpha + c^{-4}\xi_{(4)}^\alpha + \cdots\,,\\
\omega &= \omega_{(0)} + c^{-2}\omega_{(2)} + c^{-4}\omega_{(4)} + \cdots\,,
\end{aligned}
\ee 
and, since we consider closed strings, the spatial worldsheet coordinate $\sigma^1$ is periodically identified, $\sigma^1 \sim \sigma^1 + 2\pi$. The gauge parameters are periodic in $\sigma^1$. Expanding the expression for the relativistic general worldsheet gauge transformation leads to
\begin{subequations}
\be 
\delta \gamma_{(0)\alpha\beta} &=& \pounds_{\xi_{(0)}} \gamma_{(0)\alpha\beta} + 2\omega_{(0)} \gamma_{(0)\alpha\beta}\,,\label{eq:LOgauge-trafo} \\
\delta \gamma_{(2)\alpha\beta} &=& \pounds_{\xi_{(0)}}\gamma_{(2)\alpha\beta} + \pounds_{\xi_{(2)}}\gamma_{(0)\alpha\beta} + 2\omega_{(0)} \gamma_{(2)\alpha\beta} + 2\omega_{(2)} \gamma_{(0)\alpha\beta}\,,\label{eq:NLOgauge-trafo}\\
\delta \gamma_{(4)\alpha\beta} &=& \pounds_{\xi_{(0)}}\gamma_{(4)\alpha\beta} + \pounds_{\xi_{(2)}} \gamma_{(2)\alpha\beta} + \pounds_{\xi_{(4)}}\gamma_{(0)\alpha\beta} + 2\omega_{(0)}\gamma_{(4)\alpha\beta}\nn\\
&&+ 2\omega_{(2)} \gamma_{(2)\alpha\beta} + 2\omega_{(4)} \gamma_{(0)\alpha\beta}\,.\label{eq:NNLOgauge-trafo}
\ee 
\end{subequations}
The diffeomorphisms and Weyl symmetries of the worldsheet data have Ward identities associated to them. At LO, we have
\be 
\delta S_{\text{P-LO}} = \frac{T_{\text{eff}}}{2}\int d ^2\sigma\,\sqrt{-\gamma_{(0)}}\, T^{\alpha\beta}_{(0)} \delta \gamma_{(0)\alpha\beta}\,,
\ee 
where $T^{\alpha\beta}_{(0)}=\gamma_{(0)}^{\alpha\alpha'}\gamma_{(0)}^{\beta\beta'}T_{(0)\alpha'\beta'}$,
and so the Ward identity for LO Weyl transformations tells us that the energy-momentum tensor is traceless
\be 
\gamma_{(0)}^{\alpha\beta}T_{(0)\alpha\beta} = 0\,.
\ee 
For the NLO action, we have
\be 
\delta S_{\text{P-NLO}} = \frac{T_{\text{eff}}}{2}\int d ^2\sigma\,\sqrt{-\gamma_{(0)}}\left( T^{\alpha\beta}_{(2)} \delta \gamma_{(0)\alpha\beta} + T^{\alpha\beta}_{(0)} \delta \gamma_{(2)\alpha\beta}\right)\,,
\ee 
where $T^{\alpha\beta}_{(2)}=\gamma_{(0)}^{\alpha\alpha'}\gamma_{(0)}^{\beta\beta'}T^{(2)}_{\alpha'\beta'}$. The LO Ward identities are reproduced by the subleading Weyl transformations and subleading diffeomorphisms, while the Ward identity for LO Weyl transformations $\omega_{(0)}$ now takes the form
\be 
T^{(2)}_{\alpha\beta}\gamma^{\alpha\beta}_{(0)} + T^{(0)}_{\alpha\beta}\gamma^{\alpha\beta}_{(2)}= 0\,.\label{eq:NLO-Weyl-WI}
\ee 

There are similar Ward identities at the NNLO. Furthermore, there are also Ward identities for the gauge symmetries associated with the $c^{-2}$ expansion of the generator of worldsheet diffeomorphism invariance. We will refrain from writing them down as we will not need them explicitly.

\subsection{Partial gauge fixing}
The symmetries of the Lorentzian LO worldsheet metric $\gamma_{(0)\alpha\beta}$ are exactly identical to those of the relativistic Polyakov action, so we can gauge fix the LO gauge redundancy by locally going to flat gauge
\be 
\gamma_{(0)\alpha\beta} = \eta_{\alpha\beta}\,.\label{eq:LOgauge}
\ee 
The residual gauge transformations at LO are those diffeomorphisms that can be undone by a Weyl transformation $\pounds_{\xi_{(0)}} \eta_{\alpha\beta} + 2\omega_{(0)} \eta_{\alpha\beta} = 0$, and in lightcone coordinates these take the familiar form
\begin{equation}\label{eq:LOresidualgauge}
\xi_{(0)}(\sigma) = \xi_{(0)}^-(\sigma^-)\D_- + \xi_{(0)}^+(\sigma^+)\D_+\,,  
\end{equation}
where $\xi_{(0)}^\pm(\sigma^\pm)$ are periodic in their argument. For the LO Weyl transformation, this corresponds to $\omega_{(0)} = -\frac{1}{2}\D_\alpha \xi_{(0)}^\alpha$.

Turning our attention to the NLO gauge redundancies \eqref{eq:NLOgauge-trafo}, we now have 
\be 
\delta \gamma_{(2)\alpha\beta} =  2\D_{(\alpha}\xi_{(2)\beta)}  + 2\omega_{(2)} \eta_{\alpha\beta}\,,
\ee 
where $\xi_{(2)\beta} = \eta_{\beta\delta}\xi_{(2)}^\delta$, and where we left out the transformations under LO residual gauge transformations $\xi_{(0)}^\pm$. The NLO gauge transformations are independent of $\gamma_{(2)}$ and act as local shifts, allowing us (in 2 dimensions) to locally gauge fix 
\be \label{eq:NLOgaugechoice}
\gamma_{(2)\alpha\beta} = 0\,.
\ee 
The residual gauge transformations satisfy
\be 
2\D_{(\alpha}\xi_{(2)\beta)}  + 2\omega_{(2)} \eta_{\alpha\beta} =0\,,
\ee 
and so take exactly the same form as those at LO, namely 
\begin{equation}\label{eq:NLOresidualgauge}
\xi_{(2)}(\sigma) = \xi_{(2)}^-(\sigma^-)\D_- + \xi_{(2)}^+(\sigma^+)\D_+ \,, 
\end{equation}
with $\omega_{(2)} = -\frac{1}{2}\D_\alpha \xi^\alpha_{(0)}$.
This pattern repeats itself at all orders: in particular, at NNLO, we get
\be
\delta \gamma_{(4)\alpha\beta} &=&  2\D_{(\alpha}\xi_{(4)\beta)}  + 2\omega_{(4)} \eta_{\alpha\beta}\,,
\ee 
where, again, the NNLO gauge transformations act as local shifts, so that we may locally set
\be \label{eq:NNLOgauge}
\gamma_{(4)\alpha\beta} = 0\,,
\ee 
leaving once more the residual gauge transformations 
\be\label{eq:NNLOresidualgauge}
\xi_{(4)}(\sigma) = \xi_{(4)}^-(\sigma^-)\D_- + \xi_{(4)}^+(\sigma^+)\D_+\,.
\ee

The relativistic embedding field $X^M$ transforms as a scalar under worldsheet diffeomorphisms $\xi^\alpha$, $\delta_\xi X^M = \xi^\alpha \D_\alpha X^M$. We can expand these diffeomorphisms to get
\begin{subequations}
\be 
\delta x^M &=& \xi_{(0)}^\alpha\D_\alpha x^M\,,\label{eq:LOembeddingdiffeo}\\
\delta y^M &=& \xi_{(2)}^\alpha\D_\alpha x^M + \xi_{(0)}^\alpha\D_\alpha y^M\,,\label{eq:NLOembeddingdiffeo}\\
\delta z^M &=& \xi_{(4)}^\alpha\D_\alpha x^M + \xi_{(2)}^\alpha\D_\alpha y^M + \xi_{(0)}^\alpha\D_\alpha z^M\,.\label{eq:NNLOembeddingdiffeo}
\ee 
\end{subequations}
These expressions will prove useful when we fix the residual gauge invariances discussed above.

\section{Relating the NLO theory to the Gomis--Ooguri string on a curved target space}
\label{sec:Matching-with-SNC-string}

We will show that for appropriate target spacetimes, the P-NLO Lagrangian \eqref{eq:NLOPLagrangian} can be recast as the string Newton--Cartan (SNC) Polyakov Lagrangian of \cite{Bergshoeff:2018yvt,Bergshoeff:2019ctr}. To do so, we first write $\gamma_{(0)\alpha\beta}$ in terms of vielbeine 
\be 
\gamma_{(0)\alpha\beta} = \eta_{ab}e_\alpha{^a}e_\beta{^b} = -\frac{1}{2}( e_\alpha{^+}e_\beta{^-} + e_\alpha{^-}e_\beta{^+}   )\,,
\ee 
where $a,b = 0,1$ are worldsheet tangent space indices. We also defined the null combinations
\be 
e_\alpha{^\pm} = e_\alpha{^0} \pm e_\beta{^1}\,,
\ee 
which have inverses given by
\be 
e^\alpha{_\pm} = (e^\alpha{_0} \pm e^\alpha{_1})/2\,.
\ee
Note that the Minkowski metric $\eta_{ab}$ and its inverse $\eta^{ab}$ in lightcone coordinates $\sigma^\pm = \sigma^0 \pm \sigma^1$ are
\be 
\begin{aligned}
\eta_{-+} &=& \eta_{+-} = -\frac{1}{2}\,,\qquad \eta_{--} = \eta_{++} = 0\,,\\
\eta^{-+} &=& \eta^{+-} = -2\,,\qquad \eta^{--} = \eta^{++} = 0\,.
\end{aligned}
\ee 
Similarly, the Levi-Civita symbol $\varepsilon^{ab}$ in lightcone coordinates is
\be 
\label{eq:levicivitalightcone}
\begin{aligned}
\varepsilon^{-+} &=& -\varepsilon^{+-} = 2\,,\qquad \varepsilon^{++} = \varepsilon^{--} = 0\,,\\
\varepsilon_{+-} &=& -\varepsilon_{-+} = \frac{1}{2}\,,\qquad \varepsilon_{++} = \varepsilon_{--} = 0\,,
\end{aligned}
\ee 
where we use the convention $\varepsilon^{01}=1=-\varepsilon_{01}$. Note furthermore the useful identity
\begin{equation}
    \varepsilon^{\alpha\gamma}\varepsilon_{\gamma\beta} = \delta^\alpha_\beta\,,
\end{equation}
which we will use extensively in the present section.

Consider the NLO Polyakov Lagrangian in \eqref{eq:NLOPLagrangian}. We define the Lagrange multiplier Lagrangian
\be 
\mathcal{L}_{\text{LM}} = \frac{T_{\text{eff}}}{4}\sqrt{-\gamma_{(0)}} \,G_{(0)}^{\alpha\beta\gamma\delta} \tau_{\alpha\beta}\gamma_{(2)\gamma\delta}\,,
\ee 
with Lagrange multiplier $\gamma_{(2)\alpha\beta}$.
We can write $\gamma_{(2)\alpha\beta}$ in terms of worldsheet vielbeine as follows 
\be 
\gamma_{(2)\alpha\beta}&=& e_{\alpha}{^a}e_{\beta}{^b}A_{ab}\,,
\ee
where $A_{ab}=A_{ba}$.
This leads to
\be 
\mathcal{L}_{\text{LM}} = \frac{T_{\text{eff}}}{2}  \sqrt{-\gamma_{(0)}} G_{(0)}^{abcd}e^\alpha{}_a\left(-\tau_\alpha{}^0+\tau_\alpha{}^1 \right)e^\beta{_b}\left(\tau_\beta{^0} + \tau_\beta{^1}  \right) A_{cd}\,,
\ee 
where $G_{(0)}^{abcd} = e_\alpha{^a}e_\beta{^b}e_\gamma{^c}e_\delta{^d}G_{(0)}^{\alpha\beta\gamma\delta}$. We factorised $\tau_{\alpha\beta}$ using the symmetry properties of the WDW metric (the cross term $\tau_\alpha{^0}\tau_\beta{^1}-\tau_\beta{^0}\tau_\alpha{^1}$ drops out due to its antisymmetry in $\alpha$ and $\beta$). The flat WDW metric in lightcone coordinates takes the form
\be 
G_{(0)}^{abcd} = \eta^{ac}\eta^{bd} + \eta^{ad}\eta^{bc}- \eta^{ab}\eta^{cd}\,,
\ee 
and has only two nonzero components
\be 
G_{(0)}^{++--} = G_{(0)}^{--++} = 8\,.
\ee 
This means that the Lagrange multiplier Lagrangian involves only two constraints imposed by $A_{++}$ and $A_{--}$, i.e., $A_{+-}$ does not contribute. In other words, we can write the Lagrange multiplier Lagrangian as
\be 
\begin{aligned}
\mathcal{L}_{\text{LM}} =& 4T_{\text{eff}}\, \sqrt{-\gamma_{(0)}}\, e^\alpha{_-}\left(-\tau_\alpha{^0}+\tau_\alpha{^1} \right)e^\beta{_-}\left(\tau_\beta{^0} + \tau_\beta{^1}  \right) A_{++}\\
&+4T_{\text{eff}}\, \sqrt{-\gamma_{(0)}}\, e^\alpha{_+}\left(-\tau_\alpha{^0}+\tau_\alpha{^1} \right)e^\beta{_+}\left(\tau_\beta{^0} + \tau_\beta{^1}  \right) A_{--}\,.
\end{aligned}
\ee 

The constraints imposed by $A_{++}$ and $A_{--}$ are
\be
\begin{aligned}
A_{++}:\qquad 0&= e^\alpha{_-}\left(-\tau_\alpha{^0}+\tau_\alpha{^1} \right)e^\beta{_-}\left(\tau_\beta{^0} + \tau_\beta{^1}  \right)\,,\\
A_{--}:\qquad 0&=e^\alpha{_+}\left(-\tau_\alpha{^0}+\tau_\alpha{^1} \right)e^\beta{_+}\left(\tau_\beta{^0} + \tau_\beta{}^1  \right)\,.
\end{aligned}
\ee 
The $e^\alpha{}_\pm$ projections of  $\left(-\tau_\alpha{}^0+\tau_\alpha{^1} \right)$ cannot be both zero and likewise for $\left(\tau_\alpha{^0}+\tau_\alpha{^1} \right)$ because this would imply that $\tau_\alpha{^A}$ is not invertible which contradicts our assumption that the pullback metric $\tau_{\alpha\beta}$ is non-degenerate. Without loss of generality we choose the constraints to be
\be 
e^\alpha{_+}\left(-\tau_\alpha{}^0+\tau_\alpha{}^1 \right) = 0 = e^\beta{_-}\left(\tau_\beta{}^0 + \tau_\beta{^1} \right)\,,\label{eq:ConstraintChoice}
\ee 
with the other projections nonzero, i.e.,
\be 
e^\alpha{_-}\left(-\tau_\alpha{}^0+\tau_\alpha{}^1 \right) \neq 0\,,\qquad e^\beta{_+}\left(\tau_\beta{}^0 + \tau_\beta{^1} \right)\neq 0\,.\label{eq:ConstraintChoice2}
\ee 
Now, using 
\be 
e^\alpha{_a} = \frac{1}{ \sqrt{-\gamma_{(0)}}}\varepsilon^{\alpha\beta}e_{\beta}{^b}\varepsilon_{ba}\,,
\ee 
as well as our previous results for the Levi-Civita symbol in lightcone coordinates, we find that
\be 
e^\beta{_-} = \frac{1}{2 \sqrt{-\gamma_{(0)}}}\varepsilon^{\beta\gamma}e_\gamma{^+}\,,\qquad e^\beta{_+} = -\frac{1}{2  \sqrt{-\gamma_{(0)}} }\varepsilon^{\beta\gamma}e_\gamma{^-}\,.
\ee 
If we substitute this into the Lagrange multiplier Lagrangian we obtain
\be 
\mathcal{L}_{\text{LM}} &=& 2T_{\text{eff}} \,e^\alpha{_-}\left(-\tau_\alpha{^0}+\tau_\alpha{^1} \right)\varepsilon^{\beta\gamma}e_\gamma{^+} \left(\tau_\beta{^0} + \tau_\beta{}^1  \right) A_{++}\nn\\
&&+2T_{\text{eff}}\, \varepsilon^{\alpha\gamma}e_\gamma{^-}\left(\tau_\alpha{}^0-\tau_\alpha{^1} \right)e^\beta{_+}\left(\tau_\beta{^0} + \tau_\beta{^1}  \right) A_{--}\,.
\ee 
Using \eqref{eq:ConstraintChoice2} we can make the following field redefinitions
\be 
\tilde\lambda_{++} = 4e^\alpha{_-}\left(-\tau_\alpha{^0}+\tau_\alpha{^1} \right)A_{++}\,,\qquad \tilde\lambda_{--} = 4e^\beta{_+}\left(\tau_\beta{^0} + \tau_\beta{^1}  \right) A_{--}\,,
\ee 
which leads to
\be 
\mathcal{L}_{\text{LM}} = -\frac{T_{\text{eff}}}{2}\left[ \tilde\lambda_{++}\varepsilon^{\alpha\beta}e_\alpha{^+}\tau_\beta{^+}  +\tilde\lambda_{--}\varepsilon^{\alpha\beta}e_\alpha{^-}\tau_\beta{^-} \right]\,,
\ee 
where we have introduced the lightcone combinations
\be 
\tau_\beta{^\pm} &=& \tau_\beta{^0} \pm \tau_\beta{^1}\,.
\ee 
The P-NLO Lagrangian \eqref{eq:NLOPLagrangian} can thus be written as
\be 
\hspace{-1cm}\mathcal{L}_{\text{P-NLO}} &=&
-\frac{T_{\text{eff}}}{2}  \sqrt{-\gamma_{(0)}} \gamma^{\alpha\beta}_{(0)}H_{\alpha\beta}
-\frac{T_{\text{eff}}}{2}\left[ \tilde\lambda_{++}\varepsilon^{\alpha\beta}e_\alpha{^+}\tau_\beta{^+} +\tilde\lambda_{--}\varepsilon^{\alpha\beta}e_\alpha{^-}\tau_\beta{^-} \right]\nn\\
&&+y^M\frac{\delta \mathcal{L}_{\text{P-LO}}}{\delta x^M}
\,.\label{eq:NLOPLagrangian-SNC}
\ee
A somewhat tedious calculation shows that the last term in the Lagrangian above can be recast as
\begin{align}
    \begin{split}
        y^M\frac{\delta \mathcal{L}_{\text{P-LO}}}{\delta x^M} & = 
        y^M\frac{\delta\mathcal{L}_{\text{NG-LO}}}{\delta x^M}\\
 &\quad-T_{\text{eff}}\varepsilon^{\alpha\gamma}\tau_\alpha{^+}e_\gamma{^+}e^\beta{_+}\left(\tau_M{^-}\partial_\beta y^M+y^M\partial_\beta x^N \D_M\tau_N{^-}\right)\\
 &\quad+T_{\text{eff}}\varepsilon^{\alpha\gamma}\tau_\alpha{^-}e_\gamma{^-}e^\beta{_-}\left(\tau_M{^+}\partial_\beta y^M+y^M\partial_\beta x^N \D_M\tau_N{^+}\right)\,.
    \end{split}
\end{align}
With this, the P-NLO Lagrangian becomes
\be 
\hspace{-0.5cm}\mathcal{L}_{\text{P-NLO}} 
& = & -\frac{T_{\text{eff}}}{2}\left( \sqrt{-\gamma_{(0)}} \gamma^{\alpha\beta}_{(0)} H_{\alpha\beta}(x)
+ \lambda_{++} \varepsilon^{\alpha\beta}e_\alpha{^+}\tau_\beta{^+} +\lambda_{--}\varepsilon^{\alpha\beta}e_\alpha{^-}\tau_\beta{^-} \right)\nn\\
&&+y^M\frac{\delta\mathcal{L}_{\text{NG-LO}}}{\delta x^M}\,,
\ee 
where we defined the combinations
\be 
\begin{aligned}
\lambda_{++} &= \tilde\lambda_{++}  - 2 e^\gamma{_+}(\tau_N{^-}\D_\gamma y^N + \D_\gamma x^M y^N \D_N \tau_M{^-} )\,,\\
\lambda_{--} &= \tilde\lambda_{--} + 2e^\gamma{_-}(\tau_N{^+}\D_\gamma y^N + \D_\gamma x^M y^N \D_N\tau_M{^+})\,.
\end{aligned}
\ee 
This Lagrangian reduces to that of the SNC string of \cite{Bergshoeff:2018yvt} whenever the strong foliation constraint holds. This is because as shown in appendix~\ref{sec:gauge-structure} the target space geometry becomes type I or ordinary string Newton--Cartan geometry when $\alpha^A{}_B$ is proportional to $\varepsilon^A{}_B$, i.e. when the strong foliation constraint holds. In this case the $y$-term vanishes identically. However the $y$-term is also identically zero under the weaker condition that the target space admits a co-dimension-2 foliation for which $\alpha^A{}_B$ is traceless. In this more general case the target space is still a type II SNC geometry that is not equivalent to a type I SNC geometry. We thus see that the Gomis--Ooguri string can be generalised to a string moving in type II SNC geometry for which $\alpha^A{}_B$ is traceless.

\section{The Wess--Zumino term}\label{Sec:WZ}
The Wess--Zumino (WZ) Lagrangian density reads 
\be 
\mathcal{L}_{\text{WZ}}=-\frac{cT}{2}\varepsilon^{\alpha\beta} \partial_\alpha X^M \partial_\beta X^N B_{MN}(X)\,.
\ee 
We expand the Kalb--Ramond field according to
\be 
B_{MN} = c^2B_{(-2)MN} + B_{(0)MN} + c^{-2}B_{(2)MN} + \mathcal{O}(c^{-4})\,.\label{eq:KalbRamondExp}
\ee
We do not include terms of order $c^{4}$ as these would be more divergent than the LO terms in the NG action. The WZ Lagrangian acquires the following expansion
\be 
\mathcal{L}_{\text{WZ}} = c^2 \mathcal{L}_{\text{WZ-LO}} + \mathcal{L}_{\text{WZ-NLO}} + c^{-2}\mathcal{L}_{\text{WZ-NLO}} + \mathcal{O}(c^{-4})\,,\label{eq:expansion-of-WZ}
\ee 
where the Lagrangians that appear at each order in $c$ are given by
\begin{subequations}
\be 
\mathcal{L}_{\text{WZ-LO}} &=& -\frac{T_{\text{eff}}}{2}\varepsilon^{\alpha\beta} B_{(-2)\alpha\beta}\,,\\
\mathcal{L}_{\text{WZ-NLO}} &=& -\frac{T_{\text{eff}}}{2}\varepsilon^{\alpha\beta} B_{(0)\alpha\beta} + y^M\frac{\delta \mathcal{L}_{\text{WZ-LO}}}{\delta x^M}\,,\\
\mathcal{L}_{\text{WZ-NNLO}} &=& -\frac{T_{\text{eff}}}{2}\varepsilon^{\alpha\beta} B_{(2)\alpha\beta} + z^M\frac{\delta \mathcal{L}_{\text{WZ-LO}}}{\delta x^M} + y^M\frac{\delta \mathcal{L}_{\text{WZ-NLO}}}{\delta x^M} - \frac{1}{2}y^M y^N\frac{\delta^2 \mathcal{L}_{\text{WZ-LO}}}{\delta x^M\delta x^N}\,.\nn\\
\ee 
\end{subequations}

\subsection{Cancelling the LO order Nambu--Goto Lagrangian}
The addition of a WZ term modifies the LO $x^M$ equation of motion because we have
\be 
\frac{\delta \mathcal{L}_{\text{WZ-LO}}}{\delta x^M} &=& T_{\text{eff}}\varepsilon^{\alpha\beta}\D_\alpha x^N \D_\beta B_{(-2)NM} + \frac{T_{\text{eff}}}{2}\varepsilon^{\alpha\beta}\D_\alpha x^N \D_\beta x^L \D_MB_{(-2)LN}\nn\\
&=& \frac{T_{\text{eff}}}{2}\varepsilon^{\alpha\beta}\D_\alpha x^N \D_\beta x^L H_{(-2)MLN}\,,
\ee 
where 
\be 
H_{(-2)MNL} = 3\D_{[M}B_{(-2)NL]} = \D_{M}B_{(-2)NL} + \D_{N}B_{(-2)LM} + \D_{L}B_{(-2)MN}\,,
\ee 
is the $3$-form field strength of $B_{(-2)MN}$. Combining this with the LO equation of motion in the absence of a Kalb--Ramond field \eqref{eq:finalLOxEOM}, we now find that the LO $x^M$ equation of motion of the combined action $\mathcal{L}_{\text{NG-LO}}+\mathcal{L}_{\text{WZ-LO}}$ becomes
\be 
\frac{\delta \mathcal{L}_{\text{NG-LO}}}{\delta x^M}+\frac{\delta \mathcal{L}_{\text{WZ-LO}}}{\delta x^M}=0\,,
\ee
which we can write as
\be 
\varepsilon^{\alpha\beta}\D_\alpha x^K\D_\beta x^L\D_{[M}(F_{KL]} + B_{(-2)KL]})=0\,,
\ee
where we recall that $F_{MN} = \varepsilon_{AB}\tau_M{^A}\tau_N{^B}$. Hence, if $B_{(-2)MN} = -F_{MN}$, the LO $x^M$ equation of motion is identically satisfied without imposing any conditions on $\tau_M{}^A$. In fact, as we will see, for this choice the LO Lagrangian $\mathcal{L}_{\text{NG-LO}}+\mathcal{L}_{\text{WZ-LO}}$ is zero. This corresponds to a $B$-field of the form (see also~\cite{Harmark:2019upf})
\be 
B_{MN} &=& -c^2(T_M{^0}T_N{^1} - T_M{^1}T_N{^0}) + \bar B_{MN}\,,\label{eq:RelatingWZ}
\ee 
where $\bar B_{MN} = \bar B_{(0)MN} + c^{-2}\bar B_{(2)MN}$, and where the components at each order in $1/c^2$ are given by
\be
\begin{aligned}
B_{(-2)MN} &= 2\tau_{[M}{^1}\tau_{N]}{^0}\,,\\
B_{(0)MN} &= 2\tau_{[M}{^1}m_{N]}{^0} - 2\tau_{[M}{^0}m_{N]}{^1} + \bar B_{(0)MN}\,,\\
B_{(2)MN} &= 2\tau_{[M}{^1}B_{N]}{^0} - 2\tau_{[M}{^0}B_{N]}{^1} + 2m_{[M}{^1}m_{N]}{^0} + \bar B_{(2)MN}\,,
\end{aligned}
\ee 
so that the WZ Lagrangians of \eqref{eq:expansion-of-WZ} become
\begin{subequations}
\be 
\mathcal{L}_{\text{WZ-LO}} &=& T_{\text{eff}} \sqrt{-\tau}\,,\label{eq:WZLO}\\
\mathcal{L}_{\text{WZ-NLO}} &=& -\frac{T_{\text{eff}}}{2}\varepsilon^{\alpha\beta}\bar B_{(0)\alpha\beta} + T_{\text{eff}}\varepsilon^{\alpha\beta}( \tau_\alpha{^0}m_\beta{^1} - \tau_\alpha{^1}m_\beta{^0} ) + y^M\frac{\delta \mathcal{L}_{\text{WZ-LO}} }{\delta x^M}\nn\\
&=&-\frac{T_{\text{eff}}}{2}\varepsilon^{\alpha\beta}\bar B_{(0)\alpha\beta} + T_{\text{eff}}\sqrt{-\tau}\tau^{\alpha\beta}( -\tau_\alpha{^0}m_\beta{^0} + \tau_\alpha{^1}m_\beta{^1} ) + y^M\frac{\delta \mathcal{L}_{\text{WZ-LO}} }{\delta x^M}\,,\nn\\\\
\mathcal{L}_{\text{WZ-NNLO}} &=& -\frac{T_{\text{eff}}}{2}\varepsilon^{\alpha\beta}\bar B_{(2)\alpha\beta} + T_{\text{eff}}\varepsilon^{\alpha\beta}( \tau_\alpha{^0}B_\beta{^1} - \tau_\alpha{^1}B_\beta{^0} ) + T_{\text{eff}}\varepsilon^{\alpha\beta}m_\alpha{^0}m_\beta{^1}\nn\\
&&+ z^M\frac{\delta \mathcal{L}_{\text{WZ-LO}} }{\delta x^M} + y^M\frac{\delta \mathcal{L}_{\text{WZ-NLO}} }{\delta x^M} - \frac{1}{2}y^M y^N\frac{\delta^2 \mathcal{L}_{\text{WZ-LO}}}{\delta x^M\delta x^N}\nn\\
&=& -\frac{T_{\text{eff}}}{2}\varepsilon^{\alpha\beta}\bar B_{(2)\alpha\beta} + T_{\text{eff}}\sqrt{-\tau}\tau^{\alpha\beta}( -\tau_\alpha{^0}B_\beta{^0} + \tau_\alpha{^1}B_\beta{^1} )+ T_{\text{eff}}\varepsilon^{\alpha\beta}m_\alpha{^0}m_\beta{^1}\nn\\
&&+ z^M\frac{\delta \mathcal{L}_{\text{WZ-LO}} }{\delta x^M} + y^M\frac{\delta \mathcal{L}_{\text{WZ-NLO}} }{\delta x^M} - \frac{1}{2}y^M y^N\frac{\delta^2 \mathcal{L}_{\text{WZ-LO}}}{\delta x^M\delta x^N}\,,
\ee 
\end{subequations}
where we used
\be
\tau^\beta{_0} = -\tau^{\alpha\beta}\tau_\alpha{^0}\,,\qquad \tau^\beta{_1} = \tau^{\alpha\beta}\tau_\alpha{^1}\,.
\ee 
Using \eqref{eq:WZLO} and \eqref{eq:LONGLagrangian} we see that this choice identically cancels the LO Lagrangian $\mathcal{L}_{\text{NG-LO}}+\mathcal{L}_{\text{WZ-LO}}$. As was pointed out in \cite{Harmark:2019upf}, this particular choice also removes the field $m_M{^A}$ at NLO, due to the emergence of a St\"uckelberg symmetry as we now discuss.

\subsection{A St\"uckelberg symmetry}
The NLO Lagrangian in the presence of the Kalb--Ramond field \eqref{eq:KalbRamondExp}, where we keep $B_{(-2)MN}$ arbitrary, is
\be 
\mathcal{L}_{\text{NG+WZ-NLO}} &=& -\frac{T_{\text{eff}}}{2}\left( \sqrt{-\tau(x)}\tau^{\alpha\beta}(x)H_{\alpha\beta}(x) + \varepsilon^{\alpha\beta} B_{(0)\alpha\beta} \right) + y^M\frac{\delta \mathcal{L}_{\text{NG+WZ-LO}}}{\delta x^M} \,.\nn\\\label{eq:NLOwithWZ}
\ee 
As pointed out in \cite{Harmark:2019upf}, this Lagrangian possesses the following St\"uckelberg symmetry
\be 
H_{MN} \rightarrow H_{MN} + 2C_{(0)(M}{^A}\tau_{N)}{^B}\eta_{AB}\,,\qquad B_{(0)MN} \rightarrow B_{(0)MN} + 2C_{(0)[M}{^A}\tau_{N]}{^B}\varepsilon_{AB}\,.\nn\\\label{eq:stuckelberg}
\ee 
This follows from the fact that \eqref{eq:stuckelberg} leaves $\sqrt{-\tau}\tau^{\alpha\beta}H_{\alpha\beta} + \varepsilon^{\alpha\beta} B_{(0)\alpha\beta}$ in \eqref{eq:NLOwithWZ} invariant (as well as trivially the $y^M$ terms). To see this note that
\be 
&&\sqrt{-\tau}\tau^{\alpha\beta}H_{\alpha\beta} + \varepsilon^{\alpha\beta} B_{(0)\alpha\beta}\nn\\
&&\hspace{-.5cm}\rightarrow \sqrt{-\tau}\tau^{\alpha\beta}H_{\alpha\beta} + \varepsilon^{\alpha\beta} B_{(0)\alpha\beta} + 2\sqrt{-\tau}\tau^{\alpha\beta}\D_\alpha x^M\D_\beta x^N C_{(0)(M}{^A}\tau_{N)}{^B}\eta_{AB} \nn\\
&&+ 2\varepsilon^{\alpha\beta}\D_\alpha x^M\D_\beta x^NC_{(0)[M}{^A}\tau_{N]}{^B}\varepsilon_{AB}=\sqrt{-\tau}\tau^{\alpha\beta}H_{\alpha\beta} + \varepsilon^{\alpha\beta} B_{(0)\alpha\beta} \,,\nn
\ee 
since
\be 
&&\sqrt{-\tau}\tau^{\alpha\beta}\D_\alpha x^M\D_\beta x^N C_{(0)(M}{^A}\tau_{N)}{^B}\eta_{AB} + \varepsilon^{\alpha\beta}\D_\alpha x^M\D_\beta x^NC_{(0)[M}{^A}\tau_{N]}{^B}\varepsilon_{AB}\nn\\
=&& \sqrt{-\tau}\tau^\alpha{_A}C_{(0)\alpha}{^A} - \sqrt{-\tau}\tau^\alpha{_A}C_{(0)\alpha}{^A} =0\,,\label{eq:should-be-zero}
\ee 
where we used \eqref{eq:invtau}.
In this way, we can remove the field $m_M{^A}$ from the description of the NLO string with the choice
\be 
C_{(0)M}{^A} = m_M{^A}\,. \label{eq:choice-for-C}
\ee 
This St\"uckelberg symmetry exists for choice of $B_{(-2)MN}$ and thus also for the choice \eqref{eq:RelatingWZ}.

At NNLO, the NG action including the WZ term is
\be 
\mathcal{L}_{\text{NG+WZ-NNLO}} &=& -\frac{T_{\text{eff}}}{8} \sqrt{-\tau (x)}\Big[4\tau^{\alpha\beta}(x)\Phi_{\alpha\beta}(x) + (\tau^{\alpha\beta}(x)H_{\alpha\beta}(x))^2 \nn\\
&&- 2\tau^{\alpha\beta}(x)\tau^{\gamma\delta}(x)H_{\alpha\gamma}(x)H_{\delta\beta}(x) \Big]-\frac{T_{\text{eff}}}{2}\varepsilon^{\alpha\beta} B_{(2)\alpha\beta}
\\
&&+z^M\frac{\delta \mathcal{L}_{\text{NG+WZ-LO}}}{\delta x^M} + y^M\frac{\delta \mathcal{L}_{\text{NG+WZ-NLO}}}{\delta x^M} - \frac{1}{2}y^M y^N\frac{\delta^2 \mathcal{L}_{\text{NG+WZ-LO}}}{\delta x^M \delta x^N}\,.\nn
\ee 
This Lagrangian again has a St\"uckelberg symmetry acting as
\be 
\Phi_{MN} \rightarrow \Phi_{MN} + 2C_{(2)(M}{^A}\tau_{N)}{^B}\eta_{AB}\,,\qquad B_{(2)MN} \rightarrow B_{(0)MN} - 2C_{(2)[M}{^A}\tau_{N]}{^B}\varepsilon_{AB}\,,\nn\\
\ee 
and choosing
\be 
C_{(2)M}{^A} = B_M{^A}\,,
\ee 
removes all terms involving $B_M{^A}$ from NNLO Lagrangian. This means that the NNLO can be written as
\be 
\mathcal{L}_{\text{NG+WZ-NNLO}} &=& -\frac{T_{\text{eff}}}{8} \sqrt{-\tau }\left[4\tau^{\alpha\beta}(\Phi^\perp_{\alpha\beta} + \eta_{AB}m_\alpha{^A}m_\beta{^B} ) + (\tau^{\alpha\beta}H_{\alpha\beta})^2 - 2\tau^{\alpha\beta}\tau^{\gamma\delta}H_{\alpha\gamma}H_{\delta\beta} \right]\nn\\
&&-\frac{T_{\text{eff}}}{2}\varepsilon^{\alpha\beta} \bar B_{(2)\alpha\beta} + T_{\text{eff}}\varepsilon^{\alpha\beta}m_\alpha{^0}m_\beta{^1} + y^M\frac{\delta \mathcal{L}_{\text{NG+WZ-NLO}}}{\delta x^M}\,.
\ee 
Note that at NNLO we cannot remove $m_M{^A}$.

\section{The spectrum}\label{sec:spectrum}
In this section, we compute the spectrum of the $1/c^2$ expanded string theories on flat target space. Flat target space corresponds to
\begin{align}
\label{eq:flatspace}
\begin{split}
    \tau_M{^0} &= \delta_M^t\,,\qquad \tau_M{^1} = \delta^v_M\,,\qquad H^\perp_{MN} = \delta^i_M\delta^i_N\,,\qquad m_M{^A} = B_M{^A} = \phi_{MN}^\perp = 0\,,
\end{split}    
\end{align}
where the spatial index ranges over $i=1,\dots,24$ and where $v$ is a compact direction. The spectrum of the string theories matches order by order the expansion of the relativistic string spectrum~\eqref{eq:energy-expansion}; in other words, the $1/c^2$ expansion and the computation of the spectrum commute.

\subsection{Mode expansions and spectrum}
\label{sec:ModeExp}

Since we consider closed strings, the string embedding scalars (up to NNLO) --- with the exception of the leading order field\footnote{Since we do not expand the winding number, only the leading order field carries winding.} $x^v$ --- are periodic in $\sigma^1$,
\be 
x^t(\sigma^0,\sigma^1+2\pi) &=& x^t(\sigma^0,\sigma^1)\,,\qquad x^i(\sigma^0,\sigma^1+2\pi) = x^i(\sigma^0,\sigma^1)\,,\nn\\
y^t(\sigma^0,\sigma^1+2\pi) &=& y^t(\sigma^0,\sigma^1)\,,\qquad y^v(\sigma^0,\sigma^1+2\pi) = y^v(\sigma^0,\sigma^1)\,,\nn\\
y^i(\sigma^0,\sigma^1+2\pi) &=& y^i(\sigma^0,\sigma^1)\,,\nn\\
z^t(\sigma^0,\sigma^1+2\pi) &=& z^t(\sigma^0,\sigma^1)\,,\qquad z^v(\sigma^0,\sigma^1+2\pi) = z^v(\sigma^0,\sigma^1)\,,
\ee 
while the leading order embedding scalar in the compact direction satisfies
\be \label{eq:bdrycondition}
x^v(\sigma^0,\sigma^1 + 2\pi) = x^v(\sigma^0,\sigma^1) + 2\pi wR_{\text{eff}}\,,
\ee 
where $w\in \mathbb{Z}$ is the winding number, and where we remind the reader that $R_{\text{eff}}$ has dimensions of time and is independent of $c$. The P-LO Lagrangian \eqref{eq:LOPLagrangian} in flat space with the worldsheet gauge choice \eqref{eq:LOgauge} is
\be 
\mathcal{L}_{\text{P-LO}} = \frac{T_{\text{eff}}}{2}\eta^{\alpha\beta}\D_\alpha x^t\D_\beta x^t - \frac{T_{\text{eff}}}{2}\eta^{\alpha\beta} \D_\alpha x^v\D_\beta x^v\,.\label{eq:LOflatspace}
\ee 
The equations of motion are
\be 
\D_-\D_+ x^t =0\,,\qquad \D_-\D_+x^v = 0\,,\label{eq:LOEoM}
\ee 
while the LO Virasoro constraint with 
\begin{eqnarray}
\tau_{\alpha\beta}=-\partial_\alpha x^t\partial_\beta x^t+\partial_\alpha x^v\partial_\beta x^v\,,
\end{eqnarray}
from integrating out $\gamma_{(0)\alpha\beta}$ \eqref{eq:LOVirasoro} reduces to $\tau_{++}=0=\tau_{--}$, i.e.,
\begin{equation}\label{eq:LOVir}
\partial_+ x^+\partial_+ x^-=0\,,\qquad \partial_- x^+\partial_- x^-=0\,,    
\end{equation}
where 
\be 
x^\pm = x^t \pm x^v\,.
\ee 

Without loss of generality we can choose the following conditions\footnote{There are 4 solutions to \eqref{eq:LOVir} but two imply either $x^+$ is constant or $x^-$ is constant which is not allowed by our choice of boundary conditions while the other two solutions are related by interchanging $\sigma^+$ with $\sigma^-$.}
\be 
\D_- x^+ = 0\,,\qquad \D_+ x^-=0\,.\label{eq:LOConstraints}
\ee 
Note that this combination makes sense because $x^v$ was defined as having dimensions of time even though it is a spatial direction. Since $x^t = \frac{1}{2}(x^+ + x^-)$ and $x^v = \frac{1}{2}(x^+ - x^-)$, the constraints \eqref{eq:LOConstraints} imply the LO equations of motion \eqref{eq:LOEoM}, which thus are not required. The LO Nambu--Goto Lagrangian can be written as
\be 
\mathcal{L}_{\text{NG-LO}} &=& T_{\text{eff}} \D_+(x^+\D_-x^-) - T_{\text{eff}}\D_-(x^+\D_+x^-) \,,
\ee 
and is a total derivative. In deriving this result we assumed that the map from the worldsheet to the 2-dimensional Lorentzian submanifold of the target space (described by $x^t$ and $x^v$) is orientation preserving, i.e.,
\begin{equation}
    \dot x^t {x^v}'-{x^t}'\dot x^v>0\,.
\end{equation}
The Lagrangian is linear in the velocities and so the Hamiltonian is minus the potential energy, which means that the LO contribution to the energy is a constant.

With the $v$-direction compact, the following mode expansions for $x^\pm$ are compatible with the LO Virasoro constraints \eqref{eq:LOConstraints} and the boundary condition \eqref{eq:bdrycondition},
\be 
\begin{aligned}
x^- &= x_0^- + w R_{\text{eff}} \sigma^-  + \frac{1}{\sqrt{4\pi T_{\text{eff}}}}\sum_{k\neq 0} \frac{i}{k}\alpha_k^- e^{-ik\sigma^-}\,,\\
x^+ &= x_0^+ + w R_{\text{eff}} \sigma^+  + \frac{1}{\sqrt{4\pi T_{\text{eff}}}}\sum_{k\neq 0} \frac{i}{k}\tilde{\alpha}^+_k e^{-ik\sigma^+}\,,
\end{aligned}
\ee 
where $w$ is the winding number and where $ R_{\text{eff}}$ is the radius of the compact $v$-direction that we introduced in Section \ref{Sec:Decompact}. Note that this in particular implies that
\be 
x^v = x_0^v + wR_{\text{eff}}\sigma^1 + \text{oscillations}\,,\label{eq:xvExp}
\ee 
where, in agreement with the results of \cite{Harmark:2018cdl,Harmark:2019upf}, the LO embedding field of our theory carries no momentum but only winding. 
We have yet to fix the residual LO gauge transformations \eqref{eq:LOresidualgauge} with parameters $\xi_{(0)}^\pm(\sigma^\pm)$, which act on $x^\pm$ as 
\be 
\delta_{\xi_{(0)}}x^\pm &=& \xi_{(0)}^\pm(\sigma^\pm)\D_\pm x^\pm\,,
\ee 
where we used \eqref{eq:LOConstraints}, and where $\xi_{(0)}^\pm(\sigma^\pm)$ are periodic. Therefore, we can fix $\xi_{(0)}^-(\sigma^-)$ by removing all oscillations from $x^-$ and $\xi_{(0)}^+(\sigma^+)$ by removing all oscillations from $x^+$, which means that the fully gauge fixed mode expansions read
\be 
x^\pm &=& x_0^\pm + w R_{\text{eff}} \sigma^\pm\,.\label{eq:LOmodeexpansion}
\ee

As we show in appendix \ref{app:energymomentumexp} the LO energy is given by
\be 
E_{\text{LO}} &=& -\oint d \sigma^1\frac{\D \mathcal{L}_{\text{P-LO}}}{\D(\D_0x^t)} = \frac{wR_{\text{eff}}}{\alpha'_{\text{eff}}}\,,\label{eq:LOenergy}
\ee 
and is the stringy analogue of the rest mass of a point particle: the NG-LO Lagrangian is a total derivative, as is the LO point point particle Lagrangian which is responsible for producing the rest mass term $mc^2$ (see, e.g., \cite{Gomis:2019sqv}).

Now, at NLO, the Polyakov Lagrangian \eqref{eq:NLOPLagrangian} in a flat target space takes the form
\be \label{eq:LagP-NLO}
\mathcal{L}_{\text{P-NLO}} &=& -\frac{T_{\text{eff}}}{2}\eta^{\alpha\beta}\D_\alpha x^i\D_\beta x^i+T_{\text{eff}}\eta^{\alpha\beta}\D_\alpha y^t\D_\beta x^t - T_{\text{eff}}\eta^{\alpha\beta}\D_\alpha y^v\D_\beta x^v\,.
\ee 
where we used the worldsheet gauge choices \eqref{eq:LOgauge} and \eqref{eq:NLOgaugechoice}.
The equations of motion for $y^t$ and $y^v$ are \eqref{eq:LOEoM}, while the equations of motion for $x^t$, $x^v$, and $x^i$ give 
\be 
\D_-\D_+ y^t = 0\,,\qquad \D_-\D_+y^v = 0\,,\qquad \D_-\D_+ x^i = 0\,.\label{eq:NLOEoM}
\ee
These equations of motion imply the following mode expansions
\begin{subequations}
\be 
x^i &=& x_0^i + \frac{1}{2\pi T_{\text{eff}}}p_{(0)i} \sigma^0+ \frac{1}{\sqrt{4\pi T_{\text{eff}}}}\sum_{k\neq 0} \frac{i}{k}\left[\alpha_k^i e^{-ik\sigma^-}+\tilde{\alpha}^i_k e^{-ik\sigma^+} \right]\,,\\
y^t &=& y^t_0 - \frac{1}{2\pi T_{\text{eff}}}p_{(0)t}\sigma^0 + \frac{1}{\sqrt{4\pi T_{\text{eff}}}}\sum_{k\neq 0} \frac{i}{k}\left[\beta_k^t e^{-ik\sigma^-}+\tilde{\beta}^t_k e^{-ik\sigma^+} \right]\,,\\
y^v &=& y^v_0 + \frac{1}{2\pi T_{\text{eff}}}p_{(0)v}\sigma^0  +  \frac{1}{\sqrt{4\pi T_{\text{eff}}}}\sum_{k\neq 0} \frac{i}{k}\left[\beta_k^v e^{-ik\sigma^-}+\tilde{\beta}^v_k e^{-ik\sigma^+} \right]\,,\label{yvExp}
\ee 
\end{subequations}
where we have not included a term linear in $\sigma^1$ in the mode expansion for $y^v$. Had we kept such a term, it would have corresponded to expanding the winding number of the relativistic parent theory in powers $1/c^2$, but since the winding number is an integer, we refrain from doing so. The momenta $p_{(0)i}$, $p_{(0)t}$, and $p_{(0)v}$ featuring in the mode expansions above correspond to the zero modes of the canonical momenta, i.e. $p_{(0)i} = \oint d \sigma^1\frac{\D\mathcal{L}_{\text{P-NLO}}}{\D\dot x^i}$, $p_{(0)t} = \oint d \sigma^1\frac{\D\mathcal{L}_{\text{P-NLO}}}{\D\dot x^t}$, and $p_{(0)v} = \oint d \sigma^1\frac{\D\mathcal{L}_{\text{P-NLO}}}{\D\dot x^v}$, respectively.

The equation of motion for $\gamma_{(2)\alpha\beta}$ in the P-NLO Lagrangian \eqref{eq:NLOPLagrangian} in flat space gives the LO Virasoro constraints \eqref{eq:LOConstraints}, while the equation of motion for $\gamma_{(0)\alpha\beta}$ leads to the NLO Virasoro constraints \eqref{eq:NLOVirasoro}, which in flat space can be written as
\be 
\D_+ y^- = \frac{1}{w R_{\text{eff}}}\D_+ x^i\D_+ x^i\,,\qquad\D_- y^+ = \frac{1}{w R_{\text{eff}}}\D_- x^i\D_- x^i\,,\label{eq:NLOconstraints}
\ee
where we, as above, defined
\be 
y^\pm = y^t \pm y^v\,,
\ee 
with the following mode expansions
\be 
\begin{aligned}
y^- &= y^-_0 - \frac{1}{2\pi T_{\text{eff}}}p_{(0)+}(\sigma^++\sigma^-) + \frac{1}{\sqrt{4\pi T_{\text{eff}}}}\sum_{k\neq 0} \frac{i}{k}\left[\beta_k^- e^{-ik\sigma^-}+\tilde{\beta}^-_k e^{-ik\sigma^+} \right]\,,\\
y^+ &= y^+_0 - \frac{1}{2\pi T_{\text{eff}}}p_{(0)-}(\sigma^++\sigma^-)  +  \frac{1}{\sqrt{4\pi T_{\text{eff}}}}\sum_{k\neq 0} \frac{i}{k}\left[\beta_k^+ e^{-ik\sigma^-}+\tilde{\beta}^+_k e^{-ik\sigma^+} \right]\,,
\end{aligned}
\ee 
where 
\be \label{eq:p0pm}
p_{(0)\pm} &=& \frac{1}{2}(p_{(0)t}\pm p_{(0)v})\,,
\ee 
are the canonical momenta in lightcone coordinates, $p_{(0)\pm} = \oint d \sigma^1\frac{\D\mathcal{L}_{\text{P-NLO}}}{\D\dot x^\pm}$, while the oscillator modes are
\be 
\begin{aligned}
\beta^+_k &= \beta^t_k + \beta^v_k\,,\qquad \beta^-_k = \beta^t_k - \beta^v_k\,,\\
\tilde \beta^+_k &= \tilde\beta^t_k + \tilde\beta^v_k\,,\qquad \tilde\beta^-_k = \tilde\beta^t_k - \tilde\beta^v_k\,.
\end{aligned}
\ee 
Using \eqref{eq:NLOembeddingdiffeo} and the leading order mode expansion \eqref{eq:LOmodeexpansion}, the NLO residual gauge transformations \eqref{eq:NLOresidualgauge} with $\xi_{(2)}^\pm(\sigma^\pm)$ act on $y^\pm$ as
\be \label{eq:deltaypm}
\delta_{\xi_{(2)}} y^- = wR_{\text{eff}} \xi^-_{(2)}(\sigma^-)\,,\qquad \delta_{\xi_{(2)}} y^+ = wR_{\text{eff}} \xi^+_{(2)}(\sigma^+)\,.
\ee 
Hence, the quantities $\D_\pm y^\mp$, which feature prominently in the NLO Virasoro constraints \eqref{eq:NLOconstraints}, remain invariant under the subleading residual gauge symmetries \eqref{eq:NLOresidualgauge}. Thus we can use $\xi^-_{(2)}$ to set $\beta_k^- = 0$ for $k\neq 0$ and $\xi^+_{(2)}$ to set $\tilde\beta^+_k=0$ for $k\neq 0$, which fixes all residual gauge transformations. Now, the constraints \eqref{eq:NLOconstraints} imply that the zero modes satisfy
\be 
p_{(0)-} &=& -\frac{N_{(0)}}{wR_{\text{eff}}} - \frac{\alpha'_{\text{eff}}}{4wR_{\text{eff}}}(p_{(0)})^2\,,\qquad p_{(0)+} = -\frac{\tilde N_{(0)}}{wR_{\text{eff}}} - \frac{\alpha'_{\text{eff}}}{4wR_{\text{eff}}}(p_{(0)})^2\,,\label{eq:Expsforp-andp+}
\ee 
where $(p_{(0)})^2 = p_{(0)i}p_{(0)i}$, and where we defined 
\be 
N_{(0)} = \sum_{n=1}^\infty \alpha_{-n}^i \alpha_n^i\,,\qquad \tilde N_{(0)} = \sum_{n=1}^\infty \tilde\alpha_{-n}^i \tilde\alpha_n^i\,.
\ee
We ignored a normal ordering constant (see section \ref{sec:quantisation} for a discussion of the normal ordering constant). The higher oscillator modes satisfy
\be 
\tilde\beta_k^- = \frac{1}{\sqrt{4\pi T_{\text{eff}}}wR_{\text{eff}} }\sum_{n\in \mathbb Z}\alpha^i_{k-n}\alpha^i_n\,,\qquad \beta_k^+ = \frac{1}{\sqrt{4\pi T_{\text{eff}}}wR_{\text{eff}} }\sum_{n\in \mathbb Z}\tilde\alpha^i_{k-n}\tilde\alpha^i_n\,,
\ee
where the zero mode is given by 
\be 
\alpha_{(0)}^i = \tilde\alpha_{(0)}^i = \frac{1}{\sqrt{4\pi T_{\text{eff}}}}p_{(0)i}\,.
\ee 

In Appendix \ref{app:energymomentumexp}, we show that the relativistic energy $E$ can be expanded as
\be
\begin{aligned}
E &= c^2 E_{\text{LO}} + E_{\text{NLO}} + c^{-2}E_{\text{NNLO}} +\mathcal{O}(c^{-4})\\
&= -c^2\oint d \sigma^1\pd{\mathcal{L}_{\text{P-LO}}}{\D_0x^t} - \oint d \sigma^1\pd{\mathcal{L}_{\text{P-NLO}}}{\D_0x^t} - c^{-2}\oint d \sigma^1\pd{\mathcal{L}_{\text{P-NNLO}}}{\D_0x^t} +\mathcal{O}(c^{-4})\,.\label{eq:ExpansionOfSpectrum}
\end{aligned}
\ee 
This means that the contribution to the spectrum from the NLO Lagrangian becomes
\be 
E_{\text{NLO}} &=&  -\oint d \sigma^1\pd{\mathcal{L}_{\text{P-NLO}}}{\dot x^t} = -p_{(0)-} - p_{(0)+} =  \frac{N_{(0)} + \tilde N_{(0)}}{wR_{\text{eff}}} + \frac{\alpha'_{\text{eff}}}{2w R_{\text{eff}}}(p_{(0)})^2\,.\nn\\\label{eq:NLOspectrum}
\ee 
Note also that 
\be 
E_{\text{LO}} &=&  -\oint d \sigma^1\pd{\mathcal{L}_{\text{P-NLO}}}{(\D_0 y^t)}\,,
\ee 
which follows from $\pd{\mathcal{L}_{\text{P-NLO}}}{(\D_0 y^t)}=\pd{\mathcal{L}_{\text{P-LO}}}{(\D_0 x^t)}$. Thus, by \eqref{eq:ExpansionOfSpectrum}, the energy up to NLO is 
\be 
E = c^2E_{\text{LO}} + E_{\text{NLO}} + \mathcal{O}(c^{-2}) = \frac{c^2wR_{\text{eff}}}{\alpha'_{\text{eff}}} + \frac{N + \tilde N}{wR_{\text{eff}}} + \frac{\alpha'_{\text{eff}}}{2w R_{\text{eff}}}p^2 + \mathcal{O}(c^{-2})\,,
\ee 
where
\be 
N = N_{(0)} + \mathcal{O}(c^{-2})\,,\qquad \tilde N = \tilde N_{(0)} + \mathcal{O}(c^{-2})\,,\qquad p_i = p_{(0)i} + \mathcal{O}(c^{-2})\,,
\ee 
in agreement with~\eqref{eq:NLO-energy-sec-2}

The momentum $p_{(0)v}$ is quantised. To see this, note that the momentum of the relativistic string in the $v$-direction is $P_v = \oint d \sigma^1 \frac{\partial{\mathcal{L}_P}}{\partial\partial_0 X^v} = p_{(0)v} + \mathcal{O}(c^{-2})$, where we used \eqref{eq:ExpansionOfvMom} and the mode expansions \eqref{eq:xvExp} and \eqref{yvExp}. The string wave function $\Psi \sim e^{\frac{i}{\hbar}P\cdot X}$ in particular includes a term of the form $e^{\frac{i}{\hbar}p_{(0)v} x^v}$, and since the string wave function is single-valued, $p_{(0)v}$ must be quantised according to
\be 
p_{(0)v} = \frac{\hbar n}{R_{\text{eff}}}\,.
\ee 
Furthermore, since 
\be 
p_{(0)v} = p_{(0)+} - p_{(0)-} = \frac{N_{(0)}-\tilde N_{(0)}}{wR_{\text{eff}}}\,,
\ee 
we obtain the level matching condition
\be 
N_{(0)}-\tilde N_{(0)} = \hbar w n\,.\label{eq:LOLevelMatching}
\ee 

The NNLO Lagrangian \eqref{eq:LagPNNLO} with worldsheet gauge choices \eqref{eq:LOgauge}, \eqref{eq:NLOgaugechoice} and \eqref{eq:NNLOgauge} for a flat target space is
\begin{eqnarray}
\begin{aligned}
\mathcal{L}_{\text{P-NNLO}}  = &  \frac{T_{\text{eff}}}{2}\eta^{\alpha\beta}\left( \D_\alpha y^t \D_\beta y^t - \D_\alpha y^v \D_\beta y^v \right) + T_{\text{eff}}\eta^{\alpha\beta}\left( \D_\alpha x^t \D_\beta z^t - \D_\alpha x^v \D_\beta z^v \right)\\
& - T_{\text{eff}}\eta^{\alpha\beta} \D_\alpha x^i\D_\beta y^i\,.
\end{aligned}
\end{eqnarray}
In addition to the LO and NLO equations of motion, \eqref{eq:LOEoM} and \eqref{eq:NLOEoM}, which arise as the equations of motion for $z^t,z^v,y^t,y^v,y^i$, the equations of motion for $x^t$, $x^v$, $x^i$ are
\be 
\D_-\D_+ z^t = 0\,,\qquad \D_-\D_+z^v = 0\,,\qquad \D_-\D_+ y^i = 0\,.\label{eq:NNLOEoM}
\ee 
As above, this leads to the following mode expanions
\begin{subequations}
\be 
y^i &=& y_0^i + \frac{1}{2\pi T_{\text{eff}}}p_{(2)i} \sigma^0+ \frac{1}{\sqrt{4\pi T_{\text{eff}}}}\sum_{k\neq 0} \frac{i}{k}\left[\beta_k^i e^{-ik\sigma^-}+\tilde{\beta}^i_k e^{-ik\sigma^+} \right]\,,\\
z^t &=& z^t_0 - \frac{1}{2\pi T_{\text{eff}}}p_{(2)t}\sigma^0 + \frac{1}{\sqrt{4\pi T_{\text{eff}}}}\sum_{k\neq 0} \frac{i}{k}\left[\chi_k^t e^{-ik\sigma^-}+\tilde{\chi}^t_k e^{-ik\sigma^+} \right]\,,\\
z^v &=& z^v_0   +  \frac{1}{\sqrt{4\pi T_{\text{eff}}}}\sum_{k\neq 0} \frac{i}{k}\left[\chi_k^v e^{-ik\sigma^-}+\tilde{\chi}^v_k e^{-ik\sigma^+} \right]\,,
\ee 
\end{subequations}
where, as before, we do not include a term linear in $\sigma^1$ in the mode expansion of $z^v$, while we now also choose to exclude a term linear in $\sigma^0$, since this would correspond to expanding the integer-valued momentum mode $n$. As above, we have $p_{(2)i} = \oint  d \sigma^1\frac{\D\mathcal{L}_{\text{P-NNLO}}}{\D\dot x^i}$ as well as $p_{(2)t} = \oint  d \sigma^1\frac{\D\mathcal{L}_{\text{P-NNLO}}}{\D\dot x^t}$.
 
The constraints from both the LO \eqref{eq:LOConstraints} and NLO \eqref{eq:NLOconstraints} Virasoro constraints arise as the constraints from integrating out $\gamma_{(4)\alpha\beta}$ and $\gamma_{(2)\alpha\beta}$, respectively, in \eqref{eq:LagPNNLO}, while the novel constraint at NNLO is given by equation \eqref{eq:NNLOVirasoro} which becomes
\begin{eqnarray}
\label{eq:NNLOConstraints}
\begin{aligned}
0&= 2\D_-x^i \D_- y^i - \D_-y^+\D_-y^-  - wR_{\text{eff}} \D_-z^+ \,,\\
0&= 2\D_+x^i \D_+ y^i -\D_+y^+\D_+ y^- - wR_{\text{eff}}\D_+ z^- \,,
\end{aligned}
\ee 
where we used the LO Virasoro constraints, and where we defined $z^\pm = z^t \pm z^v$. We can write the mode expansions for the $z^t$ and $z^v$ fields as 
\be 
\begin{aligned}
z^- &= z^-_0 - \frac{1}{4\pi T_{\text{eff}}}p_{(2)t}(\sigma^++\sigma^-) + \frac{1}{\sqrt{4\pi T_{\text{eff}}}}\sum_{k\neq 0} \frac{i}{k}\left[\chi_k^- e^{-ik\sigma^-}+\tilde{\chi}^-_k e^{-ik\sigma^+} \right]\,,\\
z^+ &= z^+_0 - \frac{1}{4\pi T_{\text{eff}}}p_{(2)t}(\sigma^++\sigma^-)  +  \frac{1}{\sqrt{4\pi T_{\text{eff}}}}\sum_{k\neq 0} \frac{i}{k}\left[\chi_k^+ e^{-ik\sigma^-}+\tilde{\chi}^+_k e^{-ik\sigma^+} \right]\,,
\end{aligned}
\ee 
where
\be 
\begin{aligned}
\chi^+_k &= \chi^t_k + \chi^v_k\,,\qquad \chi^-_k = \chi^t_k - \chi^v_k\,,\\
\tilde \chi^+_k &= \tilde\chi^t_k + \tilde\chi^v_k\,,\qquad \tilde\chi^-_k = \tilde\chi^t_k - \tilde\chi^v_k\,.
\end{aligned}
\ee 

We can now fix residual gauge transformations \eqref{eq:NNLOresidualgauge} at NNLO $\xi^\pm_{(4)}(\sigma^\pm)$ by setting $\chi_k^- = 0$ for $k\neq 0$ and $\tilde\chi^+_k = 0$ for $k\neq 0$, which fixes all residual gauge transformations. This is entirely analogous to what happened at NLO. The gauge fixing of the subleading residual gauge transformations, discussed below \eqref{eq:deltaypm}, imply that $\D_-y^-$ and $\D_+y^+$ contain no oscillations, and so the NLO Virasoro constraints give us
\be
\begin{aligned}
\D_- y^- &= 
\frac{\alpha'_{\text{eff}} \tilde N_{(0)}}{wR_{\text{eff}}} + \frac{(\alpha'_{\text{eff}})^2}{4wR_{\text{eff}}}(p_{(0)})^2 \,,\\
\qquad  \D_+ y^+ &= 
\frac{\alpha'_{\text{eff}} N_{(0)}}{wR_{\text{eff}}} + \frac{(\alpha'_{\text{eff}})^2}{4wR_{\text{eff}}}(p_{(0)})^2\,,
\end{aligned}
\ee 
where we used the NLO Virasoro constraints in the guise of \eqref{eq:Expsforp-andp+}. Using the relation
\be 
4\tilde N_{(0)} N_{(0)} = (N_{(0)} + \tilde N_{(0)})^2 - {\hbar^2 w^2 n^2} \,,
\ee 
where we imposed the level matching condition of \eqref{eq:LOLevelMatching}, the NNLO Virasoro constraints \eqref{eq:NNLOConstraints} then imply the following two expressions for the zero mode $p_{(2)t}$
 \begin{subequations}
 \be 
 -p_{(2)t} &=& \frac{\alpha'_{\text{eff}}}{wR_{\text{eff}}}p_{(0)i}p_{(2)i} + \frac{2}{wR_{\text{eff}}}\sum_{k\neq 0}\alpha^i_{-k}\beta^i_k - \frac{(\alpha'_{\text{eff}})^3}{8w^3R_{\text{eff}}^3}(p_{(0)})^4 - \alpha'_{\text{eff}}\frac{(N_{(0)} + \tilde N_{(0)})^2}{2w^3R^3_{\text{eff}}}\nn\\
&&- (\alpha'_{\text{eff}})^2 (p_{(0)})^2 \frac{N_{(0)} + \tilde N_{(0)}}{2w^3R^3_{\text{eff}}} + \alpha'_{\text{eff}}\frac{\hbar^2 n^2}{2wR^3_{\text{eff}}}\label{eq:ptone}\\
 &=& \frac{\alpha'_{\text{eff}}}{wR_{\text{eff}}}p_{(0)i}p_{(2)i} + \frac{2}{wR_{\text{eff}}}\sum_{k\neq 0}\tilde\alpha^i_{-k}\tilde\beta^i_k - \frac{(\alpha'_{\text{eff}})^3}{8w^3R_{\text{eff}}^3}(p_{(0)})^4 - \alpha'_{\text{eff}}\frac{(N_{(0)} + \tilde N_{(0)})^2}{2w^3R^3_{\text{eff}}}\nn\\
&&- (\alpha'_{\text{eff}})^2 (p_{(0)})^2 \frac{N_{(0)} + \tilde N_{(0)}}{2w^3R^3_{\text{eff}}} + \alpha'_{\text{eff}}\frac{\hbar^2 n^2}{2wR^3_{\text{eff}}}\,,\label{eq:pttwo}
 \ee 
 \end{subequations}
 where $p_{(2)i} = \sqrt{4\pi T_{\text{eff}}}\beta^i_0= \sqrt{4\pi T_{\text{eff}}}\tilde\beta^i_0$. The contribution to the spectrum from the NNLO Lagrangian is obtained by adding the two expressions above (and dividing by two)
\be 
E_{\text{NNLO}} &=&  -\oint d \sigma^1\pd{\mathcal{L}_{\text{NNLO-P}}}{(\D_0 x^t)} = -p_{(2)t}\nn\\
&=& \frac{\alpha'_{\text{eff}}}{wR_{\text{eff}}}p_{(0)i}p_{(2)i} + \frac{N_{(2)} + \tilde N_{(2)}}{wR_{\text{eff}}} - \frac{(\alpha'_{\text{eff}})^3}{8w^3R_{\text{eff}}^3}(p_{(0)})^4 - \alpha'_{\text{eff}}\frac{(N_{(0)} + \tilde N_{(0)})^2}{2w^3R^3_{\text{eff}}}\nn\\
&&- (\alpha'_{\text{eff}})^2 (p_{(0)})^2 \frac{N_{(0)} + \tilde N_{(0)}}{2w^3R^3_{\text{eff}}} + \alpha'_{\text{eff}}\frac{\hbar^2 n^2}{2wR^3_{\text{eff}}}\,,\label{eq:NNLO-spectrum}
\ee 
where we defined the subleading number operators
\be
\begin{aligned}
N_{(2)} &= \sum_{k\neq 0}\alpha^i_{-k}\beta_k^i= \sum_{k=1}^\infty\alpha^i_{-k}\beta_k^i + \sum_{k=1}^\infty\alpha^i_{k}\beta_{-k}^i \,,\\
\tilde N_{(2)} &= \sum_{k\neq 0}\tilde \alpha^i_{-k}\tilde \beta_k^i= \sum_{k=1}^\infty\tilde \alpha^i_{-k}\tilde \beta_k^i + \sum_{k=1}^\infty\tilde \alpha^i_{k}\tilde \beta_{-k}^i \,,
\end{aligned}
\ee  
in agreement with~\eqref{eq:NNLO-energy-sec-2}. In terms of the NNLO Lagrangian, the LO and NLO contributions to the spectrum are the canonical momenta conjugate to $z^t$ and $y^t$, respectively,
\be
E_{\text{LO}} &=&  -c^2\oint d \sigma^1\pd{\mathcal{L}_{\text{NNLO-P}}}{(\D_0 z^t)}\,,\qquad E_{\text{NLO}} =  -\oint d \sigma^1\pd{\mathcal{L}_{\text{NNLO-P}}}{(\D_0 y^t)}\,.
\ee 
The means that the energy up to NNLO becomes 
\be 
E &=& c^2E_{\text{LO}} + E_{\text{NLO}} + c^{-2}E_{\text{NNLO}} + \mathcal{O}(c^{-4})\nn\\
&=& \frac{c^2wR_{\text{eff}}}{\alpha'_{\text{eff}}} + \frac{N + \tilde N}{wR_{\text{eff}}} + \frac{\alpha'_{\text{eff}}}{2w R_{\text{eff}}}p^2 - \frac{(\alpha'_{\text{eff}})^3}{8w^3c^2R_{\text{eff}}^3}p^4 - \alpha'_{\text{eff}}\frac{(N + \tilde N)^2}{2w^3c^2R^3_{\text{eff}}}\nn\\
&&- (\alpha'_{\text{eff}})^2 p^2 \frac{N + \tilde N}{2w^3c^2R^3_{\text{eff}}} + \alpha'_{\text{eff}}\frac{\hbar^2 n^2}{2wc^2R^3_{\text{eff}}}+ \mathcal{O}(c^{-4})\,,
\ee 
where at NNLO, we have
\be 
\begin{aligned}
N &= N_{(0)} + c^{-2}N_{(2)} \mathcal{O}(c^{-4})\,,\qquad \tilde N = \tilde N_{(0)} +c^{-2}\tilde N_{(2)} + \mathcal{O}(c^{-4})\,,\\
p_i &= p_{(0)i} + c^{-2}p_{(2)i} \mathcal{O}(c^{-4})\,.
\end{aligned}
\ee

Subtracting the two expressions for $p_{(2)t}$ in \eqref{eq:ptone} and \eqref{eq:pttwo} gives the subleading level matching condition
\be \label{eq:NLOLevelMatching}
N_{(2)} &=& \tilde N_{(2)}\,.
\ee 
Note that this is a consequence of our choice to not expand the momentum mode $n$ and the winding number $w$. The level matching conditions \eqref{eq:LOLevelMatching} and  \eqref{eq:NLOLevelMatching}
are compatible with the $1/c^2$ expansion of the relativistic level matching condition
\be 
N-\tilde N = \hbar w n\,.
\ee

\subsection{The Gomis--Ooguri spectrum in the presence of a $B$-field}
In flat target space \eqref{eq:flatspace}, we take the Kalb--Ramond field to have the following expansion
\be 
B_{MN} = 2c^2\delta^t_{[M}\delta^v_{N]}B_{(-2)tv} + B_{(0)MN} + c^{-2}B_{(2)MN} + \mathcal{O}(c^{-4})\,,
\ee 
where the LO term only involves $x^t$ and $x^v$. 
If we now choose $B_{(-2)tv} = B = \text{const.}$ and $B_{(0)MN} =B_{(2)MN}= 0$, we get the following expression for the spectrum
\be 
E &=& c^2E_{\text{LO}} + E_{\text{NLO}} + c^{-2}E_{\text{NNLO}} + \mathcal{O}(c^{-4}) \nn\\
&=& \frac{c^2 wR_{\text{eff}}}{\alpha'_{\text{eff}}}(1-B)  +  \frac{N + \tilde N}{wR_{\text{eff}}} + \frac{\alpha'_{\text{eff}}}{2w R_{\text{eff}}}p^2 - \frac{(\alpha'_{\text{eff}})^3}{8w^3c^2R_{\text{eff}}^3}p^4 - \alpha'_{\text{eff}}\frac{(N + \tilde N)^2}{2w^3c^2R^3_{\text{eff}}}\nn\\
&&- (\alpha'_{\text{eff}})^2 p^2 \frac{N + \tilde N}{w^3c^2R^3_{\text{eff}}} + \alpha'_{\text{eff}}\frac{\hbar^2 n^2}{2wc^2R^3_{\text{eff}}}+ \mathcal{O}(c^{-4})\,,\label{eq:NLOspectrum+WZ}
\ee 
which gives rise to the modification of $E_{\text{LO}}$ discussed in \eqref{eq:LO-modification-due-to-WZ}. Up to $\mathcal{O}(1)$, this is the result of \cite{Gomis:2000bd} when taking $B=1/2$ (see also \cite{Danielsson:2000mu}).

\section{Target space symmetries and algebra of charges}
\label{sec:target-space-symmetries}

The target space symmetries manifest themselves as global symmetries for the embedding fields, i.e., in the case of Poincar\'e symmetry for the relativistic string we have $X^M \rightarrow L^M{_N}X^N + a^M$ for constant $L^M{}_N$ and $a^M$, and the Noether charges corresponding to these global symmetries generate the Poincar\'e algebra under the Poisson bracket. The string theories we consider in this work arise as $1/c^2$ expansions of relativistic string theory, and hence it is natural to expect that the charge algebra of the string theory at a given order in $1/c^2$ corresponds to an expansion of the Poincar\'e algebra. As we demonstrate in this section, this is indeed the case. In particular, the algebra of charges at NLO gives rise to the string Bargmann algebra. 

\subsection{Expansion of target space symmetries}
We now write $X^M = (cX^A,X^i)$, having dimensions $[X^A]=\text{time}$ and $[X^i]=\text{length}$, and where, with a slight abuse of notation\footnote{Earlier in section \ref{sec:string-1/c^2-exp} we introduced captial indices $A, B$ to indicate longitudinal tangent space indices. Here we are on flat space and we will not distinguish between tangent space and spacetime indices.}, we have introduced longitudinal indices $A,B =(t,v)$. The (infinitesimal) Poincar\'e transformations act in the following way on $X^A$ and $X^i$
\be
\label{eq:poincare-trafo}
\begin{aligned}
\delta X^A &= \Lambda^A{_B}X^B + c^{-2}\Lambda^A{_i}X^i + a^A\,,\\
\delta X^i &= \Lambda^i{_j}X^j + \Lambda^i{_A}X^A + a^i\,,
\end{aligned}
\ee 
where $\Lambda^A{_B}$ are longitudinal Lorentz transformations (which are dimensionless), $\Lambda^i{_j}$ are transverse rotations (again dimensionless), while 
\be\label{eq:Lambda^A_i}
\Lambda^A{_i} = -\delta_{ij}\Lambda^j{_B}\eta^{AB}\,,
\ee
are string Galilei boosts with dimensions of velocity, where $\eta^{tt}=-1=-\eta^{vv}$ and $\eta^{tv}=0$. Finally, the $a^A$ and $a^i$ are longitudinal and transverse translations, respectively.

We now expand the embedding fields as in~\eqref{eq:EmbeddingExp} and the transformation parameters as
\begin{align}
    \begin{split}
        a^A &= a^A_{(0)} + c^{-2}a^A_{(2)} + c^{-4}a^A_{(4)} + \mathcal{O}(c^{-6})\,,\\
        \Lambda^A{_B} &= \lambda^A_{(0)B} + c^{-2}\lambda^A_{(2)B} + c^{-4}\lambda^A_{(4)B} + \mathcal{O}(c^{-6})\,,\\
        \Lambda^A{_i} &= \lambda^A_{(0)i} + c^{-2}\lambda^A_{(2)i} + \mathcal{O}(c^{-4})\,,\\
        a^i &= a^i_{(0)} + c^{-2}a^i_{(2)} + \mathcal{O}(c^{-4})\,,\\
        a^A &= a^A_{(0)} + c^{-2}a^A_{(2)} + c^{-4}a^A_{(4)} + \mathcal{O}(c^{-6})\,,\\
        \Lambda^i{_j} &= \lambda^i_{(0)j} + c^{-2}\lambda^i_{(2)j} + \mathcal{O}(c^{-4})\,.
    \end{split}
\end{align}
This leads to
\begin{subequations}
\be 
\delta x^A &=& \lambda^A_{(0)B}x^B + a^A_{(0)}\,,\label{eq:sym-nr-1}\\
\delta x^i &=& \lambda^i_{(0)j}x^j+ \lambda^i_{(0)A}x^A + a^i_{(0)}\,,\\
\delta y^A &=& \lambda^A_{(0)B}y^B + \lambda^A_{(2)B}x^B + \lambda^A_{(0)i}x^i + a^A_{(2)}\,,\\
\delta y^i &=& \lambda^i_{(0)j}y^j + \lambda^i_{(2)j}x^j + \lambda_{(0)A}^iy^A + \lambda_{(2)A}^ix^A + a_{(2)}^i\,,\\
\delta z^A &=& \lambda^A_{(0)B}z^B + \lambda^A_{(2)B}y^B + \lambda^A_{(4)B}x^B + \lambda^A_{(0)i}y^i + \lambda^A_{(2)i}x^i + a^A_{(4)}\,.\label{eq:sym-nr-sidst}
\ee 
\end{subequations}
In the expanded string theory, each of the transformations above will have an associated Noether charge. These in turn correspond to expanded components of the relativistic Noether charges, which arise from the symmetries~\eqref{eq:poincare-trafo} of the relativistic Polyakov Lagrangian~\eqref{eq:rel-polyakov}, which on flat target space and in conformal gauge is
\begin{equation}
    \mathcal{L}_{\text{P}} = -\frac{T_{\text{eff}}}{2}\eta^{\alpha\beta} \D_\alpha X^M\D_\beta X^M\eta_{MN}\,.
\end{equation}
The Noether currents corresponding to the translations $a^A$ and $a^i$ are
\begin{align}
    \begin{split}
        \Pi^\alpha_A &= \pd{\mathcal{L}_{\text{P}}}{\D_\alpha X^A} = -c^2T_{\text{eff}}\D^\alpha X^B\eta_{BA}\,,\\
        \Pi^\alpha_i &= \pd{\mathcal{L}_{\text{P}}}{\D_\alpha X^i} = -T_{\text{eff}}\D^\alpha X^j\delta_{ji}\,,
    \end{split}
\end{align}
while the Noether currents for longitudinal Lorentz transformations $\Lambda^A{_B}$ and transverse rotations $\Lambda^i{_j}$ take the form
\begin{align}
\begin{split}
    J^\alpha_{AB} &= X_A\Pi^\alpha_B - X_B\Pi^\alpha_A\,,\\
    J^\alpha_{ij} &= X_i\Pi^\alpha_j - X_j\Pi^\alpha_i\,.
\end{split}
\end{align}
Finally, the Noether current for the transformations $\Lambda^A{_i}$ is
\begin{equation}
    J^\alpha_{Ai} = X_A \Pi^\alpha_i - c^{-2}X_i \Pi^\alpha_A\,.
\end{equation}
Expanding the Noether currents in powers of $1/c^2$, we get 
\begin{align}
\label{eq:expansion-of-Noether}
    \begin{split}
        \Pi^\alpha_A &= c^2\pi^\alpha_{(-2)A} + \pi^\alpha_{(0)A} + c^{-2}\pi^\alpha_{(-2)A} + \mathcal{O}(c^{-4})\,,\\
        \Pi^\alpha_i &= \pi^\alpha_{(0)i} + c^{-2}\pi^\alpha_{(-2)i} + \mathcal{O}(c^{-4})\,,\\
        J^\alpha_{AB} &=  c^2j^\alpha_{(-2)AB} + j^\alpha_{(0)AB} + c^{-2}j^\alpha_{(-2)AB} + \mathcal{O}(c^{-4})\,,\\
        J^\alpha_{ij} &= j^\alpha_{(0)ij} + c^{-2}j^\alpha_{(-2)ij} + \mathcal{O}(c^{-4})\,,\\
        J^\alpha_{Ai} &= j^\alpha_{(0)Ai} + c^{-2}j^\alpha_{(-2)Ai} + \mathcal{O}(c^{-4})\,, 
    \end{split}
\end{align}
where 
\begin{align}
\label{eq:Noether-expansion}
\pi^\alpha_{(-2)A} &= -T_{\text{eff}} \D^\alpha x_A\,,& \pi^\alpha_{(0)A} &= -T_{\text{eff}} \D^\alpha y_A\,,\nn\\
        \pi^\alpha_{(0)i} &= -T_{\text{eff}} \D^\alpha x_i\,,& \pi^\alpha_{(2)i} &= -T_{\text{eff}} \D^\alpha y_i\,, & \nn\\
        j^\alpha_{(-2)AB} &= 2x_{[A}\pi^\alpha_{(-2)B]}\,,& j^\alpha_{(0)AB} &= 2x_{[A}\pi^\alpha_{(0)B]} + 2y_{[A}\pi^\alpha_{(-2)B]}\,,  \nn\\
        j^\alpha_{(0)ij} &= 2x_{[i}\pi^\alpha_{(0)j]}\,,& j^\alpha_{(2)ij} &= 2x_{[i}\pi^\alpha_{(2)j]} + 2y_{[i}\pi^\alpha_{(0)j]}\,,  \nn\\
        j^\alpha_{(0)Ai} &= x_A\pi^\alpha_{(0)i} - x_i\pi^\alpha_{(-2)A}\,,& j^\alpha_{(2)Ai} &= x_A\pi^\alpha_{(2)i} - x_i \pi^\alpha_{(0)A} \nn\\
        &&&\hspace{.4cm}+ y_A\pi^\alpha_{(0)i} - y_i \pi^\alpha_{(-2)A}\,, \nn\\
        j^\alpha_{(2)AB} &= 2x_{[A}\pi^\alpha_{(2)B]} + 2y_{[A}\pi^\alpha_{(0)B]} + 2z_{[A}\pi^\alpha_{(-2)B]}\,, & \pi^\alpha_{(2)A} &= -T_{\text{eff}} \D^\alpha z_A\,.
\end{align}
As we will see below, these coefficients of the $1/c^2$ expanded Noether currents are precisely the Noether currents associated to the expanded transformations. In this sense, the $1/c^2$ expansion commutes with the Noether procedure.

\subsection{Expansion of the Poincar\'e algebra}

Before we study the algebra of Noether charges associated with the global symmetries of the LO, NLO and NNLO Lagrangians we first discuss the string $1/c^2$ expansion of the Poincar\'e algebra.

The $(d+2)$-dimensional Poincar\'e algebra $\mathfrak{iso}(d+1,1)$ in the basis $(\texttt{J}_{MN},\texttt{P}_M)$ has the brackets
\begin{align}
    \begin{split}
        [\texttt{J}_{MN},\texttt{J}_{KL}] &= \eta_{MK}\texttt{J}_{NL} - \eta_{NK}\texttt{J}_{ML} - \eta_{ML}\texttt{J}_{NK} + \eta_{NL}\texttt{J}_{MK}\,,\\
        [\texttt{J}_{MN},\texttt{P}_K] &= 2\eta_{K[M}\texttt{P}_{N]}\,.
    \end{split}
\end{align}
As explained in~\cite{Harmark:2019upf}, we set up the string $1/c^2$ expansion of the Poincar\'e algebra by splitting the index $M = (A,i)$ and reinstating factors of $c$, which amounts to
\begin{equation}
    \texttt{J}_{Ai} = c\texttt{B}_{Ai}\qquad\text{and}\qquad \texttt{P}_A = \texttt{H}_A/c\,,
\end{equation}
for the stringy boosts and the longitudinal translations, respectively. 
We then expand the generators according to either
\begin{align}
\label{eq:Noether-exp}
\begin{split}
    \texttt{X} &=  \sum_{k\in \mathbb{N}_0}c^{-2k}\texttt{X}^{(2k)}\,,\\
    \texttt{Y} &=  \sum_{k\in \mathbb{N}_0}c^{-2k}\texttt{Y}^{(2k-2)}\,,
\end{split}
\end{align}
where $\texttt{X}$ consists of the generators $\{\texttt{J}_{ij},\texttt{P}_{i},\texttt{B}_{Ai}  \}$, while $\texttt{Y}$ consists of the generators $\{\texttt{J}_{AB},\texttt{H}_{A}\}$. We define the level of a generator to be the (even) integer in parentheses in the superscript; for example $\texttt{P}^{(10)}_i$ is a ``level-10 generator''.
Thus, $\texttt{Y}$ begins at level $-2$, while $\texttt{X}$ begins at level $0$. The reasoning behind this off-set for $\texttt{Y}$ will become clear below. This means that we obtain the following brackets (where $m,n\in \mathbb{N}_0$)
\begin{subequations}
\begin{align}
[ \texttt{J}_{AB}^{(2m-2)}, \texttt{H}_C^{(2n-2)} ]&= 2 \eta_{C [A } \texttt{H}_{B]}^{(2m+2n-2)}\,,&
[ \texttt{J}_{ij}^{(2m)}, \texttt{P}_{k}^{(2n)} ]&=2 \delta_{k[i } \texttt{P}_{j]}^{(2m+2n)} \, ,\\
 [ \texttt{H}_{A}^{(2m-2)}, \texttt{B}_{Bi} ^{(2n)} ]&= - \eta_{A B } \texttt{P}_{i}^{(2m+2n)}\,,&
 [ \texttt{P}_{i}^{(2m)}, \texttt{B}_{Aj} ^{(2n)} ]&=  \delta_{ij } \texttt{H}_{A}^{(2m+2n)}  \, ,\\
[ \texttt{B}_{A i}^{(2m)}, \texttt{J}_{BC}^{(2n-2)} ] &=  - 2\eta_{A [B } \texttt{B}_{C]i}^{(2m+2n)} \,,&
[ \texttt{B}_{A i}^{(2m)}, \texttt{J}_{jk}^{(2n)} ]&=  -2 \delta_{i[j |} \texttt{B}_{A | k]}^{(2m+2n)} \,,\\
[ \texttt{J}_{AB}^{(2m-2)}, \texttt{J}_{CD}^{(2n-2)} ] &= 0 \,,& [ \texttt{B}_{Ai}^{(2m)}, \texttt{B}_{Bj} ^{(2n)} ]&= \eta_{A B } \texttt{J}_{ij}^{(2m+2n+2)} + \delta_{ij} \texttt{J}_{AB}^{(2m+2n)}\, ,\nn\\
 [ \texttt{J}_{ij}^{(2m)}, \texttt{J}_{kl}^{(2n)} ]&=  4\delta_{{\color{red}[}j{\color{blue} [} k} \texttt{J}_{l{\color{blue} ]}i{\color{red} ]}}^{(2m+2n)}\,.&
\end{align}
\end{subequations}

We will denote the $1/c^2$ expanded Poincar\'e algebra, which contains infinitely many generators, by $\mathfrak{iso}_{1/c^2}(d+1,1)$. Now, notice that for a given, fixed integer $n\geq 0$, the set of all generators of level $2k\geq 2n$ forms an (infinite dimensional) ideal in $\mathfrak{iso}_{1/c^2}(d+1,1)$ that we call $\mathfrak{i}_n$. The biggest ideal is $\mathfrak{i}_{0}$, and in general we have the 
filtration
\begin{equation}
    \mathfrak{i}_{0}\supset \mathfrak{i}_1 \supset \mathfrak{i}_2\supset\dots
\end{equation}
This means that for each integer $n\geq 0$, we can form the quotient algebra
\begin{equation}
\label{eq:symmetry-algebra}
    \mathfrak{q}_{n} = \mathfrak{iso}_{1/c^2}(d+1,1)/\mathfrak{i}_n\,.
\end{equation}
For $n=0$, only the level $-2$ 
generators $\texttt{J}_{AB}^{(-2)}$ and $\texttt{H}_{A}^{(-2)}$ remain and generate the two-dimensional Poincar\'e algebra, i.e., $\mathfrak{q}_{0}\cong \mathfrak{iso}(1,1)$. 
For $n=1$, the algebra $\mathfrak{q}_1$ has the following nonzero brackets 
\begin{subequations}
\begin{align}
[ \texttt{J}_{AB}^{(-2)}, \texttt{H}_C^{(-2)} ]&= 2 \eta_{C [A } \texttt{H}_{B]}^{(-2)}\,,& [ \texttt{J}_{AB}^{(0)}, \texttt{H}_C^{(-2)} ]&= 2 \eta_{C [A } \texttt{H}_{B]}^{(0)}\,,\\
[ \texttt{J}_{AB}^{(-2)}, \texttt{H}_C^{(0)} ]&= 2 \eta_{C [A } \texttt{H}_{B]}^{(0)}\,,&
[ \texttt{J}_{ij}^{(0)}, \texttt{P}_{k}^{(0)} ]&=2 \delta_{k[i } \texttt{P}_{j]}^{(0)} \, ,\\
 [ \texttt{H}_{A}^{(-2)}, \texttt{B}_{Bi} ^{(0)} ]&= - \eta_{A B } \texttt{P}_{i}^{(0)}\,,&
 [ \texttt{P}_{i}^{(0)}, \texttt{B}_{Aj} ^{(0)} ]&=  \delta_{ij } \texttt{H}_{A}^{(0)}  \, ,\\
[ \texttt{B}_{Ai}^{(0)}, \texttt{B}_{Bj} ^{(0)} ]&= \delta_{ij} \texttt{J}_{AB}^{(0)}\, ,& [ \texttt{B}_{A i}^{(0)}, \texttt{J}_{BC}^{(-2)} ]&=  - 2\eta_{A [B } \texttt{B}_{C]i}^{(0)} \,,\\
[ \texttt{B}_{A i}^{(0)}, \texttt{J}_{jk}^{(0)} ] &=  -2 \delta_{i[j |} \texttt{B}_{A | k]}^{(0)} \,,&
[ \texttt{J}_{ij}^{(0)}, \texttt{J}_{kl}^{(0)} ] &= 4\delta_{{\color{red}[}j{\color{blue} [} k} \texttt{J}_{l{\color{blue} ]}i{\color{red} ]}}^{(0)}\,.
\end{align}
\end{subequations}
This algebra is known as the string Bargmann algebra~\cite{Bergshoeff:2019pij}. Below we will show that the algebra of Noether charges at order N$^n$LO of the string $1/c^2$ expansion is isomorphic to $\mathfrak{q}_n$ for $n=0,1,2$, and we expect this pattern to persist at all orders.\footnote{An identical picture exists for the particle: there, the first non-trivial quotient algebra is the Bargmann algebra.}

The expansion~\eqref{eq:Noether-exp} of the Poincar\'e algebra gives rise to the algebra generated by the Noether charges of the string $1/c^2$ expansion of string theory. However, this is \textit{not} the expansion of the Poincar\'e algebra that underlies the geometry to which these theories couple, cf., Table~\ref{table:SNC-geometries}. The expansion of the generators that, upon gauging, leads to the geometries of Table~\ref{table:SNC-geometries} is (see also~\cite{Harmark:2019upf})
\begin{align}
\label{eq:geometry-exp}
\begin{split}
    \texttt{X} &=  \sum_{k\in \mathbb{N}_0}c^{-2k}\texttt{X}^{(2k)}\,,\\
    \texttt{Y} &=  \sum_{k\in \mathbb{N}_0}c^{-2k}\texttt{Y}^{(2k)}\,.
\end{split}
\end{align}
This leads to larger algebras at each level than the expansion~\eqref{eq:Noether-exp}. The particle version of this expansion was worked out in \cite{Hansen:2020pqs}.

\subsection{Noether charge algebra}

The LO Polyakov Lagrangian on flat space in flat conformal gauge reads
\be 
\mathcal{L}_{\text{P-LO}} = -\frac{T_{\text{eff}}}{2}\eta_{AB}\eta^{\alpha\beta}\D_\alpha x^A\D_\beta x^B\,,\label{eq:LOflatspace-1}
\ee 
As we have already remarked, the LO Polyakov Lagrangian is identical to the relativistic Polyakov Lagrangian with two target space dimensions. The conserved current corresponding to translations $a_{(0)}^A$ is
\be 
\pi^\alpha_{(-2)A} = \pd{\mathcal{L}_{\text{P-LO}}}{\D_\alpha x^A} = -T_{\text{eff}}\D^\alpha x_A\,,\label{eq:LO-transl-current}
\ee 
while the current for longitudinal Lorentz transformations is 
\be 
j_{(-2)AB}^\alpha = x_A\pi_{(-2)B}^\alpha - x_B\pi_{(-2)A}^\alpha\,.\label{eq:LO-Lorentz-current}
\ee 

The equal-time Poisson brackets\footnote{We will have more to say about Poisson brackets in Section~\ref{sec:phase-space}.} between the canonically conjugate variables $x^A$ and $\pi^0_{(-2)A}$ at LO are
\be 
\{x^A(\sigma^1),\pi^0_{(-2)B}(\tilde\sigma^1) \} = \delta^A_B\delta(\sigma^1 - \tilde\sigma^1)\,.\label{eq:LO-brackets}
\ee 
The charges at LO are
\be 
\label{eq:LO-charges}
\mathcal{P}_{(-2)A} = \oint d \sigma^1\,\pi^0_{(-2)A} \,,\qquad
\mathcal{J}_{(-2)AB} =  \oint d \sigma^1\,j^0_{(-2)AB} \,,
\ee 
which, using the LO brackets \eqref{eq:LO-brackets} generate the $2$-dimensional Poincar\'e algebra, which we called $\mathfrak{q}_0$ above. The specific map between charges and generators is
\begin{equation}
\mathcal{P}_{(-2)A} \leftrightarrow \texttt{H}^{(-2)}_A\,,\qquad \mathcal{J}_{(-2)AB} \leftrightarrow \texttt{J}^{(-2)}_{AB}\,.
\end{equation}
This is as it should be: since the LO Lagrangian is the standard Polyakov action with a two-dimensional target space, the charge algebra better be the two-dimensional Poincar\'e algebra.

We now turn our attention to the NLO Polyakov Lagrangian, which we write as
\be 
\mathcal{L}_{\text{P-NLO}} &=& -\frac{T_{\text{eff}}}{2}\eta^{\alpha\beta}\D_\alpha x^i\D_\beta x^i-T_{\text{eff}}\eta_{AB}\eta^{\alpha\beta}\D_\alpha x^A\D_\beta y^B \,.
\ee 
Invariance under stringy boosts $\lambda^A_{(0)i}$ follows from 
\be 
\delta_{\lambda_{(0)}}\mathcal{L}_{\text{P-NLO}} = -T_{\text{eff}}\eta^{\alpha\beta}\D_\alpha x^i \D_\beta x^A(\lambda^i_{(0)A} + \eta_{AB}\lambda^B_{(0)i}) = 0\,,
\ee 
and equation \eqref{eq:Lambda^A_i},
while invariance under subleading longitudinal Lorentz transformations $\lambda_{(2)B}^A$ is a consequence of
\be 
\delta_{\lambda_{(2)}}
\mathcal{L}_{\text{P-NLO}} = -T_{\text{eff}}\eta^{\alpha\beta}\D_\alpha x^A \D_\beta x^B\lambda_{(2)AB}= 0\,.
\ee 
Invariance under the reamining symmetries follows from identical arguments. The transverse translations $a_{(0)}^i$ and rotations $\lambda^i_{(0)j}$  act only on $x^i$, and the Noether currents are
\be 
\begin{aligned}
\pi^\alpha_{(0)i} &= \pd{\mathcal{L}_{\text{P-NLO}}}{\D_\alpha x^i} = -T_{\text{eff}}\D^\alpha x_i\,,\\
j_{(0)ij}^\alpha &= x_i\pi_{(0)j}^\alpha - x_j\pi_{(0)i}^\alpha\,.
\end{aligned}
\ee 
The LO longitudinal translations $a_{(0)}^A$ now give rise to the current
\be 
\pi^\alpha_{(0)A} = \pd{\mathcal{L}_{\text{P-NLO}}}{\D_\alpha x^A} = -T_{\text{eff}}\D^\alpha y_A\,,\label{eq:NLO-transl-current}
\ee 
while the LO longitudinal Lorentz transformations $\lambda^A_{(0)B}$ produce the current
\be
j^\alpha_{(0)AB} = 2x_{[A}\pi^\alpha_{(0)B]} + 2y_{[A}\pi^\alpha_{(-2)B]}\,.
\ee 
The currents of the LO theory arise now as the currents for subleading longitudinal transformation. Specifically, NLO translations $a^A_{(2)}$, acting on $y^A$, produce the current $\pi^\alpha_{(-2)A}$ given in \eqref{eq:LO-transl-current}, and the current for NLO longitudinal Lorentz transformations $\lambda^A_{(2)B}$ is $j_{(-2)AB}^\alpha$ given in \eqref{eq:LO-Lorentz-current}. Finally, the current for stringy boosts $\Lambda^A_{(0)i}$ is
\be 
j_{(0)Ai}^\alpha = x_A\pi_{(0)i}^\alpha - x_i\pi_{(-2)A}^\alpha\,.
\ee 
The charges at NLO are (see Appendix~\ref{sec:noether-details} for further details)
\begin{align}
\mathcal{P}_{(-2)A} &= \oint d\sigma^1\,\pi^0_{(-2)A}   \,,&
\mathcal{J}_{(-2)AB} &=  \oint d\sigma^1\,j^0_{(-2)AB}\,,\nn\\
\mathcal{P}_{(0)i} &= \oint d\sigma^1\,\pi^0_{(0)i} \,,&
\mathcal{P}_{(0)A} &= \oint d\sigma^1\,\pi^0_{(0)A}\,,\\
\mathcal{J}_{(0)AB} &=  \oint d\sigma^1\,j_{(0)AB} \,,&
\mathcal{J}_{(0)ij} &=  \oint d\sigma^1\,j^0_{(0)ij} \,,\nn\\
\mathcal{J}_{(0)Ai} &=  \oint d\sigma^1\,j^0_{(0)Ai} \,.&\nn
\end{align}
The Poisson brackets between canonically conjugate variables at NLO change compared to the LO brackets:
\be 
\{ y^A(\sigma^1),\pi^0_{(-2)B}(\tilde\sigma^1) \} &=& \delta^A_B\delta(\sigma^1-\tilde\sigma^1)\,,\qquad \{ x^A(\sigma^1), \pi^0_{(0)B}(\tilde\sigma^1) \} = \delta^A_B\delta(\sigma^1-\tilde\sigma^1)\,,\nn\\
\{x^i(\sigma^1), \pi^0_{(0)j}(\tilde\sigma^1) \} &=& \delta^i_j\delta(\sigma^1-\tilde\sigma^1)\,,\label{eq:NLO-brackets}
\ee 
and using these brackets, the charges at NLO generate the string Bargmann algebra $\mathfrak{q}_1$. The dictionary between charges and generators at NLO is
\begin{align}
        \mathcal{P}_{(-2)A} &\leftrightarrow \texttt{H}^{(0)}_A\,,& \mathcal{P}_{(0)A} &\leftrightarrow \texttt{H}^{(-2)}_A\,,\nn\\
        \mathcal{J}_{(-2)AB} &\leftrightarrow \texttt{J}^{(0)}_{AB}\,,& \mathcal{J}_{(0)AB} &\leftrightarrow \texttt{J}^{(-2)}_{AB}\,,\\
        \mathcal{P}_{(0)i} &\leftrightarrow \texttt{P}^{(0)}_i\,,& \mathcal{J}_{(0)ij} &\leftrightarrow \texttt{J}^{(0)}_{ij}\,,&
        \mathcal{J}_{(0)Ai} &\leftrightarrow \texttt{B}_{Ai}^{(0)}\,.&\nn
\end{align}
Note that the charges $\mathcal{P}_{(-2)A}$ and $\mathcal{J}_{(-2)AB}$ now map to different generators compared to what we found at LO. However, the charges corresponding to the transformations $a_{(0)}^A$ and $\lambda^A_{(0)B}$ still correspond to the same generators; a pattern that persists to higher orders, as we shall see.

Moving on to NNLO, we may write the Lagrangian as
\be 
\mathcal{L}_{\text{P-NNLO}} = -\frac{T_{\text{eff}}}{2}\eta^{\alpha\beta}\eta_{AB}\D_\alpha y^A \D_\beta y^B - T_{\text{eff}}\eta^{\alpha\beta}\eta_{AB}\D_\alpha x^A \D_\beta z^B- T_{\text{eff}}\eta^{\alpha\beta} \D_\alpha x^i \D_\beta y^i\,.\nn
\ee
The symmetries are given in \eqref{eq:sym-nr-1}--\eqref{eq:sym-nr-sidst}, and we find that, again, the most subleading transformations give rise to the most leading charges: i.e., $a_{(4)}^A$ and $\lambda^A_{(4)B}$ give rise to $\mathcal{P}_{(-2)A}$ and $\mathcal{J}_{(-2)AB}$, $a_{(2)}^A$ and $\lambda^A_{(2)B}$ give rise to $\mathcal{P}_{(0)A}$ and $\mathcal{J}_{(0)AB}$, $a_{(2)}^i$ and $\lambda^i_{(2)j}$ give rise to $\mathcal{P}_{(0)i}$ and $\mathcal{J}_{(0)ij}$, and $\lambda^A_{(0)i}$ gives rise to $\mathcal{J}_{(0)Ai}$. The new charges at NNLO are those associated to the most leading transformations and are given by (see again Appendix~\ref{sec:noether-details} for their explicit expressions)
\begin{align}
\mathcal{P}_{(2)A} &= \oint d \sigma^1\,\pi^0_{(2)A} \,,&
\mathcal{J}_{(2)AB} &=  \oint d \sigma^1\,j_{(0)AB}\,,\nn\\
\mathcal{P}_{(2)i} &= \oint d \sigma^1\,\pi^0_{(0)i} \,,&
\mathcal{J}_{(2)ij} &=  \oint d \sigma^1\,j^0_{(0)ij} \,,\\
\mathcal{J}_{(2)Ai} &=  \oint d \sigma^1\,j^0_{(0)Ai} \,.\nn
\end{align}
The Poisson brackets at NNLO are
\be 
\label{eq:NNLO-brackets}
\begin{aligned}
\{ z^A(\sigma^1),\pi^0_{(-2)B}(\tilde\sigma^1) \} &= \delta^A_B\delta(\sigma^1-\tilde\sigma^1)\,,\\
\{ y^A(\sigma^1), \pi^0_{(0)B}(\tilde\sigma^1) \} &= \delta^A_B\delta(\sigma^1-\tilde\sigma^1)\,,\\
\{ x^A(\sigma^1), \pi^0_{(2)B}(\tilde\sigma^1) \} &= \delta^A_B\delta(\sigma^1-\tilde\sigma^1)\,,\\
\{y^i(\sigma^1), \pi^0_{(0)j}(\tilde\sigma^1) \} &=\delta^i_j\delta(\sigma^1-\tilde\sigma^1)\,,\\
\{x^i(\sigma^1), \pi^0_{(2)j}(\tilde\sigma^1) \} &= \delta^i_j\delta(\sigma^1-\tilde\sigma^1)\,,
\end{aligned}
\ee 
and with these we find that the charges at NNLO generate the algebra $\mathfrak{q}_2$ with the following identification (see Appendix~\ref{sec:noether-details} for more details)
\begin{align}
\label{eq:NNLO-dictionary}
\begin{aligned}
\mathcal{P}_{(2)A} &\leftrightarrow  \texttt H^{(-2)}_A\,,&  \mathcal{P}_{(0)A} &\leftrightarrow  \texttt H^{(0)}_A\,,&
\mathcal{P}_{(-2)A} &\leftrightarrow\texttt  H^{(2)}_A\,,\\
\mathcal{J}_{(2)AB} &\leftrightarrow  \texttt J^{(-2)}_{AB}\,,& \mathcal{J}_{(0)AB} &\leftrightarrow  \texttt J^{(0)}_{AB}\,,& \mathcal{J}_{(-2)AB} &\leftrightarrow  \texttt J^{(2)}_{AB}\,,\\
\mathcal{P}_{(2)i} &\leftrightarrow \texttt P^{(0)}_{i}\,,& \mathcal{P}_{(0)i} &\leftrightarrow \texttt P^{(2)}_{i}\,,\\
 \mathcal{J}_{(2)ij} &\leftrightarrow  \texttt J^{(0)}_{ij}\,,&\mathcal{J}_{(0)ij} &\leftrightarrow  \texttt J^{(2)}_{ij}\,,&\\
\mathcal{J}_{(2)Ai} &\leftrightarrow \texttt B^{(0)}_{A i }\,,&
\mathcal{J}_{(0)Ai} &\leftrightarrow \texttt B^{(2)}_{Ai } \,.&&
\end{aligned}
\end{align} 
In this way, the Noether charge corresponding to, e.g., LO longitudinal translations $a_{(0)}^A$ still plays the role as the generator $\texttt H_A^{(-2)}$, but this is a new charge compared to those that appeared at LO and NLO. We expect this pattern to continue at all orders, which leads to the following general identification between charges and generators at N$^n$LO
\begin{align}
    \begin{aligned}
        \mathcal{P}_{(-2+2k)A} &\leftrightarrow  \texttt H^{(-2+2(n-k))}_A\,,& \mathcal{J}_{(-2+2k)AB} &\leftrightarrow  \texttt J^{(-2+2(n-k))}_{AB}\,, \\
        \mathcal{P}_{(2k)i} &\leftrightarrow \texttt P^{(-2+2(n-k))}_{i}\,,& \mathcal{J}_{(2k)ij} &\leftrightarrow  \texttt J^{(-2+2(n-k))}_{ij}\,,&\mathcal{J}_{(2k)Ai} &\leftrightarrow \texttt B^{(-2+2(n-k))}_{A i }\,,
    \end{aligned}
\end{align}
where $n$ is fixed and $k\in\{0,1,\dots,n\}$ in the first line and $k\in\{0,1,\dots,n-1\}$ (for $n\geq 1$) in the second line. For example, $n=2$ corresponds to NNLO and reproduces~\eqref{eq:NNLO-dictionary}. The fact that the most subleading Noether charge is associated with the most leading generator -- implying that the counting starts at opposite ends, as it were -- is a consequence of the fact that the Poisson brackets change at each order.

\section{Phase space formulation}
\label{sec:phase-space}
As we saw in the previous section, the Poisson brackets change at each order in $1/c^2$. In this section, we develop a phase space formulation for the $1/c^2$ expansion of the closed bosonic string, and from the expansion of the symplectic form we will explicitly see how the Poisson brackets are modified at each order. In particular, this affects the brackets between the constraints.

The gauge symmetries generated by the expanded constraints can be gauged fixed using standard methods, leading to Poisson brackets for the gauge-fixed Lagrangians that can be used to quantise the string theory by passing to commutators.

\subsection{Expanding the relativistic phase space action}

The relativistic phase space Lagrangian describing a closed bosonic string is 
\be 
L = \oint d \sigma^1\left[\dot X^M P_M - \vartheta^-\mathcal{H}_- - \vartheta^+\mathcal{H}_+ \right]\,,\label{eq:BetterForm}
\ee 
where the constraints $\mathcal{H}_\pm$ are imposed by the Lagrange multipliers $\vartheta^\pm$. They are given by
\be 
\mathcal{H}_\pm &=& \frac{1}{4cT}(P\pm cTX')^M(P\pm cTX')^N\eta_{MN}\,.
\ee 
As above, we split the spacetime index $M = (A,i)$. From the expansion of the Noether currents~\eqref{eq:expansion-of-Noether} and the expansion of $X^M$ in~\eqref{eq:EmbeddingExp}, we find that the quantities involved in~\eqref{eq:BetterForm} expand as
\begin{subequations}
\begin{align}
    X^{A} &= x^{A} + c^{-2}y^{A} + c^{-4}z^{A} + \mathcal{O}(c^{-6})\,,\\
    X^i &= x^i + c^{-2}y^i + c^{-4}z^i + \mathcal{O}(c^{-6})\,,\\
    P_{A} &= c^2 P_{(-2)A} + P_{(0)A} + c^{-2}P_{(2)A} + \mathcal{O}(c^{-4})\,,\\
    P_i &= P_{(0)i} + c^{-2}P_{(2)i} + c^{-4}P_{(4)i} + \mathcal{O}(c^{-6})\,.
\end{align}
\end{subequations}
Furthermore, the Minkowski metric $\eta_{MM}$ and its inverse are given by
\begin{align}
\label{eq:minkowski-metric-exp}
    \begin{split}
    \eta_{MN} &= c^2\left( -\delta_M^t \delta_N^t + \delta_M^v\delta_N^v\right) + \delta^i_M\delta^i_N\,,\\
    \eta^{MN} &= \frac{1}{c^2}\left( -\delta^M_t\delta^N_t + \delta^M_v \delta^N_v \right) + \delta^M_i\delta^N_i\,,
    \end{split}
\end{align}
which implies that
\begin{equation}
    P^t = \eta^{tM}P_M = -\frac{1}{c^2}P_t\qquad \text{and}\qquad P^v = \eta^{vM}P_M = \frac{1}{c^2}P_v\,,
\end{equation}
leading to the following expansions for the contravariant longitudinal momenta
\begin{align}
    \begin{split}
        P^t &=  -P_{(-2)t} - c^{-2}P_{(0)t} - c^{-4}P_{(2)t} + \mathcal{O}(c^{-6})\,,\\
        P^v &=  P_{(-2)v} + c^{-2}P_{(0)v} + c^{-4}P_{(2)v} + \mathcal{O}(c^{-6})\,.
    \end{split}
\end{align}
Using the Minkowksi metric in the form~\eqref{eq:minkowski-metric-exp}, we can write the constraints as
\begin{equation}
    \mathcal{H}_\pm = \frac{c^2}{4T_{\text{eff}}}\left( -(P^t\pm T_{\text{eff}}X'^t)^2 + (P^v\pm T_{\text{eff}}X'^v)^2 \right) + \frac{1}{4T_{\text{eff}}}(P^i\pm T_{\text{eff}}X'^i)^2\,.
\end{equation}
Expanding the constraints in powers of $1/c^2$, we get
\begin{equation} 
\mathcal{H}_\pm = c^2\mathcal{H}_{(-2)\pm} + \mathcal{H}_{(0)\pm} + c^{-2}\mathcal{H}_{(2)\pm} + \mathcal{O}(c^{-4})\,,
\end{equation}
where
\begin{align}
    \begin{split}
        \mathcal{H}_{(-2)\pm} &= 
        \frac{1}{4T_{\text{eff}}}\eta_{AB}(P_{(-2)}^A\pm T_{\text{eff}}x'^A)(P_{(-2)}^B\pm T_{\text{eff}}x'^B) \,,\\
        \mathcal{H}_{(0)\pm}&= 
        \frac{1}{4T_{\text{eff}}}(P_{(0)i}\pm T_{\text{eff}}x'^i)^2+ \frac{1}{2T_{\text{eff}}}\eta_{AB}(p^A_{(-2)}\pm T_{\text{eff}}x'^A)(P_{(0)}^B\pm T_{\text{eff}}y'^B)\,,\\
        \mathcal{H}_{(2)\pm} &= \frac{1}{2T_{\text{eff}}}(P_{(0)i} \pm T_{\text{eff}} x'^i)(P_{(2)i} \pm T_{\text{eff}} y'^i) \\
        &\quad+\frac{1}{4T_{\text{eff}}}\eta_{AB}\left[ (P_{(2)}^A\pm T_{\text{eff}} y'^A)(P_{(2)}^B\pm T_{\text{eff}} y'^B) + 2(P_{(0)}^A\pm T_{\text{eff}} x'^A)(P_{(4)}^B\pm T_{\text{eff}} z'^B)\right]
    \end{split}
\end{align}
with $\eta_{AB} = \text{diag}(-1,1)$ the two-dimensional Minkowski metric.
The Lagrange multipliers are expanded according to
\be 
\vartheta_\pm = \vartheta_{(0)\pm} + c^{-2}\vartheta_{(2)\pm}  + c^{-4}\vartheta_{(4)\pm} +\cdots\,.
\ee 
Combining our findings above, the phase space Lagrangian expands as 
\begin{equation}
    L = c^2 L_{\text{LO}} + L_{\text{NLO}} + c^{-2}L_{\text{NNLO}} + \mathcal{O}(c^{-4})\,,
\end{equation}
where
\begin{subequations}
\begin{align}
    L_{\text{LO}} &= \oint d \sigma^1\left[\dot x^A P_{(-2)A}  - \vartheta^-_{(0)}\mathcal{H}_{(-2)-} - \vartheta^+_{(0)}\mathcal{H}_{(-2)+} \right]\,,\label{eq:LOActionBetterForm}\\
        L_{\text{NLO}} &= \oint d \sigma^1\left[\dot x^i P_{(0)i} + \dot x^A P_{(0)A} +\dot y^A P_{(-2)A}  - \vartheta^-_{(0)}\mathcal{H}_{(0)-} - \vartheta^+_{(0)}\mathcal{H}_{(0)+} \right.\nn\\
        &\left.- \vartheta^-_{(2)}\mathcal{H}_{(-2)-} - \vartheta^+_{(2)}\mathcal{H}_{(-2)+} \right]\,,\label{eq:NLOActionBetterForm}\\
            \begin{split}
        \label{eq:NNLO-phase-space-lagrangian}
        L_{\text{NNLO}} &= \oint d \sigma^1\big[ \dot x^A P_{(2)A} + \dot y^A P_{(0)A} + \dot z^A P_{(-2)A} + \dot x^i P_{(2)i} + \dot y^i P_{(0)i}  \\
        &\quad - \vartheta_{(0)}^- \mathcal{H}_-^{(2)} - \vartheta_{(2)}^- \mathcal{H}_-^{(0)} - \vartheta_{(4)}^- \mathcal{H}_-^{(-2)} - \vartheta_{(0)}^+ \mathcal{H}_+^{(2)} - \vartheta_{(2)}^+ \mathcal{H}_+^{(0)} - \vartheta_{(4)}^+ \mathcal{H}_+^{(-2)}  \big]\,.
    \end{split}
    \end{align}
\end{subequations}
One can read off the Poisson brackets from the phase space Lagrangian (see e.g. \cite{Faddeev:1988qp}) and we see that the Poisson brackets changes at each order in $1/c^2$; something we already made use of in Section~\ref{sec:target-space-symmetries}.

\subsection{Dirac brackets}
\label{sec:dirac-brackets}
The relativistic first class constraints $\mathcal{H}_\pm$ give, as we shall see, rise to novel first-class constraints when expanded in powers of $1/c^2$. These first class constraints generate gauge redundancies, which can be fixed by an admissible gauge fixing condition which typically restricts the canonical variables which now form the so called reduced phase space. On the reduced phase space, we may then read off the Dirac brackets~\cite{Henneaux:1992ig}, and from those we can pass to the quantum theory by promoting the Dirac brackets to quantum commutators. 

The gauge choice must satisfy the following conditions (in which case the choice is ``canonical'') 
\begin{enumerate}
    \item The gauge choice must be ``accessible'': this means that there exists a gauge transformation that transforms the canonical variables into a set that satisfies the gauge condition.
    \item The gauge choice must fix the gauge completely. This amounts to the requirement that the determinant of the brackets between the (first class) constraints and the gauge choice is nonzero; equivalently, the brackets considered as a matrix indexed by the (equal) number of constraints and gauge fixing conditions is invertible. 
\end{enumerate}
Having fixed the gauge, we then solve the constraints inside the action and read off the Dirac brackets from the terms containing the velocities of the unconstrained phase space action. 

\subsubsection*{LO}
From the first term of the LO phase space Lagrangian~\eqref{eq:LOActionBetterForm}, we read off the (equal-$\sigma^0$) Poisson brackets
\begin{equation}
\label{eq:LO-pb}
    \{x^A(\sigma^1),P_{(-2)B}(\tilde\sigma^1)\}  = \delta^A_B\delta(\sigma^1 - \tilde\sigma^1)\,.
\end{equation}
Using these brackets, the algebra of the LO constraints $\mathcal{H}_{(-2)\pm}$ becomes 
\begin{align}
\label{eq:LO-constraint-brackets}
    \begin{split}
        \{ \mathcal{H}_{(-2)+}(\sigma^1),\mathcal{H}_{(-2)+}(\tilde\sigma^1)\} &= (\mathcal{H}_{(-2)+}(\sigma^1) + \mathcal{H}_{(-2)+}(\tilde\sigma^1))\delta'(\sigma^1-\tilde\sigma^1)\,,\\
        \{ \mathcal{H}_{(-2)-}(\sigma^1),\mathcal{H}_{(-2)-}(\tilde\sigma^1)\} &= -(\mathcal{H}_{(-2)-}(\sigma^1) + \mathcal{H}_{(-2)-}(\tilde\sigma^1))\delta'(\sigma^1-\tilde\sigma^1)\,,\\
        \{ \mathcal{H}_{(-2)+}(\sigma^1),\mathcal{H}_{(-2)-}(\tilde\sigma^1)\} &= 0\,,
    \end{split}
\end{align}
where we used that $\{x'^t(\sigma^1),P_{(-2)t}(\tilde\sigma^1)\} = \delta'(\sigma^1 - \tilde\sigma^1)$, where the derivative is with respect to $\sigma^1$, as well as the identity
\begin{equation}
\pd{}{\sigma^1}\delta(\sigma^1 - \tilde\sigma^1) = -\pd{}{\tilde\sigma^1}\delta(\sigma^1 - \tilde\sigma^1)\,.
\end{equation}
The action of the gauge transformations generated by $\mathcal{H}_{(-2)\pm}$ on a function $F$ on phase space is
\begin{equation}
    \delta_{\epsilon_{(-2)}} F = \left\{F, \oint d \sigma^1[\epsilon_{(-2)}^-\mathcal{H}_{{(-2)-}} + \epsilon_{(-2)}^+\mathcal{H}_{(-2)+}]\right\}\,,
\end{equation}
where $\epsilon_{(-2)}^\pm$ are the parameters of the gauge transformation. One can check that the gauge transformations act separately on the combinations $P_{(-2)}^A\pm T_{\text{eff}}x'^A$, and since we are dealing with closed strings, these are periodic functions in $\sigma^1$ and thus we may without loss of generality express them as Fourier series
\begin{align}
\label{eq:fourier-decomp-1}
    \begin{split}
        P_{(-2)}^A - T_{\text{eff}}x'^A&= \sqrt{\frac{T_{\text{eff}}}{\pi}} \sum_{k\in\mathbb{Z}}e^{ik\sigma^1}\alpha_k^A(\sigma^0)\,, \\
        P_{(-2)}^A + T_{\text{eff}}x'^A &= \sqrt{\frac{T_{\text{eff}}}{\pi}} \sum_{k\in\mathbb{Z}}e^{-ik\sigma^1}\tilde\alpha_k^A(\sigma^0)  \,,
    \end{split}
\end{align}
where reality constrains the modes to satisfy $(\alpha_k^A)^* = \alpha^A_{-k}$ and $(\tilde\alpha_k^A)^* = \tilde\alpha^A_{-k}$. Now, since $x^v$ carries winding, we can integrate the relations above to find that the total momentum $\wp^A_{(-2)}$ (although these are on-shell equivalent to the Noether charges $\mathcal{P}^A_{(-2)}$ we considered in the previous section, we will use the symbol $\wp$ to denote the total momentum along the string in the phase space formulation) is given by
\begin{equation}
    \wp^t_{(-2)}(\sigma^0)=\oint  d  \sigma^1 p^t_{(-2)} = \sqrt{4\pi T_{\text{eff}}}\alpha_0^t(\sigma^0) = \sqrt{4\pi T_{\text{eff}}}\tilde\alpha_0^t(\sigma^0)\,,
\end{equation}
and 
\begin{equation}
    \wp^v_{(-2)}(\sigma^0)=\oint  d  \sigma^1 p^v_{(-2)} =\sqrt{4\pi T_{\text{eff}}}\alpha_0^v(\sigma^0) + 2\pi T_{\text{eff}} wR_{\text{eff}} = \sqrt{4\pi T_{\text{eff}}}\tilde\alpha_0^v(\sigma^0) - 2\pi T_{\text{eff}} wR_{\text{eff}}\,,
\end{equation}
where we assumed that $x^v$ contains a winding term linear in $\sigma^1$. Adding the expressions in \eqref{eq:fourier-decomp-1} produces the Fourier series for $p^t_{(-2)}$ and $p^v_{(-2)}$
\begin{align}
    \begin{split}
        p^t_{(-2)}(\sigma^0,\sigma^1) &= \frac{\wp^t_{(-2)}(\sigma^0)}{2\pi} + \sqrt{\frac{T_{\text{eff}}}{4\pi}}\sum_{k\neq 0}e^{ik\sigma^1}\left( \alpha_k^t(\sigma^0) + \tilde\alpha_{-k}^t(\sigma^0) \right)\,,\\
        p^v_{(-2)}(\sigma^0,\sigma^1) &= \frac{\wp^v_{(-2)}(\sigma^0)}{2\pi} + \sqrt{\frac{T_{\text{eff}}}{4\pi}}\sum_{k\neq 0}e^{ik\sigma^1}\left( \alpha_k^v(\sigma^0) + \tilde\alpha_{-k}^v(\sigma^0) \right)\,,
    \end{split}
\end{align}
while subtracting gives us 
\begin{align}
    \begin{split}
        x'^t &= -\frac{1}{\sqrt{4\pi T_{\text{eff}}}}\sum_{k\neq 0}e^{ik\sigma^1}\left( \alpha_k^t(\sigma^0) - \tilde\alpha_{-k}^t(\sigma^0) \right)\,,\\
        x'^v &= wR_{\text{eff}} -\frac{1}{\sqrt{4\pi T_{\text{eff}}}}\sum_{k\neq 0}e^{ik\sigma^1}\left( \alpha_k^t(\sigma^0) - \tilde\alpha_{-k}(\sigma^0) \right)\,,
    \end{split}
\end{align}
which integrates to
\begin{align}
    \begin{split}
        x^t &= x_0^t(\sigma^0) + \frac{1}{\sqrt{4\pi T_{\text{eff}}}}\sum_{k\neq 0}\frac{i}{k}e^{ik\sigma^1}\left( \alpha_k^t(\sigma^0) - \tilde\alpha_{-k}(\sigma^0) \right)\,,\\
        x^v &= x_0^v(\sigma^0) + wR_{\text{eff}}\sigma^1 + \frac{1}{\sqrt{4\pi T_{\text{eff}}}}\sum_{k\neq 0}\frac{i}{k}e^{ik\sigma^1}\left( \alpha_k^t(\sigma^0) - \tilde\alpha_{-k}(\sigma^0) \right)\,.
    \end{split}
\end{align}
If we then Fourier expand the constraints $\mathcal{H}_{(-2)\pm}$, we get
\begin{equation}
    \mathcal{H}_{(-2) -} = \frac{1}{2\pi} \sum_{n\in\mathbb{Z}}e^{in\sigma^1}L^{(-2)}_n\qquad\text{and}\qquad \mathcal{H}_{(-2) +} = \frac{1}{2\pi} \sum_{n\in\mathbb{Z}}e^{-in\sigma^1}\tilde L^{(-2)}_n\,,
\end{equation}
where
\begin{equation}
    L^{(-2)}_n = \frac{1}{2}\sum_{k\in\mathbb{Z}}\eta_{AB}\alpha^A_k\alpha^B_{n-k}\,,\qquad \tilde L^{(-2)}_n = \frac{1}{2}\sum_{k\in\mathbb{Z}}\eta_{AB}\tilde\alpha^A_k\tilde\alpha^B_{n-k}\,.
\end{equation}
If we also expand the Lagrange multipliers (where $\vartheta_{(0)-}$ has modes $\vartheta^{(0)}_n$ and $\vartheta_{(0)+}$ has modes $\tilde\vartheta^{(0)}_n$), the Lagrangian (up to total derivatives) takes the form
\begin{equation}
    L_{\text{LO}} = \dot x_0^A\wp_{(-2)A} + \sum_{k=1}^\infty\frac{i}{k}\eta_{AB}( \dot\alpha^A_k\alpha^B_{-k} + \dot{\tilde{\alpha}}^A_k\tilde\alpha^B_{-k} ) - \sum_{n\in\mathbb{Z}}(\vartheta^{(0)}_{-n}L^{(-2)}_n + \tilde\vartheta^{(0)}_{-n}\tilde L_n^{(-2)})\,.
\end{equation} 
From this we may read off the Poisson brackets
\begin{equation}
    \{ x_0^A,\wp_{(-2)B}\} = \delta^A_B\,,\qquad \{ \alpha^A_k,\alpha^B_{-k} \} = \{ \tilde\alpha^A_k,\tilde\alpha^B_{-k} \} = -ik\eta^{AB}\,,
\end{equation}
which lead to
\begin{equation}
    \{ L_k^{(-2)},L_n^{(-2)} \} = -i(k-n)L_{k+n}^{(-2)}\qquad \text{and}\qquad \{ \tilde L_k^{(-2)},\tilde L_n^{(-2)} \} = -i(k-n)\tilde L_{k+n}^{(-2)}\,.
\end{equation}

To fix the gauge redundancy, we set 
\begin{equation}
\label{eq:LO-gauge-fixing}
    \alpha^+_k = \tilde\alpha^+_k = 0\qquad \forall k\neq 0\,.
\end{equation}
To check that this indeed fixes the gauge, we compute
\begin{equation}
    \{ L_n^{(-2)},\alpha_{-k}^+ \} = -ik\alpha^+_{n-k} = -ik\alpha^+_0\delta_{nk}\,,\qquad \{ \tilde L_n^{(-2)},\tilde\alpha_{-k}^+ \} = -ik\tilde\alpha^+_0\delta_{nk}\,,
\end{equation}
where we used the gauge fixing conditions. Since these are invertible matrices for $k\neq 0$, we conclude that this fixes all the gauge invariances except those corresponding to $L_0^{(-2)}$ and $\tilde L_0^{(-2)}$, which in gauge fixed form are given by
\begin{align}
\begin{split}
    L_0^{(-2)} &= \frac{1}{8\pi T_{\text{eff}}}\eta_{AB}\wp^A_{(-2)}\wp^B_{(-2)} + \frac{\pi T_{\text{eff}}}{2}w^2R_{\text{eff}}^2 - \frac{1}{2}\wp^v_{(-2)}wR_{\text{eff}}\,,\\
    \tilde L_0^{(-2)} &= \frac{1}{8\pi T_{\text{eff}}}\eta_{AB}\wp^A_{(-2)}\wp^B_{(-2)} + \frac{\pi T_{\text{eff}}}{2}w^2R_{\text{eff}}^2 + \frac{1}{2}\wp^v_{(-2)}wR_{\text{eff}}\,.
\end{split}
\end{align}
Setting these constraints equal to zero implies that $\wp^v_{(-2)} = 0$ and $\wp^t_{(-2)} = 2\pi T_{\text{eff}} wR_{\text{eff}}$, in agreement with what we found previously in~\eqref{eq:LOenergy}.

For $n\neq 0$, we can solve the conditions $0 =L_n^{(-2)} = -\frac{1}{2}\alpha^+_0\alpha_n^-$ and $0 = \tilde L_n^{(-2)} = -\frac{1}{2}\tilde\alpha^+_0\tilde\alpha_n^-$ for $\alpha^-_n$ and $\tilde\alpha^-_n$, which gives us $\alpha^-_n = \tilde\alpha^-_n = 0$. Hence, the gauge fixed LO Lagrangian is
\begin{equation}
    L_{\text{LO}} = \dot x_{0}^A\wp_{(-2)A}  - \vartheta^{(0)}_{0}L^{(-2)}_0 - \tilde\vartheta^{(0)}_{0}\tilde L_0^{(-2)}\,,
\end{equation} 
leaving us with the gauge-fixed Poisson brackets 
\begin{equation}
    \{x_0^A,\wp_{(-2)B}\} = \delta^A_B\,.
\end{equation}
In agreement with our previous findings, there are no oscillations in the LO theory.

\subsubsection*{NLO}
We now repeat the above analysis for the NLO theory. We read off the following nonzero NLO Poisson brackets from the NLO Lagrangian~\eqref{eq:NLOActionBetterForm},
\begin{align}
\begin{split}
    \{x^i(\sigma^1),P_{(0)j}(\tilde\sigma^1)\} &= \delta^i_j\delta(\sigma^1 - \tilde\sigma^1)\,,\\
    \{ x^A(\sigma^1),P_{(0)B}(\tilde\sigma^1) \} &= \{ y^A(\sigma^1),P_{(-2)B}(\tilde\sigma^1) \} = \delta^A_B\delta(\sigma^1 - \tilde\sigma^1)\,.
\end{split}
\end{align}
Relative to these brackets, we may compute the algebra of the four constraints $\mathcal{H}_{(-2)\pm}$ and $\mathcal{H}_{(0)\pm}$
\begin{align}
    \begin{split}
        \{ \mathcal{H}_{(-2)\pm}(\sigma^1),\mathcal{H}_{(-2)\pm}(\tilde\sigma^1)\} &= \{ \mathcal{H}_{(-2)+}(\sigma^1),\mathcal{H}_{(-2)-}(\tilde\sigma^1)\} = 0\,,\\
        \{ \mathcal{H}_{(0)\pm}(\sigma^1),\mathcal{H}_{(-2)\pm}(\tilde\sigma^1)\} &= \pm (\mathcal{H}_{(-2)\pm}(\sigma^1) + \mathcal{H}_{(-2)\pm}(\tilde\sigma^1))\delta'(\sigma^1-\tilde\sigma^1)\,,\\
        \{ \mathcal{H}_{(0)\pm}(\sigma^1),\mathcal{H}_{(-2)\mp}(\tilde\sigma^1)\} &= 0\,,\\
        \{ \mathcal{H}_{(0)\pm}(\sigma^1),\mathcal{H}_{(0)\pm}(\tilde\sigma^1)\} &= \pm (\mathcal{H}_{(-2)\pm}(\sigma^1) + \mathcal{H}_{(-2)\pm}(\tilde\sigma^1))\delta'(\sigma^1-\tilde\sigma^1)\,,\\
        \{ \mathcal{H}_{(0)+}(\sigma^1),\mathcal{H}_{(0)-}(\tilde\sigma^1)\} &= 0\,.
    \end{split}
\end{align}
At NLO, the gauge transformations act separately on the combinations $P_{(0)i}\pm T_{\text{eff}}x'^i$ and $P_{(0)}^A\pm T_{\text{eff}}y'^A$, which we take to have the mode expansions
\begin{align}
    \begin{split}
        P_{(0)i} - T_{\text{eff}}x'^i &= \sqrt{\frac{T_{\text{eff}}}{\pi}} \sum_{k\in\mathbb{Z}}e^{ik\sigma^1}\alpha_k^i(\sigma^0)\,,\\
        P_{(0)i} + T_{\text{eff}}x'^i &= \sqrt{\frac{T_{\text{eff}}}{\pi}} \sum_{k\in\mathbb{Z}}e^{-ik\sigma^1}\tilde\alpha_k^i(\sigma^0)\,,\\
        P_{(0)}^A - T_{\text{eff}}y'^A &= \sqrt{\frac{T_{\text{eff}}}{\pi}} \sum_{k\in\mathbb{Z}}e^{ik\sigma^1}\beta_k^A(\sigma^0)\,,\\
        P_{(0)}^A + T_{\text{eff}}y'^A &= \sqrt{\frac{T_{\text{eff}}}{\pi}} \sum_{k\in\mathbb{Z}}e^{-ik\sigma^1}\tilde\beta_k^A(\sigma^0)\,.
    \end{split}
\end{align}
For the momenta and embedding fields, this leads to
\begin{align}
    \begin{split}
        p^A_{(0)}(\sigma^0,\sigma^1) &= \frac{\wp^A_{(0)}(\sigma^0)}{2\pi} + \sqrt{\frac{T_{\text{eff}}}{4\pi}}\sum_{k\neq 0}e^{ik\sigma^1}\left( \beta_k^A(\sigma^0) + \tilde\beta_{-k}^A(\sigma^0) \right)\,,\\
        p^i_{(0)}(\sigma^0,\sigma^1) &= \frac{\wp^i_{(0)}(\sigma^0)}{2\pi} + \sqrt{\frac{T_{\text{eff}}}{4\pi}}\sum_{k\neq 0}e^{ik\sigma^1}\left( \alpha_k^i(\sigma^0) + \tilde\alpha_{-k}^i(\sigma^0) \right)\,,\\
        x^i(\sigma^0,\sigma^1) &= x_0^i(\sigma^0) + \frac{1}{\sqrt{4\pi T_{\text{eff}}}}\sum_{k\neq 0}\frac{i}{k}e^{ik\sigma^1}\left( \alpha_k^i(\sigma^0) - \tilde\alpha_{-k}^i(\sigma^0) \right)\,,\\
        y^A(\sigma^0,\sigma^1) &= y_0^v(\sigma^0)  + \frac{1}{\sqrt{4\pi T_{\text{eff}}}}\sum_{k\neq 0}\frac{i}{k}e^{ik\sigma^1}\left( \beta_k^A(\sigma^0) - \tilde\beta_{-k}^A(\sigma^0) \right)\,.
    \end{split}
\end{align}
If we then Fourier expand the constraints $\mathcal{H}_{(0)\pm}$, we now get
\begin{equation}
    \mathcal{H}_{(0) -} = \frac{1}{2\pi} \sum_{n\in\mathbb{Z}}e^{in\sigma^1}L^{(0)}_n\qquad\text{and}\qquad \mathcal{H}_{(0) +} = \frac{1}{2\pi} \sum_{n\in\mathbb{Z}}e^{-in\sigma^1}\tilde L^{(0)}_n\,,
\end{equation}
where
\begin{align}
\begin{split}
    L^{(0)}_n = \sum_{k\in\mathbb{Z}}\left(\frac{1}{2}\alpha^i_k\alpha^i_{n-k} + \eta_{AB}\alpha^A_k\beta^B_{n-k}\right)\,,\\
    \tilde L^{(0)}_n = \sum_{k\in\mathbb{Z}}\left(\frac{1}{2}\tilde\alpha^i_k\tilde\alpha^i_{n-k} + \eta_{AB}\tilde\alpha^A_k\tilde\beta^B_{n-k} \right)\,.
\end{split}
\end{align}
Using the same procedure as above, we now find that the NLO Lagrangian can be written as (up to total derivatives)
\begin{align}
    \begin{split}
        L_{\text{NLO}} &= \dot x_0^i\wp_{(0)i} + \dot x_0^A\wp_{(0)A} + \dot y_0^A\wp_{(-2)A}\\
        &\quad+ \sum_{k=1}^\infty\frac{i}{k}( \dot\alpha_k^i\alpha^i_{-k} + \dot{\tilde\alpha}^i_k\tilde\alpha^i_{-k} + 2\eta_{AB}(\dot\alpha^A_k\beta^B_{-k} + \dot{\tilde{\alpha}}^A_k\tilde\beta^B_{-k} ) )\\
        &\quad- \sum_{n\in\mathbb{Z}}\left(\vartheta^{(0)}_{-n}L^{(0)}_n + \tilde\vartheta^{(0)}_{-n}\tilde L_n^{(0)}+\vartheta^{(2)}_{-n}L^{(-2)}_n + \tilde\vartheta^{(2)}_{-n}\tilde L_n^{(-2)}\right)\,,
    \end{split}
\end{align}
which allows us to read off the Poisson brackets for the modes
\begin{align}
    \begin{aligned}
            \{ x_0^i,\wp_{(0)j}\} &= \delta^i_j\,,& \{ \alpha^i_k,\alpha^j_{-k} \} &= \{ \tilde\alpha^i_k,\tilde\alpha^j_{-k} \} = -ik\delta^{ij}\,,\\
            \{ x_0^A,\wp_{(0)B}\} &= \delta^A_B\,,& \{ \alpha^A_k,\beta^B_{-k} \} &= \{ \tilde\alpha^A_k,\tilde\beta^B_{-k} \} = -\frac{ik}{2}\eta^{AB}\,,\\
             \{ y_0^A,\wp_{(-2)B}\} &= \delta^A_B\,.&
    \end{aligned}
\end{align}

We now come to the gauge fixing. In the Polyakov formaluation of section~\ref{sec:ModeExp}, we found that we could gauge fix the residual gauge invariance by removing the oscillations  $\beta_k^-$ and $\tilde\beta^+_k$
for all nonzero $k$. Hence, to find the gauge-fixed Poisson brackets for the remaining modes, we also set these to zero in the phase space formulation. In order to do so, it is convenient to gauge fix the LO modes in a different way compared to what we did at LO in~\eqref{eq:LO-gauge-fixing}, which means that the full set of gauge-fixing conditions at NLO are\footnote{At LO, we fixed the gauge by setting $\alpha^+_k = \tilde\alpha^+_k = 0$ for all $k\neq 0$. Alternatively, we could have fixed the gauge in the same way as we do at NLO, namely by setting $\alpha^-_k = \tilde\alpha^+_k = 0$ for all $k\neq 0$.}
\begin{align}
    \begin{split}
        \alpha^-_k &= \tilde\alpha^+_k = 0\qquad \forall k\neq 0\,,\\
        \beta^-_k &= \tilde\beta^+_k = 0\qquad \forall k\neq 0\,.
    \end{split}
\end{align}
We may check that this really does fix the gauge 
\begin{align}
\label{eq:gauge-fixing-101}
\begin{aligned}
    \{L^{(-2)}_n,\beta^-_{-k}\} &= -\frac{ik}{2}\alpha^-_{0}\delta_{nk}\,,& \{\tilde L^{(-2)}_n,\tilde\beta^+_{-k}\} &= -\frac{ik}{2}\tilde\alpha^+_0\delta_{nk}\,,\\
    \{L^{(0)}_n,\alpha^-_{-k}\} &= -\frac{ik}{2}\alpha^-_0\delta_{nk}\,,& \{\tilde L^{(0)}_n,\tilde\alpha^+_{-k}\} &= -\frac{ik}{2}\tilde\alpha^+_0\delta_{nk}\,,
\end{aligned}
\end{align}
where we used the gauge-fixing conditions on the right-hand side. Since these are invertible for $k\neq 0$, we have fixed all the gauge redundancy except for when $k=0$. Furthermore, from the results in~\eqref{eq:gauge-fixing-101}, we see that setting $\alpha^-_k = \tilde\alpha^+_k = 0$ for $k\neq 0$ will now fix the NLO gauge redundancy generated by $\mathcal{H}_{(0)\pm}$, while the LO gauge redundancy is now fixed by setting $\beta^-_k = \tilde\beta^+_k = 0$ for $k\neq 0$. The gauge-fixed zero-mode constraints are 
\begin{align}
    \begin{split}
            L_0^{(-2)} &= \frac{1}{8\pi T_{\text{eff}}}\eta_{AB}\wp^A_{(-2)}\wp^B_{(-2)} + \frac{\pi T_{\text{eff}}}{2}w^2R_{\text{eff}}^2 - \frac{1}{2}\wp^v_{(-2)}wR_{\text{eff}}\,,\\
    \tilde L_0^{(-2)} &= \frac{1}{8\pi T_{\text{eff}}}\eta_{AB}\wp^A_{(-2)}\wp^B_{(-2)} + \frac{\pi T_{\text{eff}}}{2}w^2R_{\text{eff}}^2 + \frac{1}{2}\wp^v_{(-2)}wR_{\text{eff}}\,,\\
        L_0^{(0)} &= N_{(0)} + \frac{1}{2}\alpha^i_{0}\alpha^i_0 + \eta_{AB}\alpha_0^A\beta_0^B\,,\\
    \tilde L_0^{(0)} &= \tilde N_{(0)} + \frac{1}{2}\tilde\alpha^i_{0}\tilde\alpha^i_0 + \eta_{AB}\tilde\alpha_0^A\tilde\beta_0^B\,.
    \end{split}
\end{align}
We may now solve the constraints for the remaining modes. At LO, we find again that all oscillations vanish (i.e., $\alpha^+_k = \tilde\alpha^-_k = 0$ for all $k\neq 0$), while at NLO we get 
\begin{equation}
\label{eq:expressions-for-beta}
    \beta^+_n = \frac{1}{\alpha^-_0}\sum_{k\in \mathbb{Z}}\alpha^i_k\alpha^i_{n-k}\,,\qquad \tilde\beta^-_n = \frac{1}{\tilde\alpha^+_0}\sum_{k\in \mathbb{Z}}\tilde\alpha^i_k\tilde\alpha^i_{n-k}\,.
\end{equation}
This means that the gauge-fixed NLO Lagrangian takes the form
\begin{align}
    \begin{split}
        L_{\text{NLO}} &= \dot x_0^i\wp_{(0)i} + \dot x_0^A\wp_{(0)A} + \dot y_0^A\wp_{(-2)A}+ \sum_{k=1}^\infty\frac{i}{k}( \dot\alpha_k^i\alpha^i_{-k} + \dot{\tilde\alpha}^i_k\tilde\alpha^i_{-k} )\\
        &\quad- \left(\vartheta^{(0)}_{0}L^{(0)}_0 + \tilde\vartheta^{(0)}_{0}\tilde L_0^{(0)}+\vartheta^{(2)}_{0}L^{(-2)}_0 + \tilde\vartheta^{(2)}_{0}\tilde L_0^{(-2)}\right)\,.
    \end{split}
\end{align}
From this, we read off the Poisson brackets of the gauge-fixed theory
\begin{align}
    \begin{split}
            \{ x_0^i,\wp_{(0)j}\} &= \delta^i_j\,,\qquad \{ x_0^A,\wp_{(0)B}\} = \delta^A_B\,,\qquad\{ y_0^A,\wp_{(-2)B}\} = \delta^A_B\,,\\
            \{ \alpha^i_k,\alpha^j_{-k} \} &= \{ \tilde\alpha^i_k,\tilde\alpha^j_{-k} \} = -ik\delta^{ij}\,.
    \end{split}
\end{align}
We can work out the remaining brackets by using the relations~\eqref{eq:expressions-for-beta}.

\subsubsection*{NNLO}
Finally, we consider the Dirac procedure at NNLO, where the Lagrangian is given by~\eqref{eq:NNLO-phase-space-lagrangian}. From this, we get the following Poisson brackets
\begin{align}
    \begin{split}
            \{x^i(\sigma^1),P_{(2)j}(\tilde\sigma^1)\} &= \{y^i(\sigma^1),P_{(0)j}(\tilde\sigma^1)\} = \delta^i_j\delta(\sigma^1 - \tilde\sigma^1)\,,\\
            \{ x^A(\sigma^1),P_{(2)B}(\tilde\sigma^1) \} &= \{ y^A(\sigma^1),P_{(0)B}(\tilde\sigma^1) \} = \{ z^A(\sigma^1),P_{(-2)B}(\tilde\sigma^1) \}= \delta^A_B\delta(\sigma^1 - \tilde\sigma^1)\,.
    \end{split}
\end{align}
The nonzero constraint brackets are
\begin{align}
    \begin{split}
        \{ \mathcal{H}_{(0)\pm}(\sigma^1),\mathcal{H}_{(2)\pm}(\tilde\sigma^1)\} &= \pm (\mathcal{H}_{(0)\pm}(\sigma^1) + \mathcal{H}_{(0)\pm}(\tilde\sigma^1))\delta'(\sigma^1-\tilde\sigma^1)\,,\\
        \{ \mathcal{H}_{(2)\pm}(\sigma^1),\mathcal{H}_{(2)\pm}(\tilde\sigma^1)\} &= \pm (\mathcal{H}_{(2)\pm}(\sigma^1) + \mathcal{H}_{(2)\pm}(\tilde\sigma^1))\delta'(\sigma^1-\tilde\sigma^1)\,,
    \end{split}
\end{align}
and so the constraints remain first-class. At NNLO, we write down mode expansions for the combinations $P_{(2)i}\pm T_{\text{eff}}y'^i$ and $P_{(2)}^A\pm T_{\text{eff}}z'^A$:
\begin{align}
    \begin{split}
        P_{(2)i} - T_{\text{eff}}y'^i &= \sqrt{\frac{T_{\text{eff}}}{\pi}} \sum_{k\in\mathbb{Z}}e^{ik\sigma^1}\beta_k^i(\sigma^0)\,,\\
        P_{(2)i} + T_{\text{eff}}y'^i &= \sqrt{\frac{T_{\text{eff}}}{\pi}} \sum_{k\in\mathbb{Z}}e^{-ik\sigma^1}\tilde\beta_k^i(\sigma^0)\,,\\
        P_{(2)}^A - T_{\text{eff}}z'^A &= \sqrt{\frac{T_{\text{eff}}}{\pi}} \sum_{k\in\mathbb{Z}}e^{ik\sigma^1}\chi_k^A(\sigma^0)\,,\\
        P_{(2)}^A + T_{\text{eff}}z'^A &= \sqrt{\frac{T_{\text{eff}}}{\pi}} \sum_{k\in\mathbb{Z}}e^{-ik\sigma^1}\tilde\chi_k^A(\sigma^0)\,.
    \end{split}
\end{align}
This leads to the following mode expansions for the momenta and the embedding fields
\begin{align}
    \begin{split}
        p^A_{(2)}(\sigma^0,\sigma^1) &= \frac{\wp^A_{(2)}(\sigma^0)}{2\pi} + \sqrt{\frac{T_{\text{eff}}}{4\pi}}\sum_{k\neq 0}e^{ik\sigma^1}\left( \chi_k^A(\sigma^0) + \tilde\chi_{-k}^A(\sigma^0) \right)\,,\\
        p^i_{(2)}(\sigma^0,\sigma^1) &= \frac{\wp^i_{(2)}(\sigma^0)}{2\pi} + \sqrt{\frac{T_{\text{eff}}}{4\pi}}\sum_{k\neq 0}e^{ik\sigma^1}\left( \beta_k^i(\sigma^0) + \tilde\beta_{-k}^i(\sigma^0) \right)\,,\\
        y^i(\sigma^0,\sigma^1) &= y_0^i(\sigma^0) + \frac{1}{\sqrt{4\pi T_{\text{eff}}}}\sum_{k\neq 0}\frac{i}{k}e^{ik\sigma^1}\left( \beta_k^i(\sigma^0) - \tilde\beta_{-k}^i(\sigma^0) \right)\,,\\
        z^A(\sigma^0,\sigma^1) &= z_0^A(\sigma^0)  + \frac{1}{\sqrt{4\pi T_{\text{eff}}}}\sum_{k\neq 0}\frac{i}{k}e^{ik\sigma^1}\left( \chi_k^A(\sigma^0) - \tilde\chi_{-k}^A(\sigma^0) \right)\,.
    \end{split}
\end{align}
If we then Fourier expand the constraints $\mathcal{H}_{(0)\pm}$, we now get
\begin{equation}
    \mathcal{H}_{(2)-} = \frac{1}{2\pi} \sum_{n\in\mathbb{Z}}e^{in\sigma^1}L^{(2)}_n\qquad\text{and}\qquad \mathcal{H}_{(2)+} = \frac{1}{2\pi} \sum_{n\in\mathbb{Z}}e^{-in\sigma^1}\tilde L^{(2)}_n\,,
\end{equation}
where
\begin{align}
\begin{split}
    L^{(2)}_n = \sum_{k\in\mathbb{Z}}\left(\alpha^i_k\beta^i_{n-k} + \frac{1}{2}\eta_{AB}(2\alpha^A_k\chi^B_{n-k} + \beta^A_k\beta^B_{n-k})\right)\,,\\
    \tilde L^{(2)}_n = \sum_{k\in\mathbb{Z}}\left(\tilde\alpha^i_k\tilde\beta^i_{n-k} + \frac{1}{2}\eta_{AB}(2\tilde\alpha^A_k\tilde\chi^B_{n-k} + \tilde\beta^A_k\tilde\beta^B_{n-k})\right)\,.
\end{split}
\end{align}
Using the same procedure as above, we now find that the NNLO Lagrangian can be written as (up to total derivatives)
\begin{align}
    \begin{split}
        L_{\text{NNLO}} &= \dot x_0^i\wp_{(2)i} + \dot y_0^i\wp_{(0)i} + \dot x_0^A\wp_{(2)A}+\dot y_0^A\wp_{(0)A} + \dot z_0^A\wp_{(-2)A}\\
        &\quad+ \sum_{k=1}^\infty\frac{i}{k}\Big( 2\dot\alpha_k^i\beta^i_{-k} + 2\dot{\tilde\alpha}^i_k\tilde\beta^i_{-k} 
        + \eta_{AB}(2\dot\alpha^A_k\chi^B_{-k} + 2\dot{\tilde{\alpha}}^A_k\tilde\chi^B_{-k} + \dot\beta^A_k\beta^B_{-k} + \dot{\tilde{\beta}}^A_k\tilde\beta^B_{-k} \Big)\\
        &\quad- \sum_{n\in\mathbb{Z}}\left(\vartheta^{(0)}_{-n}L^{(2)}_n + \tilde\vartheta^{(0)}_{-n}\tilde L_n^{(2)}+\vartheta^{(2)}_{-n}L^{(0)}_n + \tilde\vartheta^{(2)}_{-n}\tilde L_n^{(0)}+\vartheta^{(4)}_{-n}L^{(-2)}_n + \tilde\vartheta^{(4)}_{-n}\tilde L_n^{(-2)}\right)\,.
    \end{split}
\end{align}
This gives the following Poisson brackets for the modes
\begin{align}
    \begin{split}
            \{ x_0^i,\wp_{(2)j}\} &= \{ y_0^i,\wp_{(0)j}\} = \delta^i_j\,,\\
            \{ \alpha^i_k,\beta^j_{-k} \} &= \{ \tilde\alpha^i_k,\tilde\beta^j_{-k} \} = -\frac{ik}{2}\delta^{ij}\,,\\
            \{ x_0^A,\wp_{(2)B}\} &= \{ y_0^A,\wp_{(0)B}\} = \{ z_0^A,\wp_{(-2)B}\} = \delta^A_B\,,\\
            \{ \alpha^A_k,\chi^B_{-k} \} &= \{ \tilde\alpha^A_k,\tilde\chi^B_{-k} \} =-\frac{ik}{2}\eta^{AB}\,,\\
            \{\beta^A_k,\beta^B_{-k}\} &= \{\tilde\beta^A_k,\tilde\beta^B_{-k}\} =-ik\eta^{AB}\,.
    \end{split}
\end{align}
The gauge fixing follows the same pattern as what we worked out above
\begin{align}
    \begin{split}
        \alpha^-_k &= \tilde\alpha^+_k = 0\qquad \forall k\neq 0\,,\\
        \beta^-_k &= \tilde\beta^+_k = 0\qquad \forall k\neq 0\,,\\
        \chi^-_k &= \tilde\chi^+_k = 0\qquad \forall k\neq 0\,.
    \end{split}
\end{align}
Once more, we may check that this really does fix the gauge
\begin{align}
\begin{aligned}
    \{L^{(-2)}_n,\chi^-_{-k}\} &= -\frac{ik}{2}\alpha^-_{0}\delta_{nk}\,,& \{\tilde L^{(-2)}_n,\tilde\chi^+_{-k}\} &= -\frac{ik}{2}\tilde\alpha^+_0\delta_{nk}\,,\\
     \{L^{(0)}_n,\beta^-_{-k}\} &= -ik\alpha^-_0\delta_{nk}\,,& \{\tilde L^{(0)}_n,\tilde\beta^+_{-k}\} &= -ik\tilde\alpha^+_0\delta_{nk}\,,\\
     \{L^{(2)}_n,\alpha^-_{-k}\} &= -\frac{ik}{2}\alpha^-_0\delta_{nk}\,,& \{\tilde L^{(0)}_n,\tilde\alpha^+_{-k}\} &= -\frac{ik}{2}\tilde\alpha^+_0\delta_{nk}\,,
\end{aligned}
\end{align}
which are indeed invertible, which means that we have fixed all the gauge invariance except for the case corresponding to $k=0$. Furthermore, from the results above, we see that setting $\alpha^-_k = \tilde\alpha^+_k = 0$ for all $k\neq 0$ will now fix the NLO gauge redundancy generated by $\mathcal{H}_{(0)\pm}$, while the LO gauge redundancy is now fixed by setting $\beta^-_k = \tilde\beta^+_k = 0$ for all $k\neq 0$. The gauge-fixed zero-mode constraints are 
\begin{align}
    \begin{split}
            L_0^{(-2)} &= \frac{1}{8\pi T_{\text{eff}}}\eta_{AB}\wp^A_{(-2)}\wp^B_{(-2)} + \frac{\pi T_{\text{eff}}}{2}w^2R_{\text{eff}}^2 - \frac{1}{2}\wp^v_{(-2)}wR_{\text{eff}}\,,\\
    \tilde L_0^{(-2)} &= \frac{1}{8\pi T_{\text{eff}}}\eta_{AB}\wp^A_{(-2)}\wp^B_{(-2)} + \frac{\pi T_{\text{eff}}}{2}w^2R_{\text{eff}}^2 + \frac{1}{2}\wp^v_{(-2)}wR_{\text{eff}}\,,\\
        L_0^{(0)} &= N_{(0)} + \frac{1}{2}\alpha^i_{0}\alpha^i_0+ \eta_{AB}\alpha_0^A\beta_0^B\,,\\
    \tilde L_0^{(0)} &= \tilde N_{(0)} + \frac{1}{2}\tilde\alpha^i_{0}\tilde\alpha^i_0+ \eta_{AB}\tilde\alpha_0^A\tilde\beta_0^B\,,\\
    L_0^{(2)} &= N_{(2)} + \alpha^i_0\beta^i_0 + \frac{1}{2}\eta_{AB}\left( 2\alpha^A_0\chi^B_0 + \beta^A_0\beta^B_0  \right)\,,\\
    \tilde L_0^{(2)} &= \tilde N_{(2)} + \tilde \alpha^i_0\tilde \beta^i_0 + \frac{1}{2}\eta_{AB}\left( 2\tilde \alpha^A_0\tilde \chi^B_0 + \tilde \beta^A_0\tilde \beta^B_0  \right)\,.
    \end{split}
\end{align}
As before, we may now solve the constraints for the remaining modes. At LO, we find again that all oscillations vanish (i.e., $\alpha^+_k = \tilde\alpha^-_k = 0$ for all $k\neq 0$), while at NLO and NNLO, we get (for $n\neq 0$)
\begin{align}
\label{eq:expressions-for-beta-and-chi}
\begin{split}
    \beta^+_n &= \frac{1}{\alpha^-_0}\sum_{k\in \mathbb{Z}}\alpha^i_k\alpha^i_{n-k}\,,\qquad \tilde\beta^-_n = \frac{1}{\tilde\alpha^+_0}\sum_{k\in \mathbb{Z}}\tilde\alpha^i_k\tilde\alpha^i_{n-k}\,,\\
    \chi^+_n &= \frac{1}{\alpha^-_0}\sum_{k\in \mathbb{Z}}\left[2\alpha^i_k\beta^i_{n-k} - \frac{\beta^-_0}{\alpha_0^-}\alpha_k^i\alpha^i_{n-k}\right]\,,\\ 
    \tilde\chi^-_n &= \frac{1}{\tilde\alpha^+_0}\sum_{k\in \mathbb{Z}}\left[2\tilde\alpha^i_k\tilde\beta^i_{n-k} - \frac{\tilde\beta^+_0}{\tilde\alpha_0^+}\tilde\alpha_k^i\tilde\alpha^i_{n-k}\right] \,.
\end{split}
\end{align}
This means that the gauge-fixed NNLO Lagrangian takes the form
\begin{align}
    \begin{split}
        L_{\text{NNLO}} &= \dot x_0^i\wp_{(2)i} + \dot y_0^i\wp_{(0)i} + \dot x_0^A\wp_{(2)A}+\dot y_0^A\wp_{(0)A} + \dot z_0^A\wp_{(-2)A}\\
        &\quad+ \sum_{k=1}^\infty\frac{2i}{k}\Big( \dot\alpha_k^i\beta^i_{-k} + \dot{\tilde\alpha}^i_k\tilde\beta^i_{-k} \Big)\\
        &\quad- \sum_{n\in\mathbb{Z}}\left(\vartheta^{(0)}_{-n}L^{(2)}_n + \tilde\vartheta^{(0)}_{-n}\tilde L_n^{(2)}+\vartheta^{(2)}_{-n}L^{(0)}_n + \tilde\vartheta^{(2)}_{-n}\tilde L_n^{(0)}+\vartheta^{(4)}_{-n}L^{(-2)}_n + \tilde\vartheta^{(4)}_{-n}\tilde L_n^{(-2)}\right)\,.
    \end{split}
\end{align}
From this, we read off the Poisson brackets of the gauge-fixed theory
\begin{align}
    \begin{split}
            \{ x_0^i,\wp_{(2)j}\} &= \{ y_0^i,\wp_{(0)j}\} = \delta^i_j\,,\\
            \{ x_0^A,\wp_{(2)B}\} &= \{ y_0^A,\wp_{(0)B}\} = \{ z_0^A,\wp_{(-2)B}\} = \delta^A_B\,,\\
            \{ \alpha^i_k,\beta^j_{-k} \} &= \{ \tilde\alpha^i_k,\tilde\beta^j_{-k} \} = -\frac{ik}{2}\delta^{ij}\,.
    \end{split}
\end{align}
We can work out the remaining brackets by using the relations~\eqref{eq:expressions-for-beta-and-chi}.

\subsection{Commutators and the normal ordering constant}
\label{sec:quantisation}
In passing from the classical theory to the quantum theory, canonical quantisation tells us to replace Poisson brackets with commutators according to the rule
\begin{equation}
\label{eq:folklore}
    [\cdot\,,\cdot] = i\hbar \{\cdot\,,\cdot\}\,.
\end{equation}
This is based on the canonically conjugate variables having dimensions length and momentum, respectively. Explicitly, the commutator $[q,p]$ has dimensions of $\text{mass}\times\frac{\text{length}^2}{\text{time}} = [\hbar]$, while $\{q,p\} = 1$ is dimensionless (due to the definition of the Poisson bracket), and hence we must compensate with a factor of $\hbar$, leading to \eqref{eq:folklore}. 

When we expand Lagrangians in $1/c^2$ we define subleading Lagrangians as 
\begin{equation}
    \mathcal{L}=c^2\mathcal{L}_{\text{LO}}+\mathcal{L}_{\text{NLO}}+c^{-2}\mathcal{L}_{\text{NNLO}}+\mathcal{O}(c^{-4})\,.
\end{equation}
This means that for example $\mathcal{L}_{\text{NNLO}}$ does not have the dimensions of an energy density (simply because we factored out a factor of $c^{-2}$). This then means that the canonical momentum likewise does not have the usual dimensions. For this reason we replace the rule \eqref{eq:folklore} with
\begin{equation}
\label{eq:generalised-quantisation}
    [a,b] = ik_{[a][b]}\{a,b\}\,,
\end{equation}
where $k_{[a][b]}$ is some combination of fundamental constants with dimensions of $[a][b]$.
\subsubsection*{NLO}
Since there are no oscillations at LO, we jump straight to NLO, where the oscillations are $\alpha^i_k$ and $\tilde\alpha_k^i$ for $k\neq 0$. As usual, modes with $k>0$ are interpreted as annihilation operators (and vice versa), which annihilate the NLO vacuum state $\ket{0}_{\text{NLO}}$ 
\begin{equation}
    \alpha^i_k\ket{0}_{\text{NLO}} = \tilde\alpha^i_k\ket{0}_{\text{NLO}} =0 \qquad \forall k>0\,.
\end{equation}
Since 
\begin{equation}
    [\alpha^i_k] = [\tilde\alpha^i_k] = \text{length}\times\sqrt{\text{mass}/ \text{time} }\,,
\end{equation}
the relation \eqref{eq:generalised-quantisation} tells us that the modes satisfy the commutation relations
\begin{equation}
    [\alpha^i_k,\alpha_{-k}^j] = \hbar k\delta^{ij}\,,\qquad [\tilde\alpha^i_k,\tilde\alpha_{-k}^j] = \hbar k\delta^{ij}\,.
\end{equation}
In terms of these modes, the number operators are given by 
\begin{equation}
    N_{(0)} = \frac{1}{2}\sum_{k\neq 0}\alpha^i_{-k}\alpha_k^i\,,\qquad \tilde N_{(0)} = \frac{1}{2}\sum_{k\neq 0}\tilde\alpha^i_{-k}\tilde\alpha_k^i\,.
\end{equation}
Adopting normal ordering, where creation operators are moved to the left, we find that 
\begin{equation}
    \frac{1}{2}\sum_{k\neq 0}\alpha^i_{-k}\alpha_k^i = \sum_{k=1}^\infty \alpha_{-k}^i\alpha_k^i + \frac{1}{2}\sum_{k=1}^\infty [\alpha_k^i,\alpha_{-k}^i] = \sum_{k=1}^\infty \alpha_{-k}^i\alpha_k^i +  \frac{\hbar d}{2}\sum_{k=1}^\infty k = \sum_{k=1}^\infty \alpha_{-k}^i\alpha_k^i - \frac{\hbar d}{24} \,,
\end{equation}
where we used that $\sum_{k=1}^\infty k = -\frac{1}{12}$, and where $d = D -2 $ is the dimension of the transverse space. 

\subsubsection*{NNLO}
The NNLO vacuum $\ket{0}_{\text{NNLO}}$ is annihilated by both $\alpha$ and $\beta$ annihilation operators,
\begin{equation}
    \alpha^i_k\ket{0}_{\text{NNLO}} = \tilde\alpha^i_k\ket{0}_{\text{NNLO}} = \beta^i_k\ket{0}_{\text{NNLO}} = \tilde\beta^i_k\ket{0}_{\text{NNLO}} =0 \qquad \forall k>0\,.
\end{equation}
The subleading oscillator modes have dimensions of
\begin{equation}
    [\beta] = \sqrt{\text{mass}}\times\text{length}\times (\text{time})^{-5/2}\,,
\end{equation}
and hence \eqref{eq:generalised-quantisation} gives us
\begin{equation}
    [\alpha^i_k,\beta_{-k}^j] = c^2\frac{\hbar k}{2}\delta^{ij}\,,\qquad [\tilde\alpha^i_k,\tilde\alpha_{-k}^j] = c^2\frac{\hbar k}{2}\delta^{ij}\,.
\end{equation}
The LO number operators no longer suffer from an ordering ambiguity since the $\alpha$'s commute. Now, however, the operators making up the subleading number operators do not commute. The subleading number operators are given by
\begin{align}
    \begin{split}
        N_{(2)} &=  \sum_{k=1}^\infty\alpha^i_{-k}\beta_k^i + \sum_{k=1}^\infty\alpha^i_{k}\beta_{-k}^i \,,\\
\tilde N_{(2)} &= \sum_{k=1}^\infty\tilde \alpha^i_{-k}\tilde \beta_k^i + \sum_{k=1}^\infty\tilde \alpha^i_{k}\tilde \beta_{-k}^i \,,
    \end{split}
\end{align}
where the second term on each line is not normal ordered. Performing this normal ordering is entirely analogous to what we did above, for example $N_{(2)}$ can be normal ordered as follows
\begin{equation}
    \sum_{k=1}^\infty\alpha^i_{k}\beta_{-k}^i = \sum_{k=1}^\infty \beta_{-k}^i\alpha_k^i +  c^2\frac{\hbar d}{2}\sum_{k=1}^\infty k = \sum_{k=1}^\infty \beta_{-k}^i\alpha_k^i - c^2\frac{\hbar d}{24} \,.
\end{equation}
Since $N=N_{(0)}+c^{-2}N_{(2)}+\mathcal{O}(c^{-4})$ we recover the same normal ordering constant as at NLO, and in fact as in the relativistic theory. This pattern should persist to all orders in $1/c^2$.

\section{Discussion}\label{sec:discussion}
In this paper, we developed the string $1/c^2$ expansion of closed relativistic bosonic strings up to NNLO. The target space geometry of these strings is obtained from a string $1/c^2$ expansion of a Lorentzian geometry. At NLO this leads to string Newton--Cartan geometry or generalisations thereof. The string theories we obtain arise as $1/c^2$ expansions of relativistic strings in either the Nambu--Goto, Polyakov or phase space formulation. At NLO, when the target space foliation is such that $\alpha^A{_A} = 0$ in \eqref{eq:Frobenius}, the theory takes the form of the Gomis--Ooguri string. We have computed the spectrum of the string by computing the energy at every order and shown that it agrees with the $1/c^2$ expansion of the relativistic energy of a string in a flat target spacetime. In order to perform this expansion we had to assume that the target space has a circle and that string has a nonzero (positive) winding along that circle. In the phase space formulation, we were able to perform the Dirac procedure and, with the Dirac brackets in hand, quantise the theory order by order. This, in particular, led to the same normal ordering constant as in the closed bosonic relativistic string, but the way this constant appears is different at each order. Let us now conclude with a list of open questions inspired by the results of this work.

The beta functions of NR string theory were considered in \cite{Gomis:2019zyu,Gallegos:2019icg,Yan:2019xsf}, and an action has been proposed that recovers all but one of the beta functions in \cite{Bergshoeff:2021bmc}, namely the stringy counterpart of the Poisson equation. We expect that the string $1/c^2$ expansion of NS-NS gravity would lead to an action principle for all the beta functions of NR string theory including the Poisson equation. This expectation is based on an analogous situation observed in the context of gravity: the action for NR gravity obtained using the particle $1/c^2$ expansion of GR in \cite{Hansen:2018ofj,Hansen:2020pqs} is precisely able to reproduce the Poisson equation, while previous approaches based on strict limits were unable to do so.

Another natural generalisation would be to consider the inclusion of odd powers of $1/c$ in the expansion, thereby turning the string $1/c^2$ expansion into a string $1/c$ expansion. For gravity, the analogous generalisation for the particle expansions was considered in \cite{VandenBleeken:2019gqa}.

In this work, we have only considered closed strings, but open strings and branes also have a role to play in NR string theory \cite{Gomis:2020izd,Gomis:2020fui,Blair:2021ycc}. The $1/c^2$ expansion of open strings and D-branes would be a natural next step (see \cite{upcoming2}). In a related direction, it would be interesting to explore the $p$-brane $1/c^2$ expansion of Lorentzian geometry. This would presumably lead to a ``type II $p$-brane Newton--Cartan geometry'', which would generalise the geometry developed in~\cite{Pereniguez:2019eoq} in much the same way that both type II string and particle Newton--Cartan geometries generalise their ``type I'' counterparts.

From the point of view of the NS-NS sector of relativistic string theory, it is natural to include the $B$-field in the geometry on equal footing with the metric. Including the symmetries of the $B$-field in the spacetime symmetry algebra, one gets what was called the \textit{string Poincar\'e algebra} in~\cite{Bidussi:2021ujm}. As the authors of that paper show, one can take a stringy nonrelativistic limit of this to obtain what they call the \textit{F-string Galilei algebra}, which, when gauged, leads to so-called TSNC geometry. It would be interesting to apply the $1/c^2$ expansion to both the algebra and the accompanying geometry rather than taking a limit.

More generally, it has become increasingly clear that NR strings are part of a landscape of non-Lorentzian string theories. The existence of such a landscape of non-Lorentzian string theories has for example been demonstrated within double field theory, where many non-Lorentzian geometries were found within double geometry \cite{Ko:2015rha,Morand:2017fnv}. Based on these observations, it would be particularly interesting to apply the $1/c^2$ expansion in the context of double field theory, which might give another perspective on the emergence of non-Lorentzian geometries.

\section*{Acknowledgements} 

We thank José Figueroa-O'Farrill for useful discussions. The work of JH is supported by the Royal Society University Research Fellowship ``Non-Lorentzian Geometry in Holography'' (grant number UF160197). The work of EH is supported by the Royal Society Research Grant for Research Fellows 2017 “A Universal Theory for Fluid Dynamics” (grant number RGF$\backslash$R1$\backslash$180017).

\appendix

\section{Gauge structure of the $1/c^2$ expanded Lorentzian geometry}
\label{sec:gauge-structure}

In this appendix, we derive the gauge transformations that appear at each order in the string $1/c^2$ expansion of a $(d+2)$-dimensional Lorentzian geometry that we considered in Section~\ref{sec:string-1/c^2-exp}. We adopt the same strategy as in~\cite{Hansen:2020pqs}, where the analogous particle $1/c^2$ expansion of Lorentzian geometry was developed.

It is useful to define Lorentzian vielbeine $\hat{E}_M{^{\bar A}}$ with $\bar A = 0,1,\dots,d+1$ via
\begin{equation}
    \label{eq:vielbeindef}
    \begin{split}
        G_{MN} &= \eta_{\bar A\bar B}\hat E_M{^{\bar A}}\hat E_N{^{\bar B}}\,,\\ 
        \hat{E}_M{^{\bar A}} &= cT_M{^A}\delta_A^{\bar A} + \mathcal{E}_M{^{A'}}\delta_{A'}^{\bar A}\,.       
    \end{split}
\end{equation}
The Lorentzian vielbein $\hat{E}_M{^{\bar A}} = (cT_M{^A},\mathcal E_M{^{A'}})$ transforms under diffeomorphisms and local Lorentz transformations $\hat L^{\bar A}{_{\bar B}} = (L^A{_B}, L^{A'}{_{B'}},L^A{_{B'}} )$. We can write the local Lorentz transformations as 
 \be 
 \hat L^{\bar A}{_{\bar B}} &=&  \delta^{\bar A}_A\delta^B_{\bar B} L \varepsilon^A{_B}+c^{-1}\delta^{\bar A}_{A'}\delta^B_{\bar B}L^{A'}{_B}+c^{-1}\delta^{B'}_{\bar B}\delta^{\bar A}_{A}L^A{_{B'}}+\delta^{\bar A}_{A'}\delta^{B'}_{\bar B}L^{A'}{_{B'}}\,,\label{eq:local-Lorentz-transformation-parameter}
 \ee 
 where factors of $c$ above have been chosen to respect their appearance in \eqref{eq:vielbeindef}, and where we wrote $L^A{_B} = L\varepsilon^A{_B} $. Using the expression \eqref{eq:local-Lorentz-transformation-parameter} for the local Lorentz transformations, the transformation property $\delta  \hat{E}_M{^{\bar A}} =\pounds_\Xi \hat{E}_M{^{\bar A}} + \hat L^{\bar A}{_{\bar B}}\hat{E}_M{^{\bar B}}$ for the Lorentzian vielbein gives rise to
\begin{subequations}
\be 
\delta T_M{^A}&=& \pounds_\Xi T_M{^A} + L\varepsilon^A{_B}T_M{^B} + c^{-2}L^A{_{B'}}\mathcal{E}_M{^{B'}}\label{eq:trafo-of-T}\,,\\
\delta \mathcal{E}_M{^{A'}}&=& \pounds_\Xi \mathcal{E}_M{^{A'}} + L^{A'}{_{B'}}\mathcal{E}_M{^{B'}}+{L}^{A'}{_B}T_M{^B}\label{eq:trafo-of-cal-E}\,,
\ee 
\end{subequations}
where we used \eqref{eq:vielbeindef}. Note that 
\be 
L^{A'}{_B} = -\delta^{A'B'}\eta_{AB}L^{A}{_{B'}}\,.
\ee 
In terms of these vielbeine, $\Pi^\perp_{MN}$ becomes
\be 
\Pi^\perp_{MN} = \delta_{A'B'} \mathcal{E}_M{^{A'}} \mathcal{E}_N{^{B'}}\,,
\ee 
where the primed indices range over $A'=2,\dots,d+1$, and where we expand
\be 
\mathcal{E}_M{^{A'}} = E_M{^{A'}} + c^{-2}\pi_M{^{A'}} + \mathcal{O}(c^{-4})\,,\label{eq:expansiontransversevielbein}
\ee 
which means that $\Pi^\perp_{MN}=H^\perp_{MN}+c^{-2}\phi^\perp_{MN}+\mathcal{O}(c^{-4})$ where
\be 
H^\perp_{MN} = \delta_{A'B'}E_M{^{A'}}E_N{^{B'}}\,,\qquad \phi^\perp_{MN} = 2\delta_{A'B'}E_{(M}{^{A'}}\pi_{N)}{^{B'}}\,.\label{eq:HperpAndPhiperpVielbein}
\ee 
Expanding the gauge parameters above according to
\be 
\begin{aligned}
\Xi^M &= \xi^M + c^{-2}\zeta^M + c^{-4}\vartheta^M + \mathcal{O}{(c^{-6})}\,,\\
L &= \Lambda + c^{-2}\tilde\sigma + c^{-4}\Lambda_{(4)}+\mathcal{O}{(c^{-6})}\,,\\
L^{A'}{_{B'}} &= \lambda^{A'}{_{B'}} + c^{-2}\lambda_{(2)}^{A'}{_{B'}} + \mathcal{O}{(c^{-4})}\,,\\
L^{A'}{_{B}} &= \lambda^{A'}{_{B}} + c^{-2}\lambda_{(2)}^{A'}{_{B}} + \mathcal{O}{(c^{-4})}\,,\\
L^{A}{_{B'}} &= \lambda^{A}{_{B'}} + c^{-2}\lambda_{(2)}^{A}{_{B'}} + \mathcal{O}{(c^{-4})}\,,
\end{aligned}
\ee 
the expansions \eqref{eq:type-II-SNC-expansions} and \eqref{eq:expansiontransversevielbein} for $T_M{^A}$ and $\mathcal{E}_M{^{A'}}$, respectively, combined with the transformation properties in \eqref{eq:trafo-of-T} and \eqref{eq:trafo-of-cal-E} yields
\begin{subequations}
\be 
\delta \tau_M{^A} &=&\pounds_\xi\tau_M{^A} + \Lambda\varepsilon^A{_B}\tau_M{^B}\,,\label{eq:SNCtrafo1}\\
\delta E_M{^{A'}} &=& \pounds_\xi  E_M{^{A'}} + \lambda^{A'}{_{B'}} E_M{^{B'}} +  \lambda^{A'}{_{B}}\tau_M{^B}\,,\\
\delta m_M{^A} &=& \pounds_\xi m_M{^A} + \pounds_\zeta \tau_M{^A}+\Lambda\varepsilon^A{_B}m_M{^B} + \tilde\sigma\varepsilon^A{_B}\tau_M{^B}+\lambda^A{_{B'}}E_M{^{B'}}\,,\\
\delta\pi_M{^{A'}} &=& \pounds_\xi \pi_M{^{A'}} + \pounds_\zeta E_M{^{A'}} + \lambda^{A'}{_{B'}}\pi_M{^{B'}} +\lambda_{(2)}^{A'}{_{B'}}E_M{^{B'}} + \lambda^{A'}{_{B}}m_M{^{B}} + \lambda_{(2)}^{A'}{_{B}}\tau_M{^{B}}\,,\nn\\\\
\delta B_M{^A} &=& \pounds_\xi B_M{^A} + \pounds_\zeta m_M{^A} + \pounds_\vartheta \tau_M{^A} + \Lambda\varepsilon^A{_B}B_M{^B} + \tilde\sigma\varepsilon^A{_B}m_M{^B} + \Lambda_{(4)}\varepsilon^A{_B}\tau_M{^B}\nn\\
&&+\lambda^A{_{B'}}\pi_M{^{B'}}+\lambda_{(2)}^A{_{B'}}E_M{^{B'}}\label{eq:SNCtrafofinal}\,.
\ee 
\end{subequations}
These match the standard SNC transformations (for the fields that exist in SNC geometry), except for the transformation of $m_M{^A}$, which is analogous to the situation for type II torsional Newton--Cartan geometry~\cite{Hansen:2020pqs}. To make the transformations match, we decompose the subleading diffeomorphisms as
\be 
\zeta^M = \sigma^A\tau^M{_A} + E^M{_{A'}}\zeta^{A'}\,,
\ee 
and impose the strong foliation constraint \eqref{eq:Foliation1} written in the form
\be 
D_{[M}\tau_{N]}{^A} = 0\,,\label{eq:Foliation11}  
\ee 
where $D$ includes the $SO(1,1)$ ``spin connection'' $\omega_M\varepsilon^{AB}$, i.e., 
\begin{equation}
    D_M \tau_N{^A} = \D_M \tau_N{^A} - \omega_M\varepsilon^A{_B}\tau_N{}^B\,.
\end{equation}
We can rewrite the Lie derivative that appears in the transformation of $m_M{^A}$ as
\be
\begin{aligned}
\pounds_\zeta \tau_M{^A} &= \D_M\left(\zeta^N\tau_N{^A} \right)+2\zeta^ND_{[N}\tau_{M]}{^A} - 2\zeta^N \varepsilon^A{_B}\omega_{[M}\tau_{N]}{^B}\\
&= D_M \sigma^A +2\zeta^ND_{[N}\tau_{M]}{^A} + \zeta^N \varepsilon^A{_B}\omega_N\tau_{M}{^B}\,.
\end{aligned}
\ee
Defining
\be 
\sigma = \tilde\sigma + \zeta^N\omega_N
\ee 
the transformation of $m_M{^A}$ precisely reduces to that of SNC if \eqref{eq:Foliation1} holds, namely
\be 
\delta m_M{^A} &=& \pounds_\xi m_M{^A} + D_M\sigma^A +\Lambda\varepsilon^A{_B}m_M{^B} + \sigma\varepsilon^A{_B}\tau_M{^B}+\lambda^A{_{B'}}E_M{^{B'}}\,,
\ee 
and using $\lambda^{A}{_{B'}} = -\lambda_{B'}{^A}$, this becomes Eq. (A.6) of \cite{Harmark:2018cdl}.

\section{Nonrelativistic expansion of energy and momenta}\label{app:energymomentumexp}

In this appendix we discuss how derivatives of the Lagrangian can be expanded. Let $f$ be some function of the embedding coordinates $X^M$. In general, if $f(X) = f_{\text{LO}}(x) + \epsilon f_{\text{NLO}}(x,y)+ \epsilon^2 f_{\text{NNLO}}(x,y,z) + \mathcal{O}(\epsilon^3)$ is a function of $X = x + \epsilon y + \epsilon^2 z + \mathcal{O}(\epsilon^3)$, where $\epsilon$ is an expansion parameter, we find that
\be 
f'(X) &=& \lim\limits_{t\rightarrow 0}\frac{f(X+t) - f(X)}{t} \nn\\
&=& \lim\limits_{t\rightarrow 0}\frac{f(x+t + \epsilon y + \epsilon^2 z + \cdots ) - f(X+ \epsilon y + \epsilon^2 z + \cdots)}{t}\nn\\
&=& \lim\limits_{t\rightarrow 0}\frac{1}{t}\Big(f_{\text{LO}}(x+t) + \epsilon f_{\text{NLO}}(x+t,y)+ \epsilon^2 f_{\text{NNLO}}(x+t,y,z)\nn\\
&&\qquad- f_{\text{LO}}(x) - \epsilon f_{\text{NLO}}(x,y) - \epsilon^2 f_{\text{NNLO}}(x,y,z) + \mathcal{O}(\epsilon^3)  \Big)\nn\\
&=& \pd{f_{\text{LO}}(x)}{x} + \epsilon \pd{f_{\text{NLO}}(x,y)}{x} + \epsilon^2\pd{f_{\text{NNLO}}(x,y,z)}{x} + \mathcal{O}(\epsilon^3)\,.
\ee 
We can use a similar argument to show that the relativistic energy acquires an expansion of the form
\be
E &=& -\oint d \sigma^1\pd{\mathcal{L}_{\text{P}}}{\D_0X^t} \nn\\
&=& -c^2\oint d \sigma^1\pd{\mathcal{L}_{\text{P-LO}}}{\D_0x^t} - \oint d \sigma^1\pd{\mathcal{L}_{\text{P-NLO}}}{\D_0x^t} - c^{-2}\oint d \sigma^1\pd{\mathcal{L}_{\text{P-NNLO}}}{\D_0x^t} +\mathcal{O}(c^{-4})\nn\\
&=& c^2 E_{\text{LO}} + E_{\text{NLO}} + c^{-2}E_{\text{NNLO}} +\mathcal{O}(c^{-4})\,,
\ee 
while the relativistic momentum density in the $v$-direction becomes
\be
\label{eq:ExpansionOfvMom}
\begin{aligned}
P_v &=  \pd{\mathcal{L}_{\text{P}}}{\D_0X^v} = c^2\pd{\mathcal{L}_{\text{P-LO}}}{\D_0x^v} + \pd{\mathcal{L}_{\text{P-NLO}}}{\D_0x^v} + c^{-2}\pd{\mathcal{L}_{\text{P-NNLO}}}{\D_0x^v} + \mathcal{O}(c^{-4})\\
&= c^2p_{(-2)v} + p_{(0)v} + c^{-2}p_{(2)v}+ \mathcal{O}(c^{-4})\,,
\end{aligned}
\ee 
and, finally, the relativistic momentum density in the transverse space expands according to
\be 
\begin{aligned}
P_i &=  \pd{\mathcal{L}_{\text{P}}}{\D_0X^i} =  \pd{\mathcal{L}_{\text{P-NLO}}}{\D_0x^i} + c^{-2}\pd{\mathcal{L}_{\text{P-NNLO}}}{\D_0x^i} + \mathcal{O}(c^{-4})\\
&= p_{(0)i} + c^{-2}p_{(2)i}+ \mathcal{O}(c^{-4})\,,
\end{aligned}
\ee 
where we used that the P-LO Lagrangian does not depend on $x^i$.

\section{Noether Charges}
\label{sec:noether-details}
In this appendix, we provide explicit expressions for the Noether charges and their brackets that we discussed in Section~\ref{sec:target-space-symmetries}.
\subsection{The next-to leading order charges and their algebra}
At NLO, the full list of charges is
\begin{subequations}
\be 
\mathcal{P}_{(-2)A} &=&  T_{\text{eff}}\oint d \sigma^1\,\dot x_A\,,\label{eq:Charge-1}\\
\mathcal{P}_{(0)A} &=& T_{\text{eff}}\oint d \sigma^1\,\dot y_A\,,\\
\mathcal{J}_{(-2)AB} &=&   2\oint d \sigma^1\,x_{[A}p_{(-2)B]}\,,\\
\mathcal{J}_{(0)AB} &=&   2\oint d \sigma^1\,(x_{[A}p_{(0)B]} + y_{[A}p_{(-2)B]})\,,\\
\mathcal{P}_{(0)i} &=&  T_{\text{eff}}\oint d \sigma^1\,\dot x_i\,,\\
\mathcal{J}_{(0)ij} &=&   2\oint d \sigma^1\,x_{[i}p_{(0)j]}\,,\\
\mathcal{J}_{(0)Ai} &=&   \oint d \sigma^1\,(x_Ap_{(0)i} - x_ip_{(-2)A})\,.\label{eq:Charge-final}
\ee 
\end{subequations}
Using the NLO Poisson brackets~\eqref{eq:NLO-brackets}, we get 
\begin{subequations}
\be 
\{ \mathcal{P}_{(0)i},\mathcal{J}_{(0)Aj} \} &=& \delta_{ij}\mathcal{P}_{(-2)A}\,,\\
\{ \mathcal{J}_{(0)Ai},\mathcal{J}_{(0)Bj} \} &=& \delta_{ij}\mathcal{J}_{(-2)AB}\,,\\
\{\mathcal{J}_{(0)AB},\mathcal{P}_{(-2)C} \} &=& 2\eta_{C[A}\mathcal{P}_{(-2)B]}\,,\\
\{\mathcal{J}_{(-2)AB},\mathcal{P}_{(0)C} \} &=& 2\eta_{C[A}\mathcal{P}_{(0)B]}\,,\\
\{\mathcal{J}_{(0)AB},\mathcal{P}_{(0)C} \} &=& 2\eta_{C[A}\mathcal{P}_{(0)B]}\,,\\
\{ \mathcal{J}_{(0)Ai},\mathcal{P}_{(0)B} \} &=& \eta_{AB}\mathcal{P}_{(0)i}\,,\\
\{\mathcal{J}_{(0)ij},\mathcal{P}_{(0)k} \} &=& 2\delta_{k[i}\mathcal{P}_{(0)j]}\,,\\
\{\mathcal{J}_{(0)AB},\mathcal{J}_{(0)Ci} \} &=& 2\eta_{C[A}\mathcal{J}_{(0)B]i}\,,\\
\{\mathcal{J}_{(0)ij}, \mathcal{J}_{(0)Ak} \} &=& 2\mathcal{J}_{(0)A[j}\delta_{i]k}\,,\\
\{\mathcal{J}_{(0)ij}, \mathcal{J}_{(0)kl} \} &=& \delta_{ik}\mathcal{J}_{(0)jl} -\delta_{il}\mathcal{J}_{(0)jk} - \delta_{jk}\mathcal{J}_{(0)il} + \delta_{jl}\mathcal{J}_{(0)ik}\,.
\ee 
\end{subequations}
This matches the string Bargmann algebra $\mathfrak{q}_1$.

\subsection{The next-to-next-to leading order charges and their algebra}
At NNLO, we have the following charges in addition to those listed in~\eqref{eq:Charge-1}--\eqref{eq:Charge-final}
\begin{subequations}
\be 
\mathcal{P}_{(2)A} &=&  T_{\text{eff}}\oint d \sigma^1\,\dot z_A\,,\\
\mathcal{J}_{(2)AB} &=&   2\oint d \sigma^1\,(x_{[A}p_{(2)B]} + y_{[A}p_{(0)B]} + z_{[A}p_{(-2)B]})\,,\\
\mathcal{P}_{(2)i} &=&  T_{\text{eff}}\oint d \sigma^1\,\dot y_i\,,\\
\mathcal{J}_{(2)ij} &=&   2\oint d \sigma^1\,(x_{[i}p_{(2)j]} + y_{[i}p_{(0)j]})\,,\\
\mathcal{J}_{(2)Ai} &=&  \oint d \sigma^1\,(x_Ap_{(2)i} - x_ip_{(0)A} + y_Ap_{(0)i} - y_ip_{(-2)A})\,.\label{eq:Charge-final-1}
\ee 
\end{subequations}
Using the NNLO brackets~\eqref{eq:NNLO-brackets}, this leads to
\begin{align} 
\begin{aligned}
\{ \mathcal{P}_{(2)i},\mathcal{J}_{(0)Aj} \} &= \delta_{ij}\mathcal{P}_{(-2)A}\,,& 
\{ \mathcal{P}_{(0)i},\mathcal{J}_{(2)Aj} \} &= \delta_{ij}\mathcal{P}_{(-2)A}\,,\\
\{ \mathcal{P}_{(2)i},\mathcal{J}_{(2)Aj} \} &= \delta_{ij}\mathcal{P}_{(0)A}\,,&
\{ \mathcal{J}_{(2)Ai},\mathcal{J}_{(0)Bj} \} &= \delta_{ij}\mathcal{J}_{(-2)AB}\,,\\
\{\mathcal{J}_{(2)AB},\mathcal{P}_{(2)C} \} &= 2\eta_{C[A}\mathcal{P}_{(2)B]}\,,& \{\mathcal{J}_{(2)AB},\mathcal{P}_{(0)C} \} &= 2\eta_{C[A}\mathcal{P}_{(0)B]}\,,\\
\{\mathcal{J}_{(2)AB},\mathcal{P}_{(-2)C} \} &= 2\eta_{C[A}\mathcal{P}_{(-2)B]}\,,&
\{\mathcal{J}_{(0)AB},\mathcal{P}_{(2)C} \} &= 2\eta_{C[A}\mathcal{P}_{(0)B]}\,,\\
\{\mathcal{J}_{(0)AB},\mathcal{P}_{(0)C} \} &= 2\eta_{C[A}\mathcal{P}_{(-2)B]}\,,&
\{\mathcal{J}_{(-2)AB},\mathcal{P}_{(2)C} \} &= 2\eta_{C[A}\mathcal{P}_{(-2)B]}\,,\\
\{\mathcal{J}_{(2)ij},\mathcal{P}_{(2)k} \} &= 2\delta_{k[i}\mathcal{P}_{(2)j]}\,,&
\{\mathcal{J}_{(2)ij},\mathcal{P}_{(0)k} \} &= 2\delta_{k[i}\mathcal{P}_{(0)j]}\,,\\
\{\mathcal{J}_{(0)ij},\mathcal{P}_{(2)k} \} &= 2\delta_{k[i}\mathcal{P}_{(0)j]}\,,&
\{\mathcal{P}_{(2)A},\mathcal{J}_{(2)Bi} \} &= -\eta_{AB}\mathcal{P}_{(2)i}\,,\\
\{\mathcal{P}_{(0)A},\mathcal{J}_{(2)Bi} \} &= -\eta_{AB}\mathcal{P}_{(0)i}\,,&
\{\mathcal{P}_{(2)A},\mathcal{J}_{(0)Bi} \} &= -\eta_{AB}\mathcal{P}_{(0)i}\,,\\
\{\mathcal{P}_{(-2)A},\mathcal{J}_{(0)Bi} \} &= -\eta_{AB}\mathcal{P}_{(2)i}\,,&
\{ \mathcal{J}_{(2)Ai},\mathcal{J}_{(2)BC} \} &= -2\eta_{A[B}\mathcal{J}_{(2)C]i}\,,\\
\{ \mathcal{J}_{(2)Ai},\mathcal{J}_{(0)BC} \} &= -2\eta_{A[B}\mathcal{J}_{(0)C]i}\,,&
\{ \mathcal{J}_{(0)Ai},\mathcal{J}_{(2)BC} \} &= -2\eta_{A[B}\mathcal{J}_{(0)C]i}\,,\\
\{ \mathcal{J}_{(2)Ai},\mathcal{J}_{(2)jk} \} &= -2\delta_{i[j\vert}\mathcal{J}_{(2)A\vert k]}\,,&
\{ \mathcal{J}_{(2)Ai},\mathcal{J}_{(0)jk} \} &= -2\delta_{i[j\vert}\mathcal{J}_{(0)A\vert k]}\,,&\\
\{ \mathcal{J}_{(2)ij},\mathcal{J}_{(2)kl} \} &= 4\delta_{{\color{red}[}j{\color{blue} [} k} \mathcal{J}_{(2)l{\color{blue} ]}i{\color{red} ]}}^{(0)}\,,&
\{ \mathcal{J}_{(2)ij},\mathcal{J}_{(0)kl} \} &= 4\delta_{{\color{red}[}j{\color{blue} [} k} \mathcal{J}_{(0)l{\color{blue} ]}i{\color{red} ]}}^{(0)}\,,\\
\{ \mathcal{J}_{(2)Ai},\mathcal{J}_{(2)Bj} \} &= \delta_{ij}\mathcal{J}_{(0)AB} + \eta_{AB}\mathcal{J}_{(0)ij}\,,& \{ \mathcal{J}_{(0)Ai},\mathcal{J}_{(2)jk} \} &= -2\delta_{i[j\vert}\mathcal{J}_{(0)A\vert k]}\,.
\end{aligned}
\end{align}
This is the algebra $\mathfrak{q}_2$.

\addcontentsline{toc}{section}{\refname}

\providecommand{\href}[2]{#2}\begingroup\raggedright\endgroup

\end{document}